\documentclass[
 reprint,
 superscriptaddress,
 amsmath,
 amssymb,
 aps,
 prd,
 longbibliography,
 floatfix,
 letterpaper,
 10pt,
 preprintnumbers,
 nofootinbib
]{revtex4-1}

\usepackage[usenames,dvipsnames,svgnames,table]{xcolor}
\usepackage{natbib}
\usepackage{graphicx}
\usepackage{dcolumn}
\usepackage{bm}
\usepackage{longtable}
\usepackage{xcolor}
\usepackage[english]{babel}
\usepackage{helvet}
\usepackage{microtype}
\usepackage[pdftex,
  pdftitle={A Bumpy Start to a Smooth Ride: Onset of Inflation amid Backreaction from Inhomogeneities},
  pdfauthor={Jolyon K. Bloomfield, Patrick Fitzpatrick, Kiriakos Hilbert and David I. Kaiser},
  bookmarks,bookmarksopen=false,
  pdfstartview={FitH},
  linktocpage=true
]{hyperref}
\hypersetup{
    colorlinks=true,
    linkcolor=blue,
    filecolor=magenta,      
    urlcolor=cyan,
}

\newcommand{\beq}{\begin{equation}}
\newcommand{\eeq}{\end{equation}}

\newcommand{\numsim}{34}
\newcommand{\numsimflucts}{32}

\begin{document}

\title{\texorpdfstring{A Bumpy Start to a Smooth Ride: \\
Onset of Inflation amid Backreaction from Inhomogeneities}{A Bumpy Start to a Smooth Ride: Onset of Inflation amid Backreaction from Inhomogeneities}}

\author{Jolyon K. Bloomfield}
\email{jolyon@mit.edu}
\affiliation{Department of Physics, Massachusetts Institute of Technology, Cambridge, Massachusetts 02139 USA}

\author{Patrick Fitzpatrick}
\email{fitzppat@mit.edu}
\affiliation{Department of Physics, Massachusetts Institute of Technology, Cambridge, Massachusetts 02139 USA}

\author{Kiriakos Hilbert}
\email{khilbert@mit.edu}
\affiliation{Department of Physics, Massachusetts Institute of Technology, Cambridge, Massachusetts 02139 USA}

\author{David I. Kaiser}
\email{dikaiser@mit.edu}
\affiliation{Department of Physics, Massachusetts Institute of Technology, Cambridge, Massachusetts 02139 USA}

\preprint{MIT-CTP/5111}

\date{\today}

\begin{abstract}
We analyze the onset of inflation for a simple single-field model when the system begins with significant inhomogeneities on length-scales shorter than the initial Hubble radius. We incorporate certain nonlinear interactions among the coupled degrees of freedom by using the nonperturbative Hartree approximation.
Consistent with recent, more computationally intensive numerical-relativity studies, we find inflation to be robust for large-field models, even when the system begins with significant structure on sub-Hubble scales. We consider the space of initial conditions $(\varphi (t_0), \dot{\varphi} (t_0))$, where $\varphi$ is the vacuum expectation value of the quantized field that drives inflation. Although some regions of $(\varphi (t_0), \dot{\varphi} (t_0))$ that would have yielded sufficient inflation in the absence of inhomogeneities fail to do so when backreaction from inhomogeneities is incorporated, an equal volume of such regions succeeds in producing sufficient inflation which did not do so in the absence of inhomogeneities. For large-field models, in other words, the total volume of the space of initial conditions $(\varphi (t_0), \dot{\varphi} (t_0))$ that yields sufficient inflation is {\it conserved} when we incorporate nonlinear backreaction from inhomogeneities, compared to the case in which inhomogeneities are neglected.

\end{abstract}

\maketitle

\section{Introduction}

Several observable features of our Universe today, including its spatial flatness, large-scale homogeneity, and the specific pattern of temperature anisotropies in the cosmic microwave background radiation, are readily explained if our Universe underwent a brief phase of early-universe inflation. (For reviews, see Refs.~\cite{guthkaiser_05,Mukhanov,bassett_05,liddle+lyth_09,MartinRingevalVennin2014,guth+al_14,linde_15,baumann+mcallister_15,MartinTrick}.) The question remains, however, whether the onset of inflation itself is fairly generic, or whether inflation requires fine-tuned initial conditions. (For reviews, see Refs.~\cite{GoldwirthPiranReview,brandenberger_16}.) 

Many analyses have highlighted the effectiveness of an inflationary attractor: at least for single-field models, if one neglects spatial inhomogeneities,
then one can show that large classes of models will flow into inflation, across broad ranges of initial conditions and couplings, even for regions of phase space in which the system is initially dominated by the field's kinetic (rather than potential) energy. The attractor behavior is especially effective in large-field models with sufficiently flat regions of the potential, for which $\epsilon_n \equiv (M_{\rm pl} \, \partial_\phi)^n \ln V (\phi) \ll 1$, where $M_{\rm pl} \equiv 1 / \sqrt{ 8 \pi G} = 2.43 \times 10^{18}$ GeV is the reduced Planck mass \cite{SalopekBond1990,KungBrandenberger1990,Muller1990,LiddleParsonsBarrow1994,Mukhanov,Kinney2007,RemmenCarroll2013,MartinRingevalVennin2014,AzharKaiser2018,Chowdhury:2019otk}.

These analyses neglected perturbations in the spacetime metric, however, thereby leaving open the question of whether the onset of inflation remains robust even in the presence of significant inhomogeneities. Pioneering efforts began as early as the mid-1980s to numerically simulate the onset of inflation amid such inhomogeneities \cite{AlbrechtBrandenbergerMatzner1985,AlbrechtBrandenbergerMatzner1987,Matzner1987,Feldman1989a,Feldman1989b,GoldwirthPiran1989,GoldwirthPiranPRL,Goldwirth1991,Matzner1991,Matzner1993,GoldwirthPiranReview,brandenberger_16}. Yet these early studies
did not yield an unambiguous answer to the question of whether the onset of inflation required an initial patch of spacetime to be smooth across a characteristic length-scale as large as (or perhaps considerably larger than) the Hubble radius. (See also Refs.~\cite{Trodden2000,ISL2013,Trodden2015}.)

Recently two groups have conducted studies of the onset of inflation that implement full $(3 + 1)$-dimensional numerical relativity; each found that large-field inflation (in single-field models) is strongly robust, even amid significant initial inhomogeneities \cite{East:2015ggf,Clough:2016ymm,Clough:2017efm}. (They also found that the onset of small-field inflation can be significantly more robust than previous semi-analytic treatments had suggested, although small-field models generically require more finely tuned initial conditions than large-field models do; see also Ref.~\cite{Marsh2018}.) In particular, each group found that large-field inflation would start even if the spacetime initially had no Hubble-sized smooth patches. 
Given the significant computational resources required to simulate the full Einstein field equations in $(3 + 1)$-dimensions, however, the latest studies have been restricted, to date, to a handful of specific forms of the potential $V (\phi)$ and to limited regions of parameter space
\cite{East:2015ggf,Clough:2016ymm,Clough:2017efm}. 

In this work, we develop a complementary numerical approach. Although by design our approach cannot capture the full range of gravitational effects that could (in principle) affect the onset of inflation, it incorporates certain nonlinear interactions while remaining significantly more efficient than the full-bore numerical relativity studies. It therefore facilitates the study of a wide range of models, across broad regions of phase space and parameter space, while tracking the evolution of as many as $n \sim {\cal O} (10^2)$ coupled modes, spanning a wide range of length-scales. (It also offers a feasible means of generalizing to multifield models; see also Ref.~\cite{EastherMultifield}.) For the case we have investigated so far --- a single-field model that yields large-field inflation --- our numerical results are consistent with the fully relativistic simulations of Refs.~\cite{East:2015ggf,Clough:2016ymm,Clough:2017efm}.  

We work to linear order in metric perturbations, $\Psi (x^\mu)$, but incorporate nonlinear interactions among the field fluctuations, $\delta \phi (x^\mu)$, by adopting the (nonperturbative) Hartree approximation. We therefore capture some of the nonlinear backreaction effects that are absent in studies that work only to linear order in fluctuations.
Whereas an ordinary quantum loop expansion corresponds to a power series in the coupling constant $\lambda$ and in $\hbar$, in the Hartree approximation we resum a particular class of Feynman diagrams (the so-called ``cactus" or ``daisy" diagrams) to all orders. Hence the Hartree approximation moves beyond perturbative series in $\lambda$, $\hbar$, or $\delta \phi$. (See, e.g., Refs.~\cite{Jackiw:1974cv,Dolan:1973qd,Cornwall:1974vz,Chang:1975dt,Boyanovsky:1993xf}.)

By working only to linear order in $\Psi (x^\mu)$, our approach cannot capture extreme gravitational phenomena like the formation of trapped surfaces or the collapse of local regions to form black holes. But the fully relativistic simulations in Refs.~\cite{East:2015ggf,Clough:2016ymm,Clough:2017efm} confirm that the formation of such black holes does not interrupt the overall flow of the system into inflation: regions outside of the collapse continue to expand, and the density of any such collapsed regions quickly dilutes. Moreover, as discussed in Ref.~\cite{Clough:2016ymm}, no black hole that forms from pre-inflationary overdensities can grow so large as to encompass the entire Hubble sphere, so even the largest black holes that might form fail to disrupt the overall flow into inflation; see also Ref.~\cite{Kleban:2016sqm}. Therefore our  simplified treatment can complement the more computationally intensive studies.

Another feature of our approach is that we treat the origin of inhomogeneities as ultimately quantum-mechanical. One of the most significant achievements of inflationary cosmology is to provide a first-principles description of the primordial inhomogeneities that seed large-scale structure as arising from quantum-mechanical fluctuations of matter fields during inflation \cite{guthkaiser_05,Mukhanov,bassett_05,liddle+lyth_09,MartinRingevalVennin2014,guth+al_14,linde_15,baumann+mcallister_15,MartinTrick}. We aim to analyze the onset of inflation in a comparable way;
after all, if there were a priori reasons to expect some pattern of classical inhomogeneities at arbitrarily early times, we wouldn't need a mechanism like inflation to account for large-scale structure. Given the high energies involved and the possibility of nonlinear interactions among the quantized matter fields at very early times, we consider scenarios in which the initial state of the quantum fluctuations departs significantly from the minimum-energy (Bunch-Davies) state. The larger quantum fluctuations seed significant spatial inhomogeneities and affect the dynamics of the system.

Our goal in this paper is to establish our formalism, introduce details of our numerical approach, and apply our system to a simple model; we defer detailed applications to a wider range of models, including both large-field and small-field models, single-field and multifield cases, to future work. In Section \ref{sec:EOMs} we derive the coupled equations of motion for gravitational and matter degrees of freedom within the Hartree approximation. In Section \ref{sec:Parameters} we discuss initial conditions for the field fluctuations and our UV regularization scheme. 
Section \ref{sec:Results} presents our main numerical results, indicating that across broad regions of phase space, a simple model like $V (\phi) = \lambda \phi^4 / 4$ will flow into the inflationary attractor even for initial inhomogeneities as large as $\vert \Psi (x^\mu) \vert \lesssim 0.5$. In particular, we find a shift in the regions of phase space that support sufficient inflation (compared to the case in which we neglect all perturbations), but no reduction in the total volume of the space of initial conditions $(\varphi (t_0), \dot{\varphi} (t_0))$ that yields sufficient inflation, where $\varphi$ is the vacuum expectation value of the quantized field that drives inflation. (We consider $N \geq 65$ efolds of inflation to be ``sufficient" for addressing the usual shortcomings of standard hot big bang cosmology \cite{Dodelson:2003vq,Liddle:2003as,Mukhanov,bassett_05,liddle+lyth_09,MartinTrick}.) Concluding remarks follow in Section \ref{sec:Conclusions}. In Appendix \ref{sec:AppendixADiscrete} we present more details of our discrete spectrum for the fluctuations. Appendix \ref{sec:initialconditions} provides additional information about how we set initial conditions for the field fluctuations, while Appendix \ref{sec:MetricInitial} discusses the initialization of the metric perturbations. In Appendix \ref{sec:NumConvergence} we discuss various numerical convergence tests.

\section{Equations of Motion in the Hartree Approximation}
\label{sec:EOMs}

We work in $(3 + 1)$ spacetime dimensions and use units in which $c = \hbar = 1$. We consider scalar metric perturbations around a background Friedmann-Lema\^{i}tre-Robertson-Walker (FLRW) line-element, and work in longitudinal gauge,
\begin{equation}
    ds^2 = - (1 + 2 \Phi ) dt^2 + a^2 (t) (1 - 2 \Psi) h_{ij} ({\bf x}) \, dx^i \, dx^j ,
    \label{ds}
\end{equation}
where $\Phi (x^\mu)$ and $\Psi (x^\mu)$ are scalar functions. As usual, the background metric on (comoving) spatial sections may be written
\beq
h_{ij} ({\bf x} ) \, dx^i \, dx^j = \frac{ dr^2}{(1 - K r^2 )} + r^2 \left( d\theta^2 + \sin^2 \theta \, d\phi^2 \right) .
\label{hijK}
\eeq
As we will see below, within the Hartree approximation the anisotropic pressure vanishes, so that $\Phi (x^\mu) = \Psi (x^\mu)$.

As is well known, there is no straightforward generalization of longitudinal gauge beyond linear order in $\Psi (x^\mu)$. In particular, at second order in metric perturbations, first-order scalar perturbations source tensor perturbations, and so on. Gauge subtleties therefore affect any perturbative calculation that aims to move beyond ${\cal O} (\Psi)$ while incorporating only scalar degrees of freedom. (For a recent discussion, see Appendix B of Ref.~\cite{Giblin:2018ndw}.)  Hence we restrict our analysis to first order in spatially varying quantities. 

The restriction to linear order in metric perturbations, treated in longitudinal gauge, can be reasonably well motivated. As demonstrated in Ref.~\cite{Giblin:2018ndw}, corrections from a fully relativistic treatment compared to a linear treatment in longitudinal gauge typically scale as ${\cal O} (\Psi^2)$, unlike the case for synchronous gauge, in which relativistic corrections to the linearized treatment can be as large as ${\cal O} (1)$. In fact, as found in Ref.~\cite{Giblin:2018ndw}, linearized scalar metric perturbations in longitudinal gauge tend to {\it exaggerate} gravitational effects on length-scales longer than the Hubble radius, $\ell > H^{-1} (t_0)$, compared to a fully relativistic treatment. Given the prior focus on whether inflation can start amid inhomogeneities with typical length-scales $\ell \leq H^{-1} (t_0)$ \cite{Matzner1987,Matzner1991,Matzner1993,GoldwirthPiran1989,GoldwirthPiranPRL,Goldwirth1991,GoldwirthPiranReview,brandenberger_16,AlbrechtBrandenbergerMatzner1985,AlbrechtBrandenbergerMatzner1987,Feldman1989a,Feldman1989b,Trodden2000,ISL2013,Trodden2015}, we will be most interested in gravitational effects on sub-Hubble length-scales. In longitudinal gauge, meanwhile, large perturbations, $\vert \Psi (x^\mu) \vert \geq 0.5$, can lead to coordinate singularities. Hence the regime of interest is $\vert \Psi (x^\mu) \vert \lesssim {\cal O} (0.5)$.

We consider single-field models for which the action may be written
\beq
S = \int d^4 x \sqrt{-g} \left[ \frac{ M_{\rm pl}^2 }{2} R - \frac{1}{2} g^{\mu\nu} \, \partial_\mu \phi \, \partial_\nu \phi - V (\phi) \right] .
\label{action}
\eeq
Varying the action with respect to $\phi$ and $g_{\mu\nu}$ yields the coupled equations of motion,
\beq
\Box \phi - V_{, \phi} = 0
\label{eomphi}
\eeq
and
\beq
G_{\mu\nu} \equiv R_{\mu\nu} - \frac{1}{2} g_{\mu\nu} R = \frac{1}{ M_{\rm pl}^2} T_{\mu\nu} ,
\label{EFE}
\eeq
with
\beq
T_{\mu\nu} = \partial_\mu \phi \, \partial_\nu \phi - g_{\mu\nu} \left[ \frac{1}{2} g^{\alpha \beta} \, \partial_\alpha \phi \, \partial_\beta \phi + V (\phi ) \right] .
\label{Tmn}
\eeq
As usual, the covariant d'Alembertian operator is given by
\beq
\Box \phi  = \frac{1}{ \sqrt{-g}} \partial_\mu \left[ \sqrt{-g} \, g^{\mu\nu} \, \partial_\nu \phi \right] .
\label{Boxdef}
\eeq

In addition to expanding the gravitational degrees of freedom to first order in $\Psi (x^\mu)$, we also consider fluctuations in the field $\phi$. Upon quantizing the field, we have
\beq
\phi (x^\mu) \rightarrow \hat{\phi} (x^\mu) = \varphi (t) + \delta \hat{\phi} (x^\mu) ,
\label{phivarphi}
\eeq
with 
\beq
\langle \hat{\phi} (x^\mu) \rangle \equiv \langle 0 \vert \hat{\phi} (x^\mu ) \vert 0 \rangle = \varphi (t) \> , \>\> \langle \delta \hat{\phi} (x^\mu) \rangle = 0 .
\label{phivev}
\eeq
We expand the field fluctuations as
\beq
\delta \hat{\phi} (x^\mu) = \int d\tilde{\mu} (k) \left[ \delta \phi_{k \ell m} (t) \, \hat{a}_{k \ell m} \, Z_{k \ell m} ({\bf x} ) + H.c. \right] ,
\label{deltaphiexpand}
\eeq
where ``$H.c.$" denotes Hermitian conjugate, the measure $d \tilde{\mu} (k)$ is given by 
\beq
\int d\tilde{\mu} (k) = \int_0^\infty dk \sum_{\ell = 0}^\infty \sum_{m = - \ell}^\ell ,
\label{dmu}
\eeq
and $Z_{k \ell m} ({\bf x})$ is an eigenfunction of the comoving spatial Laplacian operator, 
\beq
\nabla^2 Z_{k \ell m} \equiv \frac{1}{ \sqrt{h} } \, \partial_i \left[ \sqrt{h} \, h^{ij} \partial_j Z_{k \ell m} \right] = - k^2 Z_{k \ell m} .
\label{nablaZ}
\eeq
The creation and annihilation operators obey the usual commutation relations
\beq
\begin{split}
[ \hat{a}_{k \ell m} , \hat{a}^\dagger_{k' \ell' m'} ] &= \delta (k - k') \, \delta_{\ell \ell'} \, \delta_{m m'} \> , \\
 [ \hat{a}_{k \ell m} , \hat{a}_{k' \ell' m'} ] &= [ \hat{a}^\dagger_{k \ell m} , \hat{a}^\dagger_{k' \ell' m'} ] = 0
 \end{split}
\label{acommute}
\eeq
and satisfy
\beq
\hat{a}_{k \ell m} \vert 0 \rangle = 0 \> , \> \langle 0 \vert \hat{a}^\dagger_{k \ell m} = 0 
\label{avac}
\eeq
for all $k$, $\ell$, and $m$. As we will see, upon expanding the field $\phi$ as in Eq.~(\ref{phivarphi}) and working to linear order in $\Psi$, Eqs.~(\ref{eomphi}) and (\ref{EFE}) will couple $\Psi (x^\mu)$ to terms linear in $\delta \hat{\phi} (x^\mu)$. To remain consistent, we therefore quantize $\Psi (x^\mu) \rightarrow \hat{\Psi} (x^\mu)$. Because we are working with single-field models, we only require a single set of operators $\hat{a}_{k \ell m}$ and $\hat{a}^\dagger_{k \ell m}$. We expand $\hat{\Psi}$ as
\beq
\hat{\Psi} (x^\mu) = \int d\tilde{\mu} (k) \left[ \Psi_{k \ell m} (t) \, \hat{a}_{k \ell m} \, Z_{k \ell m} ({\bf x} ) + H.c. \right] .
\label{Psihat}
\eeq

Given our interest in inflationary models, we focus on weakly coupled systems. For such systems, one expects the $S$ matrix to be dominated by forward-scattering processes. Hard-scattering events, which impart large transverse momenta to the scattered particles, should be relatively rare, such that most of the time the particles will emerge from an interaction with (nearly) the same momentum as they had prior to the interaction. The dominant processes, in other words, involve particles propagating along a given trajectory, but moving with a modified, effective mass induced by the (self-)interactions. In such cases, we may approximate the effects of nonlinear self-interactions by calculating a dressed propagator: we select the dominant subset of Feynman diagrams at each order of perturbation theory that contribute to the effective mass, and sum all members of that subclass to all orders, while neglecting the other terms that would appear in a full expansion of the $S$ matrix. (See, e.g., Section 4.7 of Ref.~\cite{Mattuck}.) Within the Hartree approximation, we construct the dressed propagator by performing an infinite resummation of the so-called ``cactus" or ``daisy" diagrams  \cite{Jackiw:1974cv,Dolan:1973qd,Cornwall:1974vz,Chang:1975dt,Boyanovsky:1993xf}.

The Hartree approximation becomes exact for an $O({\cal N})$-symmetric model of ${\cal N}$ interacting scalar fields in the limit ${\cal N} \rightarrow \infty$. In that case, the amplitudes associated with the set of Feynman diagrams picked out for resummation remain parametrically larger than all other terms in the expansion of the $S$ matrix, order by order \cite{Jackiw:1974cv,Dolan:1973qd,Cornwall:1974vz}. Yet even in the case ${\cal N} = 1$, the diagrams that contribute to the Hartree approximation dominate among the contributions to the effective mass at a given order in perturbation theory, and hence Hartree is often used to incorporate nonperturbative effects even for ${\cal N} = 1$, as we will do here. (See, e.g., Refs.~\cite{Chang:1975dt,Boyanovsky:1993xf,Khlebnikov:1996wr,Bassett:2000ha}.)

Operationally, this means that within the equations of motion, all terms that are higher order in the field fluctuations, of the form $(\delta \hat{\phi} )^n$ for $n \geq 2$, are replaced by \cite{Cornwall:1974vz,Chang:1975dt,Boyanovsky:1993xf}
\beq
\begin{split}
(\delta \hat{\phi} )^2 &\rightarrow \langle (\delta \hat{\phi} )^2 \rangle , \\
(\delta \hat{\phi} )^3 &\rightarrow 3 \langle (\delta \hat{\phi} )^2 \rangle \, \delta \hat{\phi} , \\
(\delta \hat{\phi} )^4 &\rightarrow 3 \langle (\delta \hat{\phi} )^2 \rangle^2 , \\
(\delta \hat{\phi} )^5 &\rightarrow 15 \langle (\delta \hat{\phi} )^2 \rangle^2 \, \delta \hat{\phi} ,
\end{split}
\label{HartreeSubs}
\eeq
where the particular coefficients on the right-hand side arise from the combinatorics of the various Wick contractions. (One may continue in a similar manner for $(\delta \hat{\phi} )^n$ with $n > 5$, though for the particular model of interest to us here, the higher-order terms will not be relevant.) Because the background spacetime around which we are perturbing is homogeneous and isotropic, the (dressed) two-point function $\langle (\delta \hat{\phi} )^2 \rangle$ is spatially homogeneous. 

The Hartree approximation is nonperturbative, so we make no assumption about the relative magnitude of $\langle (\delta \hat{\phi} )^2 \rangle$ compared to $\varphi^2$. In particular, when expanding Eqs.~(\ref{eomphi}) and (\ref{EFE}), we retain terms of the form $\langle (\delta \hat{\phi} )^2 \rangle \hat{\Psi}$ as well as $\varphi^2 \hat{\Psi}$. On the other hand, because we are working only to linear order in $\hat{\Psi}$, we do not include any terms of the form $\langle \hat{\Psi} \, \delta \hat{\phi} \rangle$. Within the Hartree approximation, such terms would arise from summing diagrams involving virtual $\hat{\Psi}$ quanta; yet even the bare propagator for such lines, $\Delta_{F} (x - y)$, is ${\cal O} (\hat{\Psi}^2)$, and hence remains beyond our approximation. To linear order in metric perturbations, in other words, the perturbations $\hat{\Psi} (x^\mu)$ do not contribute to the Hartree corrections, though, as we will see, the evolution of $\hat{\Psi} (x^\mu)$ is sensitive to the nonlinear evolution of $\delta \hat{\phi} (x^\mu)$.

Lastly, we note that the Hartree approximation is spherically symmetric in $k$-space; it does not include any direct mode-mode coupling, which would arise from convolutions of the sort $\int d\tilde{\mu} (k') \, d\tilde{\mu} (q) \, \delta \phi_{k - k' - q} \, \delta \phi_{k - k'} \, \delta \phi_{k}$. (This is consistent with neglecting scattering events that would impart large transverse momenta.) By neglecting such rescattering effects, the Hartree approximation is less efficient at transferring power between modes of different length-scales than a fully nonlinear analysis would be. (See, e.g., the discussion in Ref.~\cite{Bassett:1999mt}.) On the other hand, any such terms would require moving beyond linear order in the metric perturbations $\hat{\Psi} (x^\mu)$ --- since they would be higher order in spatially varying quantities --- and hence the Hartree approximation is especially well-suited for any study that is restricted to linear order in gravitational degrees of freedom.

Our procedure is to expand Eqs.~(\ref{eomphi}) and (\ref{EFE}) to linear order in $\hat{\Psi}$ and to arbitrarily higher order in $(\delta \hat{\phi} )^n$; implement the Hartree approximation to replace higher-order terms $(\delta \hat{\phi} )^n$ as in Eq.~(\ref{HartreeSubs}); and discard any remaining terms that are beyond linear order in spatially varying quantities. Expanding Eq.~(\ref{eomphi}), we find a set of terms that are spatially homogeneous, and a set of terms that are linear in spatially varying quantities. Requiring each set to vanish yields the coupled equations of motion:
\beq
\ddot{\varphi} + 3 H \dot{\varphi} + V^{(1)} (\varphi) + \frac{1}{2} V^{(3)} (\varphi) \langle (\delta \hat{\phi} )^2 \rangle = 0
\label{eomvarphi}
\eeq
and
\beq
\begin{split}
    &\delta \ddot{\phi}_{k \ell m} + 3 H \delta \dot{\phi}_{k \ell m} \\
    & \quad + \left[ \frac{ k^2}{a^2} + V^{(2)} (\varphi) + \frac{1}{2} V^{(4)} (\varphi) \langle (\delta \hat{\phi} )^2 \rangle \right] \delta \phi_{k \ell m} \\
    &\quad\quad\quad   = 2 (\ddot{\varphi} + 3 H \dot{\varphi} ) \Psi_{k \ell m} + 4 \dot{\varphi} \dot{\Psi}_{k \ell m} ,
\end{split}
\label{eomdeltaphi}
\eeq
where overdots denote derivatives with respect to cosmic time $t$, $H (t) \equiv \dot{a} / a$, and we use the notation
\beq
V^{(n)} (\varphi) \equiv \left( \frac{ d^n V (\phi) }{d \phi^n} \right) \bigg\vert_{\phi = \varphi} .
\label{Vndef}
\eeq
(Because we have in mind application to a model with $V (\phi) = \lambda \phi^4 /4$ in this paper, no terms with $V^{(n)} (\varphi)$ appear for $n \geq 5$.) The term in Eq.~(\ref{eomdeltaphi}) proportional to $V^{(4)} (\varphi) \, \langle (\delta \hat{\phi} )^2 \rangle$, which contributes to the effective mass for the modes $\delta \phi_{k \ell m}$, would not appear if we had performed a perturbative loop expansion. It appears in Eq.~(\ref{eomdeltaphi}) because the Hartree approximation yields a self-consistent gap equation for the dressed propagator \cite{Chang:1975dt,Boyanovsky:1993xf}.

From the $0i$ component of Eq.~(\ref{EFE}), we find
\beq
 \partial_i \left( \dot{\hat{\Psi} } + H \hat{\Psi} \right) = \frac{1}{ 2M_{\rm pl}^2} \left( \dot{\varphi} + \delta \dot{\hat{\phi}} \right) \, \partial_i \delta \hat{\phi} .
\label{0ia}
\eeq
The Hartree approximation stipulates that any terms quadratic in the field fluctuations $\delta \hat{\phi}$ should be replaced by the corresponding vacuum expectation value. In this case, the relevant term would be $\langle \delta \dot{\hat{\phi}} \, \partial_i \delta \hat{\phi} \rangle$, which will vanish: the resulting integrand is an odd function of $k_i$, integrated over symmetric limits. Hence this term vanishes within the Hartree approximation, and we find
\beq
\dot{\Psi}_{k \ell m} + H \Psi_{k \ell m} = \frac{1}{ 2 M_{\rm pl}^2} \dot{\varphi} \, \delta \phi_{k \ell m} .
\label{dotPsi}
\eeq
In a similar way, the anisotropic pressure that could arise from $T_{ij} = \partial_i \delta \hat{\phi} \,\partial_j \delta \hat{\phi}$ (for $i \neq j$) vanishes within the Hartree approximation, which is why the metric perturbations $\Phi (x^\mu)$ and $\Psi (x^\mu)$ in Eq.~(\ref{ds}) remain equal to each other.

We next expand $T^0_{\>\> 0} = - \rho$ to find the various contributions to the energy density. We find three distinct contributions: $\bar{\rho} (t)$, which depends only on the (homogeneous) vacuum expectation value of the field, $\varphi (t)$; $\delta \hat{\rho}_{(1)} (x^\mu)$, which includes all terms that are linear in spatially varying quantities; and $\delta \rho_{(2)} (t)$, which includes all spatially homogeneous terms that arise from the fluctuations:
\beq
\begin{split}
    \bar{\rho} (t) &\equiv \frac{1}{2} \dot{\varphi}^2 + V (\varphi) , \\
    \delta \hat{\rho}_{(1)} (x^\mu) &\equiv \dot{\varphi} \, \delta \dot{\hat{\phi}} - \dot{\varphi}^2 \hat{\Psi} - \langle (\delta \dot{\hat{\phi}} )^2 \rangle \, \hat{\Psi} + \frac{1}{a^2} \hat{\Psi} \langle ( \nabla \delta \hat{\phi} )^2  \rangle \\
    &\quad\quad + V^{(1)} (\varphi) \, \delta \hat{\phi} + \frac{1}{2} V^{(3)} (\varphi) \langle (\delta \hat{\phi} )^2 \rangle \, \delta \hat{\phi} , \\
    \delta \rho_{(2)} (t) &\equiv \frac{1}{2} \langle (\delta \dot{\hat{\phi}} )^2 \rangle + \frac{1}{ 2a^2} \langle ( \nabla \delta \hat{\phi} )^2  \rangle \\
    &\quad\quad + \frac{1}{2} V^{(2)} (\varphi) \, \langle (\delta \hat{\phi} )^2 \rangle + \frac{1}{8} V^{(4)} (\varphi) \langle (\delta \hat{\phi} )^2 \rangle^2 ,
\end{split}
\label{rhodeltarho}
\eeq
where
\beq
\langle (\nabla \delta \hat{\phi} )^2  \rangle \equiv h^{ij} \langle \partial_i \delta \hat{\phi} \, \partial_j \delta \hat{\phi} \rangle .
\label{vecnabla}
\eeq
We see that $\langle \delta \hat{\rho}_{(1)} \rangle = 0$. Upon expanding $G^\mu_{\>\> \nu}$ to first order in $\Psi$ and equating the terms from the $00$ component of Eq.~(\ref{EFE}) that are spatially homogeneous, we find the effective Friedmann equation
\beq
H^2 = \frac{1}{3 M_{\rm pl}^2} \left[ \bar{\rho} + \delta\rho_{(2)} \right] - \frac{ K}{a^2} .
\label{Friedmann1}
\eeq
Equating the terms in the $00$ component that are linear in spatially varying quantities, we have
\beq
\begin{split}
&- 6 H \left( \dot{\Psi}_{k \ell m} + H \Psi_{k \ell m} \right) + \frac{2}{a^2} \left( 3K - k^2 \right) \Psi_{k \ell m} \\
&\quad = \frac{1}{ M_{\rm pl}^2} \bigg\{ \dot{\varphi} \, \delta \dot{\phi}_{k \ell m} - \dot{\varphi}^2 \Psi_{k \ell m} - \langle (\delta \hat{\phi} )^2 \rangle \Psi_{k \ell m} \\
&\quad\quad\quad\quad\quad + \frac{1}{ a^2} \Psi_{k \ell m} \langle ( \nabla \delta \hat{\phi} )^2  \rangle + V^{(1)} (\varphi) \, \delta \phi_{k \ell m} \\
&\quad\quad\quad\quad\quad\quad\quad + \frac{1}{2} V^{(3)} (\varphi) \langle (\delta \hat{\phi} )^2 \rangle \, \delta\phi_{k \ell m} \bigg\} .
\end{split}
\label{00Psi}
\eeq
From the $ij$ component of Eq.~(\ref{EFE}), we are interested in extracting the spatially homogeneous contributions to the pressure, since these are relevant to the evolution of the background spacetime; in particular, we will use these terms to solve for $\dot{H}$. As usual we may parameterize the pressure as $T^i_{\>\> i}= 3p$, and hence, adopting notation as above, we find
\beq
\begin{split}
\bar{p} (t) &\equiv \frac{1}{2} \dot{\varphi}^2 - V (\varphi ) , \\
\delta p_{(2)} (t) &\equiv \frac{1}{2} \langle (\delta \dot{\hat{ \phi}} )^2 \rangle - \frac{1}{6 a^2} \langle ( \nabla \delta \hat{\phi} )^2  \rangle \\
&\quad\quad - \frac{1}{2} V^{(2)} (\varphi) \langle (\delta \hat{\phi} )^2 \rangle - \frac{1}{8} V^{(4)} (\varphi) \langle (\delta \hat{\phi} )^2 \rangle^2 .
\end{split}
\label{pdeltap}
\eeq
(We will not need an explicit expression for $\delta \hat{p}_{(1)}$.) Equating the spatially homogeneous terms in the $ij$ component of Eq.~(\ref{EFE}) yields
\beq
\left( 2 \dot{H} + 3 H^2 + \frac{ K}{a^2} \right) = - \frac{1}{ M_{\rm pl}^2} \left[ \bar{p} + \delta p_{(2)} \right].
\label{GijtermA}
\eeq
Combining Eqs.~(\ref{Friedmann1}) and (\ref{GijtermA}), we find
\beq
\begin{split}
\dot{H} &= - \frac{1}{ 2 M_{\rm pl}^2 } \left[ \bar{\rho} + \bar{p} + \delta \rho_{(2)} + \delta p_{(2)} \right] \\
&= - \frac{1}{ 2M_{\rm pl}^2} \left[ \dot{\varphi}^2 + \langle (\delta \dot{\hat{\phi}})^2 \rangle + \frac{1}{3a^2} \langle ( \nabla \delta \hat{\phi} )^2 \rangle  \right] .
\end{split}
\label{Hdot}
\eeq
From Eqs.~(\ref{Friedmann1}) and (\ref{Hdot}), we see that the evolution of the background spacetime depends on the homogeneous field $\varphi$ as well as on contributions from two-point functions of the fluctuations. A welcome feature of the Hartree approximation is that the Hartree-corrected energy-momentum tensor obeys the covariant conservation relation $\langle T^{\mu\nu} \rangle_{; \nu} = 0$, ensuring that these evolution equations remain mutually consistent with the equations of motion in Eqs.~(\ref{eomvarphi}) and (\ref{eomdeltaphi}).

Finally, we may combine our expressions from the $00$ and $0i$ components of Eq.~(\ref{EFE}) --- in particular, Eqs.~(\ref{dotPsi}) and (\ref{00Psi}) --- to derive a constraint equation relating the modes $\Psi_{k \ell m}$ to $\delta \phi_{k \ell m}$. Upon making algebraic substitutions from Eqs.~(\ref{eomvarphi}) and (\ref{Hdot}), we find
\beq
\begin{split}
&\left[ \dot{H} + \frac{2}{3 M_{\rm pl}^2 a^2} \langle ( \nabla \delta \hat{\phi} )^2  \rangle + \frac{1}{ a^2} \left( k^2 - 3K \right) \right] \Psi_{k \ell m} \\
&\quad\quad = \frac{1}{ 2M_{\rm pl}^2} \left[ \ddot{\varphi} \, \delta \phi_{k \ell m} - \dot{\varphi} \, \delta \dot{\phi}_{k \ell m} \right] .
\end{split}
\label{constraintPsi}
\eeq
In our numerical simulations, we use Eq.~(\ref{constraintPsi}) only to set initial conditions for the modes $\Psi_{k \ell m} (t_0)$, based on the initial conditions for $\varphi (t_0)$, $\dot{\varphi} (t_0)$, $H (t_0)$, $\delta \phi_{k \ell m} (t_0)$, and $\delta \dot{\phi}_{k \ell m} (t_0)$; we then evolve the metric perturbations over time using Eq.~(\ref{dotPsi}). Although the source term in Eq.~(\ref{dotPsi}) is linear in $\delta \phi_{k \ell m}$, the evolution of $\varphi (t)$, $H (t)$, and $\delta \phi_{k \ell m} (t)$ each incorporates nonlinear backreaction effects from the Hartree corrections.

When working to linear order in $\hat{\Psi}$ and $\delta \hat{\phi}$, it is common to study the evolution of the gauge-invariant comoving curvature perturbation, $\hat{\cal R} (x^\mu)$, which (for single-field models) takes the form \cite{bassett_05,liddle+lyth_09}
\beq
\hat{\cal R} = \hat{\Psi} + \frac{ H}{\dot{\varphi} } \, \delta \hat{\phi} .
\label{Rdef}
\eeq
(As is well-known, $\hat{\cal R}$ is proportional to the gauge-invariant Mukhanov-Sasaki variable, and is equivalent to the curvature perturbation on uniform-density hypersurfaces, $\hat{\zeta}$, in the limit $k \ll aH$ \cite{bassett_05,liddle+lyth_09}.) Although $\hat{\cal R}$ only remains gauge-invariant for linear gauge transformations, it remains a useful quantity for considering the evolution of perturbations even when we incorporate the nonlinear, nonperturbative Hartree corrections, as we will see in Section \ref{sec:Results}.

\section{Setting Parameters}
\label{sec:Parameters}

In our numerical simulations, we track the evolution of coupled modes within a finite (comoving) spatial volume; this restriction, in turn, means that for any Gaussian curvature $K$ of the background spatial sections, we have a discrete spectrum of allowable wavenumbers. Then the integral over $dk$ in the measure $d \tilde{\mu} (k)$ defined in Eq.~(\ref{dmu}) is replaced by a discrete sum, indexed by an integer $n \geq 1$:
\beq
\begin{split}
\delta \hat{\phi} (x^\mu) &= \sum_{n, \ell, m} \left[ \delta \phi_{n \ell m}(t) \, \hat{a}_{n\ell m} \, Z_{n\ell m}(r, \theta, \phi) + H.c. \right],
\end{split}
\label{discretemodesum}
\eeq
and similarly for $\hat{\Psi} (x^\mu)$. In place of Eq.~(\ref{acommute}), the creation and annihilation operators now obey $[ \hat{a}_{n \ell m} , \hat{a}^\dagger_{n' \ell' m'} ] = \delta_{n n'} \delta_{\ell \ell'} \delta_{m m'}$, with $\hat{a}_{n \ell m} \vert 0 \rangle = \langle 0 \vert \hat{a}^\dagger_{n \ell m} = 0$ for all $(n \ell m)$.

For the remainder of this paper we consider $K = 0$ and evolve the modes within a sphere of comoving radius $R$. (We defer the case of nonzero $K$ to future work.) As described in Appendix \ref{sec:AppendixADiscrete}, for $K = 0$ the normalized spatial eigenfunctions $Z_{n \ell m} (r, \theta, \phi)$ take the form
\beq
Z_{n \ell m} (r, \theta, \phi) = N_{n \ell m} \, j_\ell (k_{n \ell} r) \, Y_{ \ell m} (\theta, \phi) .
\label{Znlm}
\eeq
Here $N_{n \ell m}$ is a normalization constant, $j_\ell (x)$ is the spherical Bessel function, and $Y_{\ell m} (\theta, \phi)$ is the usual spherical harmonic. We choose to use Dirichlet boundary conditions at $r = R$, which fixes $Z_{n \ell m} (R, \theta, \phi) = 0$ for all $(\theta, \phi)$, which in turn restricts the allowable wavenumbers $k_{n \ell}$ to a discrete spectrum, namely
\beq
k_{n \ell} (R) \equiv \frac{x_{n \ell}}{R} ,
\label{knl}
\eeq
where $x_{n \ell}$ is the $n$th zero of the Bessel function $j_{\ell} (x)$, that is, $j_\ell (x_{n \ell}) = 0$ for $n \geq 1$. (For $\ell = 0$, the $k_{n0}$ take the simple form, $k_{n0} = n \pi / R$.)

Within the Hartree approximation, the evolution of $\varphi (t)$ and $H (t)$, as well as the modes $\Psi_{n \ell m} (t)$, depends on the initial conditions for the modes $\delta \phi_{n \ell m} (t_0)$ and $\delta \dot{\phi}_{n \ell m} (t_0)$. Because the Hartree approximation replaces higher-order interaction terms in the equation of motion for the fluctuations $\delta \hat{\phi}$ by an effective mass, we may follow many of the usual steps for quantizing a free scalar field in FLRW spacetime, and use this quantization procedure to parameterize initial conditions for $\delta \phi_{n \ell m} (t_0)$ and $\delta\dot{\phi}_{n \ell m} (t_0)$. 

In the regime of interest, the field fluctuations are nearly massless around $t_0$. From Eq.~(\ref{eomdeltaphi}), the effective mass is given by
\beq 
m_{\rm eff}^2 (t) = V^{(2)} (\varphi) + \frac{1}{2} V^{(4)} (\varphi) \langle (\delta \hat{\phi} )^2 \rangle,
\label{meff}
\eeq 
which is suppressed by the small coupling constant $\lambda$; hence we have $m_{\rm eff}^2 (t_0) \ll H^2 (t_0)$. Moreover, when we evolve the modes within a sphere of comoving radius $R$, we introduce an infrared cut-off given by $k_{\rm min} = \pi / R$, with $R \sim 1/ [a (t_0) \, H (t_0) ]$. Even for the longest-wavelength modes in our simulation, we therefore have $k_{n \ell}^2/ a^2 (t_0) \gtrsim m_{\rm eff}^2 (t_0)$. 

We do not assume that the system has attained the mininum energy state at the initial time $t_0$, and hence we consider initial conditions for $\delta \phi_{n \ell m} (t_0)$ and $\delta \dot{\phi}_{n \ell m} (t_0)$ that depart from the usual Bunch-Davies vacuum state. As described in Appendix \ref{sec:initialconditions}, we parameterize 
\beq 
\begin{split}
    \delta \phi_{n \ell m} (t_0) &= \frac{ \alpha_{n \ell m} }{\sqrt{ 2 k_{n \ell} } }, \\
    \delta \dot{\phi}_{n \ell m} (t_0) &= \sqrt{\frac{k_{n \ell} }{2} } \left( - i \gamma_{n \ell m} + \delta_{n \ell m} - \frac{ \alpha_{n \ell m} \bar{H}_0 }{k_{n \ell}} \right),
\end{split}
\label{deltaphiICn}
\eeq 
where $\alpha_{n \ell m}$, $\gamma_{n \ell m}$, and $\delta_{n \ell m}$ are each real-valued dimensionless constants and $\bar{H}_0$ is given by
\beq 
\bar{H}_0^2 \equiv \frac{ \bar{\rho} (t_0) }{3 M_{\rm pl}^2 } .
\label{barH}
\eeq 
From Eq.~(\ref{rhodeltarho}) we note that $\bar{\rho} (t_0)$ is the energy density associated with the vacuum expectation value at the initial time, $\varphi (t_0)$. For the quantum fluctuations, the equal-time commutation relation at $t_0$ requires 
\beq 
\alpha_{n \ell m} = \frac{1}{ \gamma_{n \ell m} }
\label{alphagamma}
\eeq 
for all $(n \ell m)$. The Bunch-Davies initial state corresponds to $\gamma_{n \ell m} = 1$ and $\delta_{n \ell m} = 0$ for all $(n \ell m)$. To depart from the Bunch-Davies initial state, we treat the coefficients $\gamma_{n \ell m}$ and $\delta_{n \ell m}$ as random variables for each mode, drawn from flat distributions within specific ranges. Once $\gamma_{n \ell m}$ is drawn for a given mode, $\alpha_{n \ell m}$ for that mode is fixed from Eq.~(\ref{alphagamma}).

In addition to selecting initial conditions for individual modes $\delta \phi_{n \ell m} (t_0)$ and $\delta \dot{\phi}_{n \ell m} (t_0)$, we also need to evaluate the various two-point functions that appear in the evolution equations for $\varphi (t)$, $\delta \phi_{n \ell m} (t)$, $H(t)$, and $\dot{H} (t)$; only then can we set initial conditions for the metric perturbation modes $\Psi_{n \ell m} (t_0)$ and evolve the coupled system forward in time. As discussed in Appendix \ref{sec:AppendixADiscrete}, when we evaluate the two-point functions within a sphere of comoving radius $R$, only modes with $\ell = 0$ contribute to $\langle (\delta \hat{\phi} )^2 \rangle$ and $\langle (\delta \dot{\hat{\phi}} )^2 \rangle$, whereas only modes with $\ell = 1$ contribute to $\langle (\nabla \delta \hat{\phi} )^2 \rangle$, and we find
\beq
\begin{split}
\langle (\delta \hat{\phi} )^2 \rangle &= \frac{ \pi}{2 R^3} \sum_n n^2 \vert \delta \phi_{n00} (t) \vert^2 ,\\
\langle (\delta \dot{\hat{\phi}} )^2 \rangle &= \frac{ \pi}{2 R^3} \sum_n n^2 \vert \delta\dot{\phi}_{n 00} (t) \vert^2 , \\
\langle (\nabla \delta \hat{\phi} )^2 \rangle &= \frac{1}{ 6 \pi R^3} \sum_n \sum_{m = -1,0,1} \frac{ k_{n1}^2 }{ \vert j_2 (k_{n1} R) \vert^2 } \vert \delta \phi_{n1m} (t) \vert^2 .
\end{split}
\label{twopointR}
\eeq
Although we have considered the $K = 0$ case here, the pattern of which $\ell$ modes contribute to which two-point functions holds for arbitrary $K$, since the $(\theta, \phi)$ portion of the background metric in Eq.~(\ref{hijK}) does not depend on $K$.

Our expressions for the various two-point functions in Eq.~(\ref{twopointR}) diverge in the limit $n \rightarrow \infty$; this is just the usual ultraviolet divergence that appears in the continuum limit for $k \rightarrow \infty$. Hence we must regularize all sums that appear in the various two-point functions. Since we expand the quantum fluctuations $\delta \hat{\phi} (x^\mu)$ as sums over modes of comoving wavenumber $k_{n \ell}$, we adopt a simple Gaussian regulator with a comoving UV scale $\kappa$:
\beq
F_{n \ell} (\kappa, R) = \exp \left[ - \frac{ k_{n\ell}^2 (R)}{(2 \kappa )^2} \right] .
\label{Fregulator}
\eeq
We insert $F_{n \ell} (\kappa, R)$ within the sums when evaluating all two-point functions. For example, 
\beq
\langle (\delta \hat{\phi} )^2 \rangle \rightarrow \langle (\delta \hat{\phi} )^2 \rangle_{\rm reg} = \frac{ \pi}{2 R^3} \sum_n n^2 \vert \delta \phi_{n 00} (t) \vert^2 F_{n0} (\kappa, R) .
\label{twopointreg}
\eeq
We use a single UV regulator scale $\kappa$, independent of $\ell$. Once $\kappa$ is fixed, the sums over $n$ may be truncated at some finite number of modes, $n_{\rm max}$.

The regularized two-point functions depend on the UV regulator scale $\kappa$. In general, the two-point function for quantum fluctuations of a nearly massless scalar field in an FLRW background scales as $\langle (\delta \hat{\phi} )^2 \rangle \propto H^2$. We therefore parameterize $\kappa = b \bar{H}_0$, with $b$ a real, dimensionless constant. To confirm the scaling of the regularized two-point function with $H$, we use the fact that the random coefficients $\alpha_{n \ell m}$ for the mode functions $\delta \phi_{n \ell m} (t_0)$ vary independently of $n$, so we may take the term $\alpha_{n 00}^2$ out of the sum in Eq.~(\ref{twopointreg}) at $t_0$ and replace it by an average value. As detailed in Appendix \ref{sec:initialconditions}, this yields
\beq
\langle (\delta \hat{\phi} (t_0) )^2 \rangle_{\rm reg} \simeq (\alpha^2)_{\rm avg} \frac{ b^2 \bar{H}_0^2}{4 \pi^2} .
\label{twopointreg2}
\eeq
(Although only modes with $\ell = m = 0$ contribute to the sum in Eq.~(\ref{twopointreg}), we draw the random variables $\gamma_{n \ell m} = 1/ \alpha_{n \ell m}$ from the same distribution for all $(n \ell m)$, and hence the average value $( \alpha^2)_{\rm avg}$ holds for any $n$, $\ell$, and $m$.) Following similar steps (see Appendix \ref{sec:initialconditions}) we find
\beq 
\delta \rho_{(2)} (t_0) \simeq \frac{ b^4 {\cal C} }{4 \pi^2} \bar{H}_0^4 .
\label{deltarhoICn}
\eeq 
with
\beq
{\cal C} \equiv \left( \alpha^2 + \gamma^2 + \delta^2  \right)_{\rm avg} .
\label{Cdef}
\eeq
The quantity ${\cal C}$ measures how strongly (on average) the amplitude of the initial state of the quantum fluctuations deviates from the minimum-energy (Bunch-Davies) state. Since $\gamma_{n \ell m} = \alpha_{n \ell m} = 1$ and $\delta_{n \ell m} = 0$ for all $(n \ell m)$ in the Bunch-Davies state, ${\cal C}_{\rm BD} = 2$. Throughout our analysis, we consider quantum fluctuations whose average initial energy density exceeds the Bunch-Davies value by an order of magnitude, with ${\cal C} \simeq 20$, by drawing the random coefficients for each mode from flat distributions within the ranges
\beq
\gamma_{n \ell m} \in \{ 0.09 , 1 \} \> , \>\> \delta_{n \ell m} \in \{ -5, 5 \} .
\label{gammadeltaranges}
\eeq
Once $\gamma_{n \ell m}$ is drawn for a given mode, $\alpha_{n \ell m}$ for that mode is fixed from Eq.~(\ref{alphagamma}). The ranges in Eq.~(\ref{gammadeltaranges}) yield $(\alpha^2)_{\rm avg} = 11.11$, $(\gamma^2)_{\rm avg} = 0.37$, and $(\delta^2)_{\rm avg} = 8.33$.

The actual value of the initial Hubble scale $H_0 \equiv H (t_0)$ will be greater than $\bar{H}_0$, because $H$ includes contributions from both $\bar{\rho}$ and $\delta \rho_{(2)}$, as in Eq.~(\ref{Friedmann1}). We define $f \equiv (H_0 / \bar{H}_0 ) > 1$; using Eqs.~(\ref{barH}) and (\ref{deltarhoICn}) we find that on average
\beq
f_{\rm avg} = \left[ 1 + \frac{  b^4 {\cal C}  }{12\pi^2} \left( \frac{\bar{H}_0}{ M_{\rm pl} } \right)^2 \right]^{1/2}  .
\label{Fdef}
\eeq
We aim to study initial conditions for the system such that $H_0 \gg H_{\rm infl}$, where $H_{\rm infl}$ is the Hubble scale of the slow-roll inflationary attractor. Yet we also need to keep $H_0$ low enough (compared to $M_{\rm pl}$) so that we can begin the simulations with significant power in fluctuations on sub-Hubble scales, $H_0 < k / a (t_0) < M_{\rm pl}$, while avoiding trans-Planckian ambiguities. Hence we focus on initial conditions such that $H_0 \sim 0.1 \, M_{\rm pl}$. For perturbations that depart from the Bunch-Davies initial state, with $2 \leq {\cal C} \leq 20$, we find $\delta \rho_{(2)} (t_0) \lesssim \bar{\rho} (t_0)$ for $\bar{H}_0 = 0.1 \, M_{\rm pl}$ if we fix $b = 5$. Eq.~(\ref{Fdef}) then yields $f_{\rm avg} \leq 1.4$, corresponding to $H_0 \lesssim 0.14 \, M_{\rm pl}$.

Because we are interested in effects of initial inhomogeneities on length-scales shorter than the initial Hubble radius, we fix the comoving radius $R = 1.5 \pi \bar{H}_0^{-1} > 1.5 \pi H_0^{-1}$. (We set $a (t_0) = 1$.) Then the longest comoving wavelength in the spectrum is $\lambda_{\rm max} = 2R > 3 \pi H_0^{-1}$, corresponding to $k_{\rm min} = \pi / R$. This ensures that our longest wavelength modes begin on a superhorizon scale, but that most of our modes are initially subhorizon. Given the form of $F_{n \ell} (\kappa, R)$ in Eq.~(\ref{Fregulator}), meanwhile, we find strong suppression of the contribution to the various two-point functions by modes with comoving wavenumber $k_{n \ell} > k_{UV} = 3 \kappa$, or $k_{UV} = 3b \bar{H}_0$. Setting $b = 5$, we find $k_{UV} \sim {\cal O} (10 H_0) \sim M_{\rm pl}$. To ensure strong numerical convergence we fix $k_{\rm max} = 4 \kappa$, which corresponds to $n_{\rm max} = 30$. Our simulations then involve $4 n_{\rm max} = 120$ coupled modes: $n_{\rm max}$ each for $\ell = 0$ and for $\ell = 1$ with $m = -1, 0, 1$. 

We implement our UV regularization via Eq.~(\ref{Fregulator}), but do not pursue formal renormalization. For one thing, we are interested in scenarios in which the system begins at high energies $H_0 \sim 0.1 \, M_{\rm pl}$, and we aim to track effects of excited modes with wavenumbers up to $k_{UV} \sim M_{\rm pl}$; hence there are no arbitrarily large hierarchies between the physical energy scales of interest and the natural cut-off scale. More important, by studying initial states for the fluctuations that depart from the usual Bunch-Davies state, as in Eq.~(\ref{deltaphiICn}), any renormalization scheme would depend {\it both} on the coupling constants in the Lagrangian {\it and} on the particular selection of initial parameters $\alpha_{n \ell m}, \gamma_{n \ell m}$, and $\delta_{n \ell m}$ for the modes. Hence any renormalization scheme would change, run by run, with the random draws for these parameters. (See, e.g., Refs.~\cite{HolmanInitiala,HolmanInitialb}.) We therefore defer questions about formal renormalization to future work.

\section{Numerical Results}
\label{sec:Results}

In this section we first consider typical behavior of the coupled system for a particular set of initial conditions, before turning to a more systematic investigation across initial conditions. We study $V(\phi) = \lambda \phi^4 / 4$ with $\lambda = 10^{-10}$, and later compare results for $\lambda = 10^{-12}$. 

When one ignores field fluctuations and metric perturbations, this model yields sufficient inflation, with $N_{\rm infl} \geq 65$ efolds, for $\varphi_0 \equiv \varphi (t_0) \sim 15 - 30 \, M_{\rm pl}$, depending on the initial value of $\dot{\varphi}_0 \equiv \dot{\varphi} (t_0)$. For $\lambda = 10^{-10}$, this corresponds to a value of the Hubble parameter once the system has reached the slow-roll inflationary attractor (with $\dot{\varphi}^2 \ll V (\varphi)$) of $H_{\rm infl} \sim 10^{-3} \, M_{\rm pl}$; for $\lambda = 10^{-12}$, we have $H_{\rm infl} \sim 10^{-4} \, M_{\rm pl}$. We are therefore interested in the behavior of this system in the vicinity of $\varphi_0 \sim 15 - 30 \, M_{\rm pl}$ when the nonlinear effects of fluctuations are incorporated into the dynamics.

For each value of $\lambda$, we varied $12 \, M_{\rm pl} \leq \varphi_0 \leq 30 \, M_{\rm pl}$ in steps of $\Delta \varphi_0 = 0.25 \, M_{\rm pl}$, and $- 0.25 \, M_{\rm pl}^2 \leq \dot{\varphi}_0 \leq 0.25 \, M_{\rm pl}^2$ in steps of $\Delta \dot{\varphi}_0 = 0.01 \, M_{\rm pl}^2$, for a $73 \times 51$ search grid. For each grid point in $(\varphi_0, \dot{\varphi}_0)$, we computed $\numsim$ simulations: one with the Hartree corrections turned off (to neglect coupled fluctuations), one with the quantum fluctuations $(\delta \phi_{n \ell m} (t_0), \delta \dot{\phi}_{n \ell m} (t_0) )$ initialized in the Bunch-Davies initial state with ${\cal C}_{\rm BD} = 2$, and $\numsimflucts$ in which we initialized the quantum fluctuations with ${\cal C} \simeq 20$ by drawing random coefficients $\gamma_{n \ell m}$ and $\delta_{n \ell m}$ for each mode from the ranges in Eq.~(\ref{gammadeltaranges}). This yielded a total of roughly 250,000 individual simulations. The simulations were performed on the Amazon Web Services EC2 service on a 16-core instance, and took a little under two days to complete. Our code was implemented in Python.

Our simulations are initialized in a number of steps. Starting from the values $(\varphi_0,\dot{\varphi}_0)$, we construct $\bar{H}_0$ as in Eq.~(\ref{barH}), from which $R$, $\kappa$, and $k_{\rm max}$ are computed. The spectrum of allowable wavenumbers $k_{n \ell} (R)$ with $\ell = 0$ and $\ell = 1$ is then constructed. Next, we construct the Bunch-Davies initial conditions for each mode $\delta \phi_{n\ell m} (t_0)$ and $\delta \dot{\phi}_{n \ell m} (t_0)$. For perturbed initial data, we draw random values for $\gamma_{k \ell m}$ and $\delta_{k \ell m}$ for each mode, and construct the initial mode data appropriately from Eq.~(\ref{deltaphiICn}). We then compute the appropriate two-point functions, initialize the modes $\Psi_{n \ell m} (t_0)$ from Eq.~(\ref{constraintPsi}), and evaluate $\delta \rho_{(2)} (t_0)$ to construct the actual initial Hubble factor $H_0$.

Our simulations evolve the quantities $a (t)$, $\varphi (t)$, and $\dot{\varphi} (t)$. For each wavenumber $k_{n \ell}$ with $\ell = 0$ and $\ell = 1$, we also evolve two modes: one initialized with $(\delta \phi_{n \ell 0} (t_0), \delta \dot{\phi}_{n \ell 0} (t_0) ) = (1,0)$ and a second with $(0,1)$ 
(along with a corresponding initialization for $\Psi_{n \ell 0} (t_0)$ for each mode). Given the time evolution of these modes and the initial conditions $\delta \phi_{n \ell m} (t_0)$, $\delta \dot{\phi}_{n \ell m} (t_0)$, and $\Psi_{k \ell m} (t_0)$, every mode can be reconstructed by exploiting the linearity of the equations of motion, since the nonlinear two-point functions are effectively functions of time that are independent of $n$, $\ell$ and $m$. Note that the modes that are evolved remain real. Doing this split makes little difference for the $\ell = 0$ modes, but reduces the computational cost threefold for the $\ell = 1$ modes. We evolve the system forward in time using a variable time-step RK45 algorithm. We declare that slow-roll begins once $\epsilon < 0.1$, and we terminate evolution thereafter at $\epsilon \ge 1$, where
\beq
\epsilon \equiv - \frac{\dot{H} } { H^2} . 
\label{epsilon}
\eeq

\subsection{Evolution of Perturbations}
\label{sec:EvolutionPerts}

In this subsection and the following, we present results from a representative sample of random initializations. We set $\lambda = 10^{-10}$ and consider the case $\varphi_0 = 25 \, M_{\rm pl}$ and $\dot{\varphi}_0 = - 0.25 \, M_{\rm pl}^2$, for which the initial kinetic energy in the field $\varphi$ exceeds the potential energy by a factor of 3200. We initialize the fluctuations $\delta \hat{\phi} (t_0, {\bf x})$ by parameterizing the mode functions as in Eq.~(\ref{deltaphiICn}) and drawing the random initial coefficients $\gamma_{n \ell m}$ and $\delta_{n \ell m}$ for each mode from the distributions in Eq.~(\ref{gammadeltaranges}). Then ${\cal C} \simeq 20$ and the energy density in fluctuations $\delta \rho_{(2)} (t_0)$ begins about ten times greater than in the Bunch-Davies initial state. With these parameters, we find $\delta \rho_{(2)} (t_0) \simeq \bar{\rho} (t_0)$ and $H_0 = 0.14 \, M_{\rm pl}$, two orders of magnitude greater than $H_{\rm infl}$.

\begin{figure}[t]
\begin{center}
\includegraphics[width=3.41in]{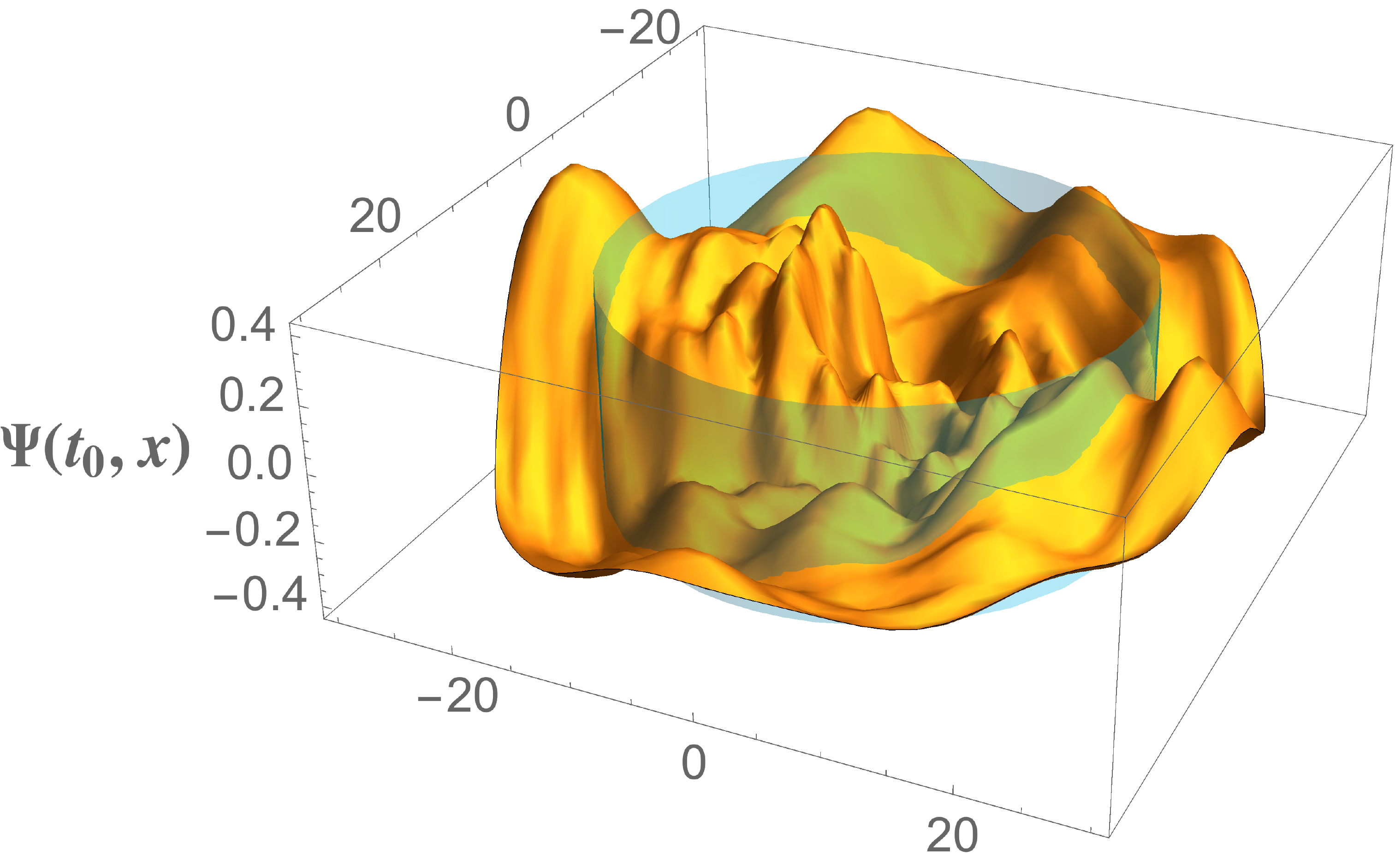}
\caption{\small Typical initial surface of $\Psi \left( t_0, \mathbf{x} \right)$ in the $x$-$y$ plane (dependence on the polar angle $\theta$ in Eq.~(\ref{hijK}) is suppressed) for $\varphi_0 = 25 \, M_{\rm pl}$ and $\dot{\varphi}_0 = - 0.25 \, M_{\rm pl}^2$. The field fluctuations $\delta \hat{\phi} (t_0, \mathbf{x})$ were initialized as in Eq.~(\ref{deltaphiICn}), with the random coefficients $\gamma_{n \ell m}$ and $\delta_{n \ell m}$ for each mode drawn from the ranges in Eq.~(\ref{gammadeltaranges}), which in turn determined the modes $\Psi_{n \ell m} (t_0)$ from Eq.~(\ref{constraintPsi}). Further details about the construction of the initial surface for $\Psi ( t_{0}, \mathbf{x} )$ are given in Appendix \ref{sec:MetricInitial}. 
The blue cylinder has radius equal to $r_H (t_0) = \pi / H (t_0)$, in units of $M_{\rm pl}^{-1}$. }
\label{PsiPos}
\end{center}
\end{figure}

For these initial conditions, the system begins with significant inhomogeneities on length-scales well within the initial Hubble radius. Fig.~\ref{PsiPos} shows $\Psi (t_0, {\bf x})$, constructed from modes $\Psi_{n \ell m} (t_0)$ whose amplitudes are set by Eq.~(\ref{constraintPsi}). (Further details of how we construct $\Psi (t_0, {\bf x})$ are given in Appendix \ref{sec:MetricInitial}.) The blue cylinder in Fig.~\ref{PsiPos} has a radius equal to $r_H (t_0) = \pi / H (t_0)$, such that modes with $k \geq a (t_0) H (t_0)$ have wavelengths that fit within the diameter $2 r_H (t_0)$. For this choice of initial conditions, the metric perturbations $\Psi (t_0, {\bf x})$ begin with substantial structure on sub-Hubble length-scales, with spatial inhomogeneities as large as $\vert \Psi (t_0, {\bf x} ) \vert \simeq 0.4$.

\begin{figure}[t]
\begin{center}
\includegraphics[width=3.41in]{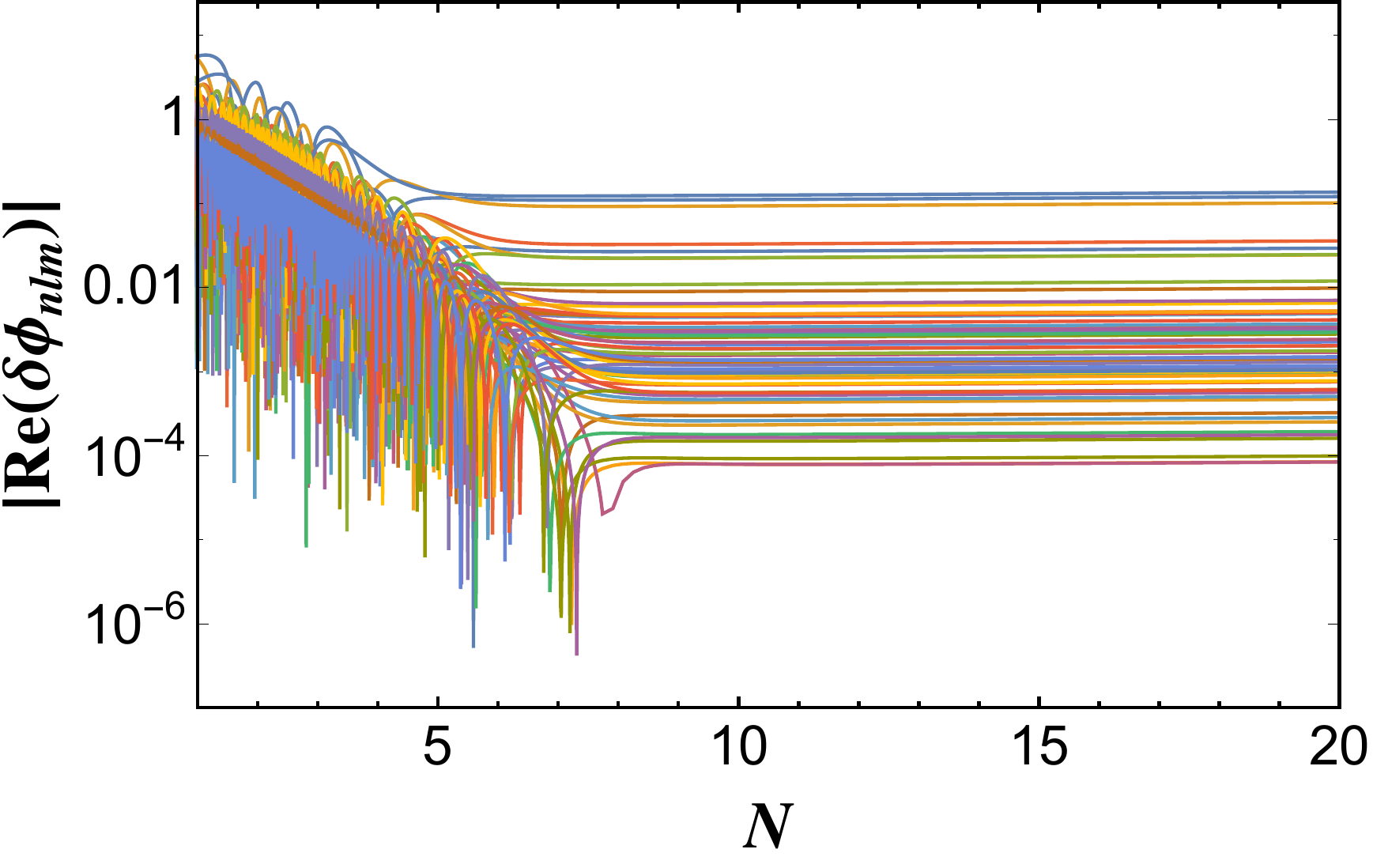}
\caption{\small $\vert {\rm Re} ( \delta \phi_{n \ell m} ) \vert $ versus $N = \ln a$ for the $\ell = 0, 1$ modes in our simulation with $k_{n \ell } \geq a_0 H_0$. }
\label{dphikPlotOn}
\end{center}
\end{figure}

\begin{figure}[t]
\begin{center}
\includegraphics[width=3.41in]{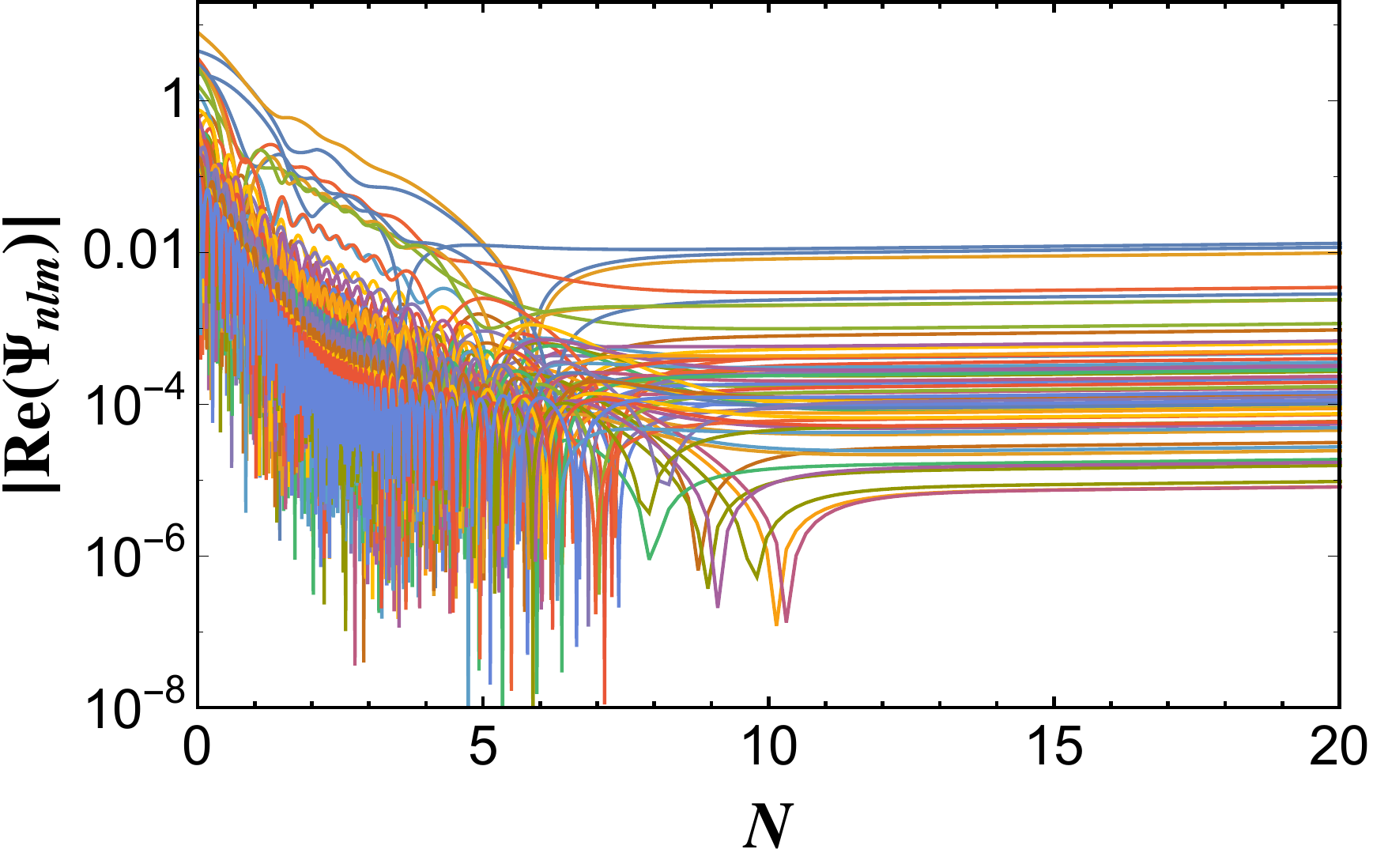}
\caption{\small $\vert {\rm Re} ( \Psi_{n \ell m}) \vert$ versus $N$ for the $\ell = 0, 1$ modes in our simulation with $k_{n \ell } \geq a_0 H_0$.    }
\label{PsikPlotOn}
\end{center}
\end{figure}

\begin{figure}[th]
\begin{center}
\includegraphics[width=3.41in]{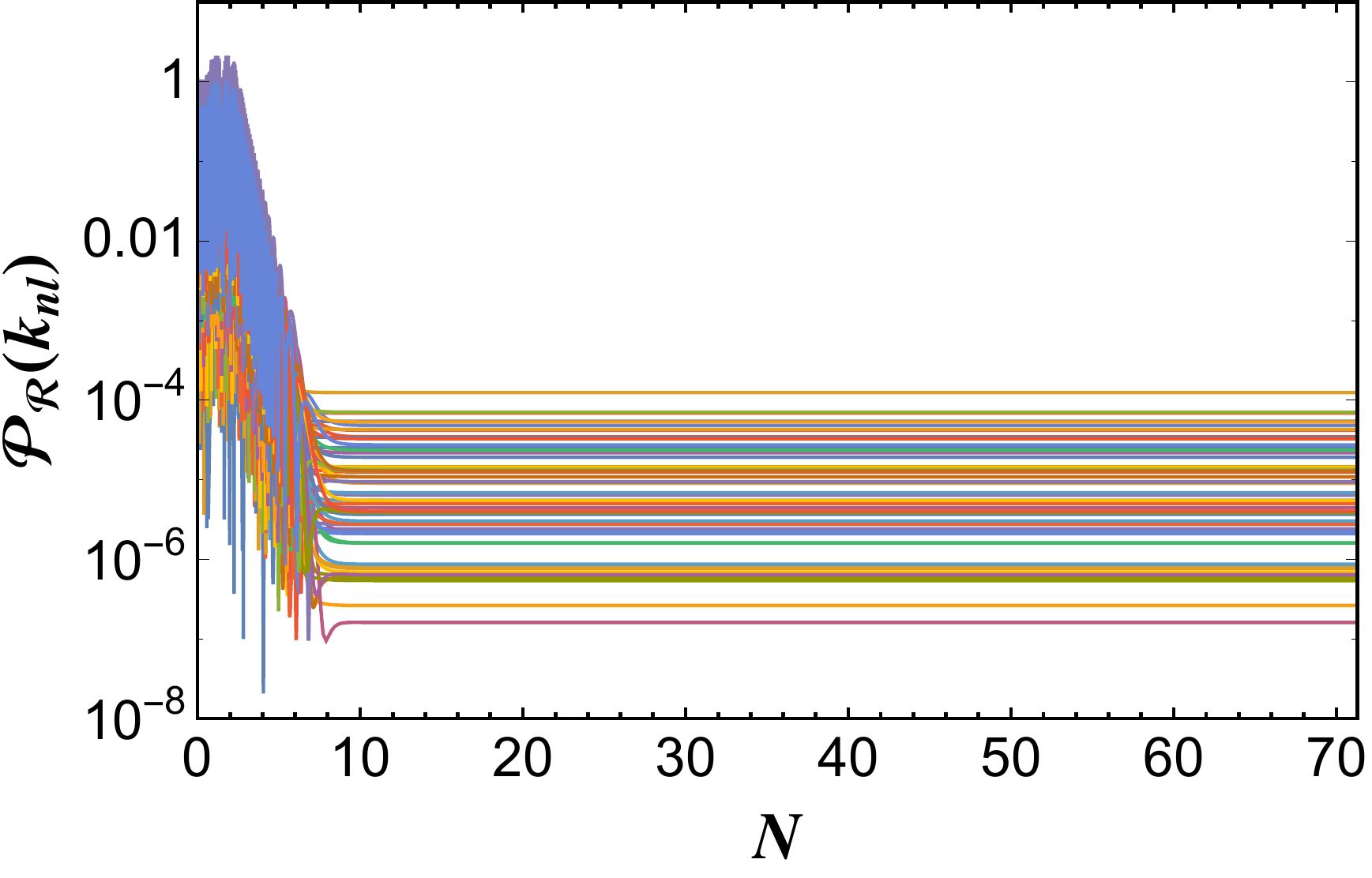}
\caption{\small  The dimensionless power spectrum of the curvature perturbation ${\cal P}_{\cal R} (k_{n\ell} )$ versus $N$ for the $\ell = 0, 1$ modes in our simulation with $k_{n \ell} \geq a_0 H_0$. }
\label{RkPPlotOn}
\end{center}
\end{figure}

Figures \ref{dphikPlotOn} and \ref{PsikPlotOn} show $\vert {\rm Re} ( \delta \phi_{n \ell m} ) \vert$ and $\vert {\rm Re} ( \Psi_{n \ell m} ) \vert$ versus $N \equiv \int H dt = \ln a$ for early times, for the $\ell = 0$ and $\ell = 1$ modes with $k_{n \ell} \geq a_0 H_0$. As expected, the modes oscillate with decaying amplitude while inside the Hubble radius, and their amplitudes freeze after Hubble crossing. At later times, after the physical wavelengths of the modes have redshifted to be exponentially larger than the Hubble radius, the amplitudes show a modest secular growth, rising as $\vert \delta \phi_{n \ell m} \vert \propto \sqrt{\epsilon}$ and $\vert \Psi_{n \ell m} \vert \propto \epsilon$, where $\epsilon$ is given in Eq.~(\ref{epsilon}). This modest late-time growth matches the well-known behavior of perturbations deep in the infrared during the slow-roll regime, as treated in linear perturbation theory. (See, e.g., Section 8.2 of Ref.~\cite{Mukhanov}, as well as Ref.~\cite{SeeryInfrared}.) Nonetheless, the curvature perturbation, $\hat{\cal R}$, defined in Eq.~(\ref{Rdef}), remains conserved once modes cross outside the Hubble radius. In Fig.~\ref{RkPPlotOn} we plot the dimensionless power spectrum for the curvature perturbation,
\beq
{\cal P}_{\cal R} (k_{n \ell}) \equiv \frac{ k_{n\ell}^3}{2 \pi^2 } \vert {\cal R}_{n \ell m} \vert^2
\label{PR}
\eeq
for $\ell = 0$ and $\ell = 1$ modes that begin with $k_{n \ell} \geq a_0 H_0$. Consistent with the analytic results in Ref.~\cite{SenatoreZaldarriaga}, we find that the curvature perturbation remains conserved on super-Hubble length-scales, even when we incorporate nonlinear self-interactions. As shown in Fig.~\ref{RkPPlotOn}, substantial structure on sub-Hubble length-scales at early times damps out before modes cross outside the Hubble radius, producing a smooth patch on horizon scales, and remains exponentially suppressed for the duration of the simulation.

\begin{figure}[t]
\begin{center}
\includegraphics[width=3in]{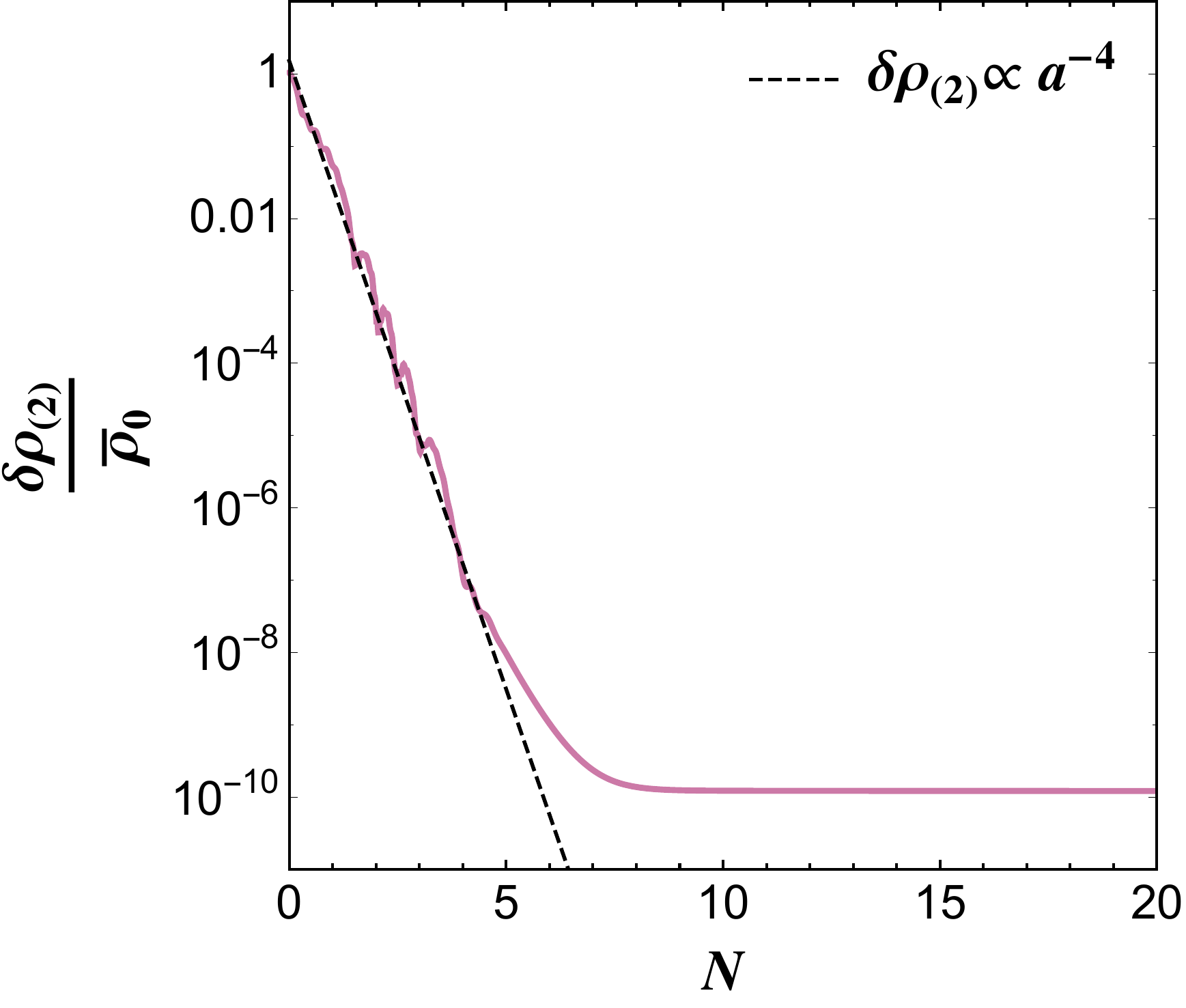}
\caption{\small The energy density in fluctuations $\delta \rho_{(2)} (t)$ (normalized by the initial value $\bar{\rho} (t_0)$) versus $N$, with $\lambda = 10^{-10}$. At early times, while most modes are still inside the Hubble radius, $\delta \rho_{(2)} (t)$ decays like radiation. }
\label{drho2PlotOn}
\end{center}
\end{figure}

The energy density in fluctuations $\delta\rho_{(2)}$ begins with $\delta \rho_{(2)} (t_0) \simeq \bar{\rho} (t_0)$ and then begins to decay, as shown in Fig.~\ref{drho2PlotOn}. 
Because of the weak coupling $\lambda$, the effective mass for the fluctuations satisfies $m_{\rm eff} (t) \ll H (t)$ at early times, where $m_{\rm eff}$ is given in Eq.~(\ref{meff}). While most modes are inside the Hubble radius, with $k / a >  H \gg  m_{\rm eff}$, their energy density therefore evolves like a gas of (nearly) massless particles, with an equation of state like radiation, $\delta \rho_{(2)} (t) \propto a^{-4} (t)$. At later times, after the modes have crossed outside the Hubble radius and their amplitudes have frozen, $\delta \rho_{(2)} (t)$ becomes constant.  

\begin{figure}[tb]
\begin{center}
\includegraphics[width=3in]{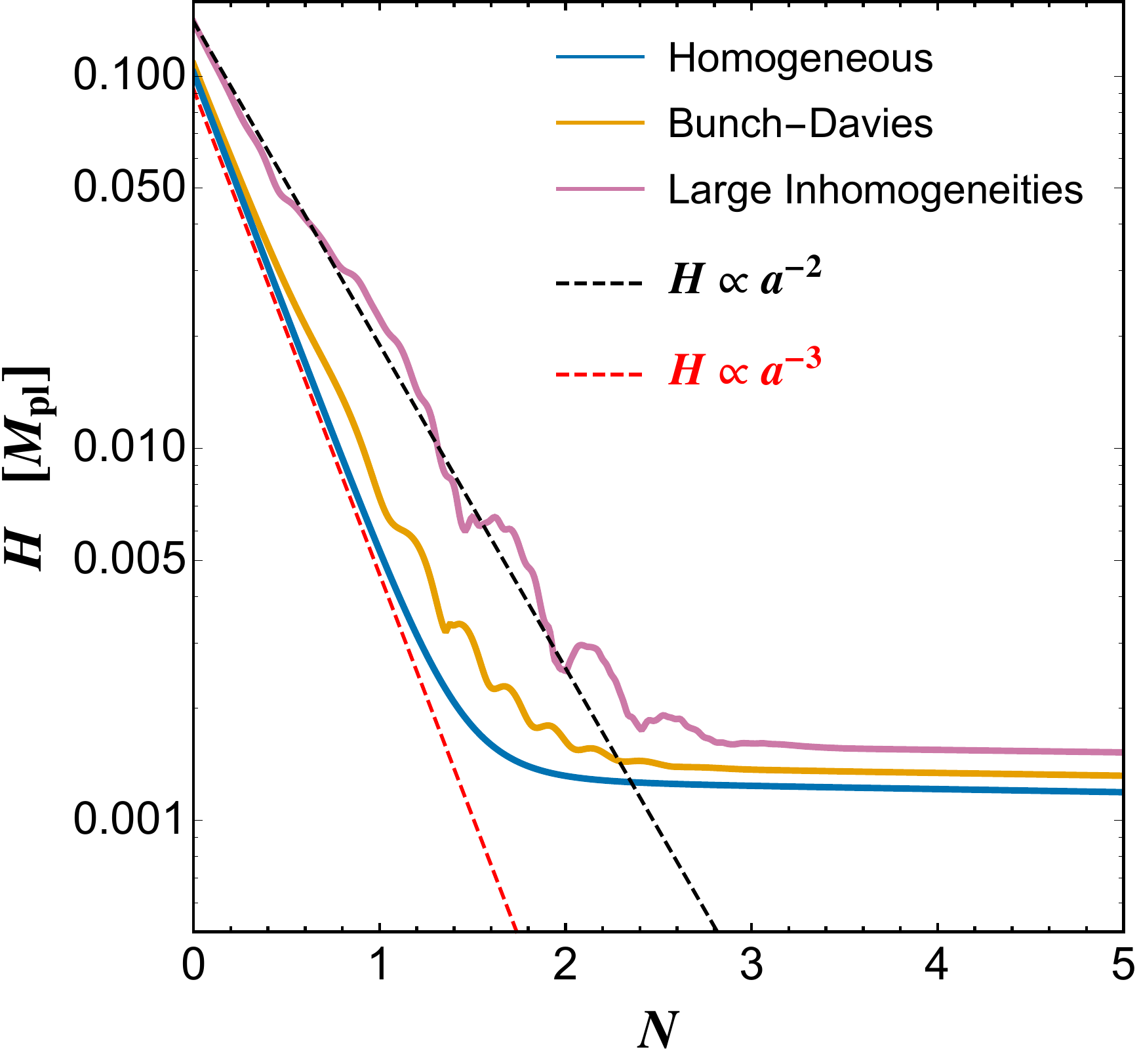}
\caption{\small The evolution of the Hubble parameter $H(t)$ versus $N$ for early times, with $\lambda = 10^{-10}$. In the absence of fluctuations (blue), $H (t) \propto a^{-3} (t)$. For large initial fluctuations (pink), $H(t) \propto a^{-2} (t)$ at early times. For fluctuations that begin in the Bunch-Davies initial state (yellow), the evolution of $H (t)$ falls between the other two cases.}
\label{HPlotOn}
\end{center}
\end{figure}

Next we consider the impact of these large initial inhomogeneities on the evolution of the Hubble parameter $H(t)$, shown in Fig.~\ref{HPlotOn}. The figure shows $H(t)$ for the same initial values $\varphi_0 = 25 \, M_{\rm pl}$ and $\dot{\varphi}_0 = - 0.25\, M_{\rm pl}^2$, when we neglect fluctuations (blue); when we initialize the fluctuations in the Bunch-Davies state, with ${\cal C}_{\rm BD} = 2$ (yellow); and for a particular simulation in which we initialized the system with large initial fluctuations, ${\cal C} = 20$ (pink). The energy density associated with $\varphi$, $\bar{\rho} (t)$, is dominated at early times by the kinetic energy of $\varphi$, and hence it decays as $\bar{\rho} (t) \propto a^{-6} (t)$. When we neglect fluctuations, we therefore find $H (t) \propto a^{-3} (t)$ at early times. On the other hand, for large initial fluctuations with ${\cal C} = 20$ and hence $\delta \rho_{(2)} (t_0) \simeq \bar{\rho} (t_0)$, we find $H (t) \propto [ \bar{\rho} (t) + \delta \rho_{(2)} (t) ]^{1/2} \propto a^{-2} (t)$ at early times, while most fluctuations remain sub-Hubble and $\delta \rho_{(2)} (t)$ decays like radiation. For fluctuations that begin in the Bunch-Davies initial state, with ${\cal C}_{\rm BD}  = 2$ and $\delta \rho_{(2)} (t_0) \sim 0.1 \, \bar{\rho} (t_0)$, we find an evolution for $H(t)$ intermediate between these two cases. (The authors of Ref.~\cite{East:2015ggf} likewise found the volume-averaged quantities $\rho_{\rm avg} (t) \propto [ a_{\rm avg} (t)]^{-4}$ and $H_{\rm avg} (t) \propto [a_{\rm avg} (t)]^{-2}$ at early times in their numerical simulations of large-field models with significant initial inhomogeneities. See also Ref.~\cite{Chowdhury:2019otk}.)

The system begins to inflate, with $\ddot{a} > 0$, once $\epsilon < 1$. For the set of initial conditions we consider here, inflation begins by $N \sim 2$, and the system enters a phase of slow-roll inflation ($\epsilon < 0.1$) by $N \sim 3$. During slow-roll, $\bar{\rho}$ is dominated by $V (\varphi)$, while $\delta \rho_{(2)}$ continues to redshift like radiation until most of the modes have crossed outside the Hubble radius, by $N \sim 7$ (as shown in Fig.~\ref{drho2PlotOn}). Hence after slow-roll inflation begins, the system evolves with $\bar{\rho} \simeq V (\varphi) \gg \delta\rho_{(2)}$ and $H ( t )$ settles onto a nearly constant value at $H_{\rm infl} \sim 10^{-3} \, M_{\rm pl}$.

\begin{figure}[t]
\begin{center}
\includegraphics[width=3.2in]{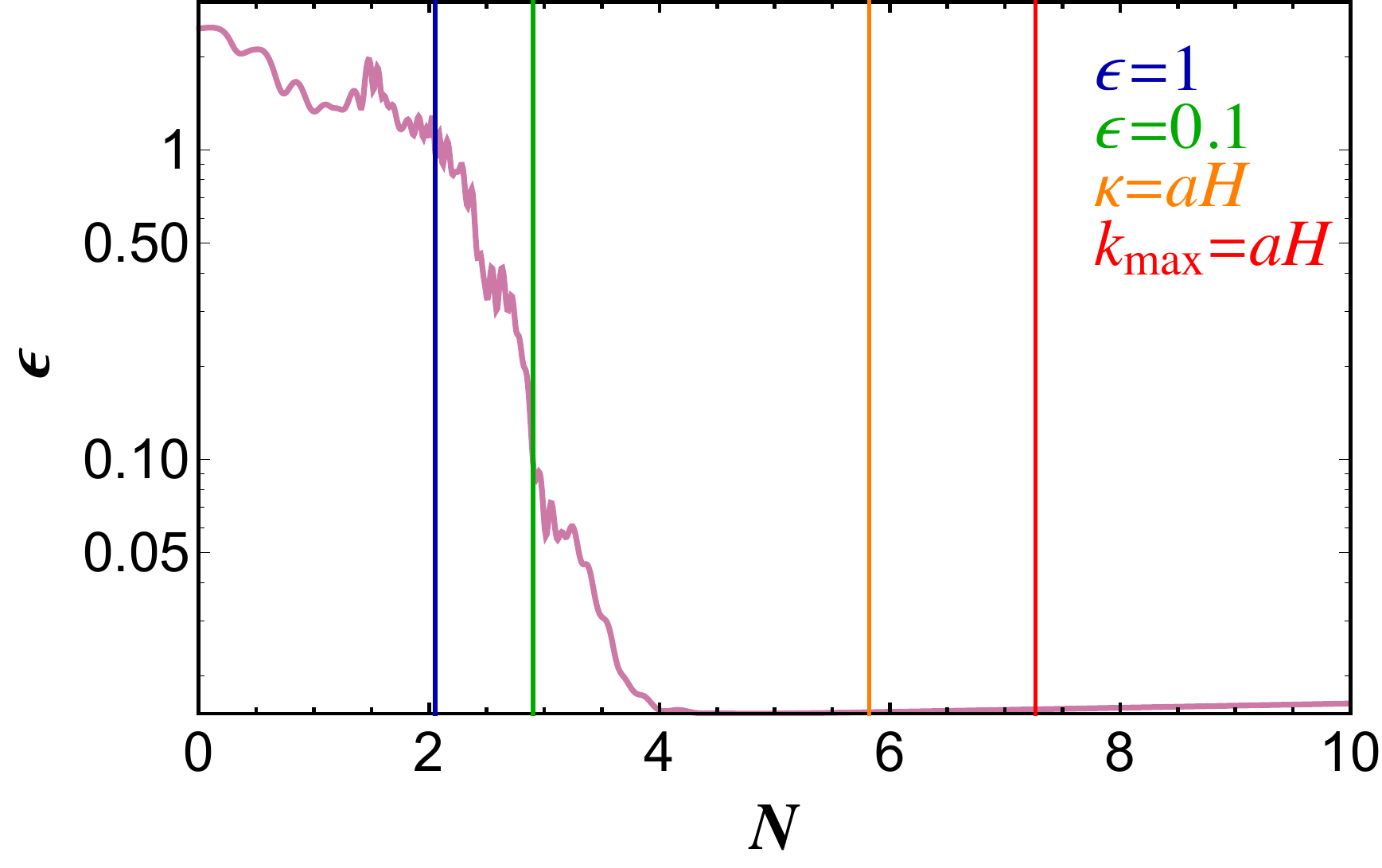}
\caption{\small The slow-roll parameter $\epsilon$ versus $N$ for early times with $\lambda = 10^{-10}$, for the system with large initial fluctuations. Inflation begins at $N \sim 2$ when $\epsilon < 1$, and slow-roll begins by $N \sim 3$ with $\epsilon < 0.1$. Note that modes with comoving wavenumber up to the UV regulator scale $\kappa$ remain within the Hubble radius until $N \sim 6$, and the shortest-wavelength mode in the spectrum, with $k_{\rm max} = 4 \kappa$, crosses outside the Hubble radius at $N \sim 7$. }
\label{TrafficLight}
\end{center}
\end{figure}

As shown in Fig.~\ref{TrafficLight}, for the case with large initial fluctuations, the system reaches the slow-roll inflationary attractor while most of the power in fluctuations remains inside the Hubble radius. In the presented simulation, when slow-roll inflation begins (with $\epsilon \leq 0.1$), all of the modes that had begun inside the Hubble radius still remain inside the Hubble radius. Modes with comoving wavenumber up to the UV regulator scale $\kappa = 5 \bar{H}_{0} = M_{\rm pl}/2$ remain inside the Hubble radius for another 3 efolds after slow-roll begins, and the shortest-wavelength mode in the simulation, with $k_{\rm max} = 4 \kappa = 2 M_{\rm pl}$, crosses outside the Hubble radius more than 4 efolds after the system reaches the slow-roll attractor. Hence the early-time dynamics, during which the system enters a phase of slow-roll inflation, occurs with substantial inhomogeneity on sub-Hubble length scales. For this set of initial conditions, in other words, inflation is robust even amid large initial inhomogeneities and with initial conditions for $\varphi (t)$ dominated by kinetic energy.

\begin{figure}[t]
\begin{center}
\includegraphics[width=2.8in]{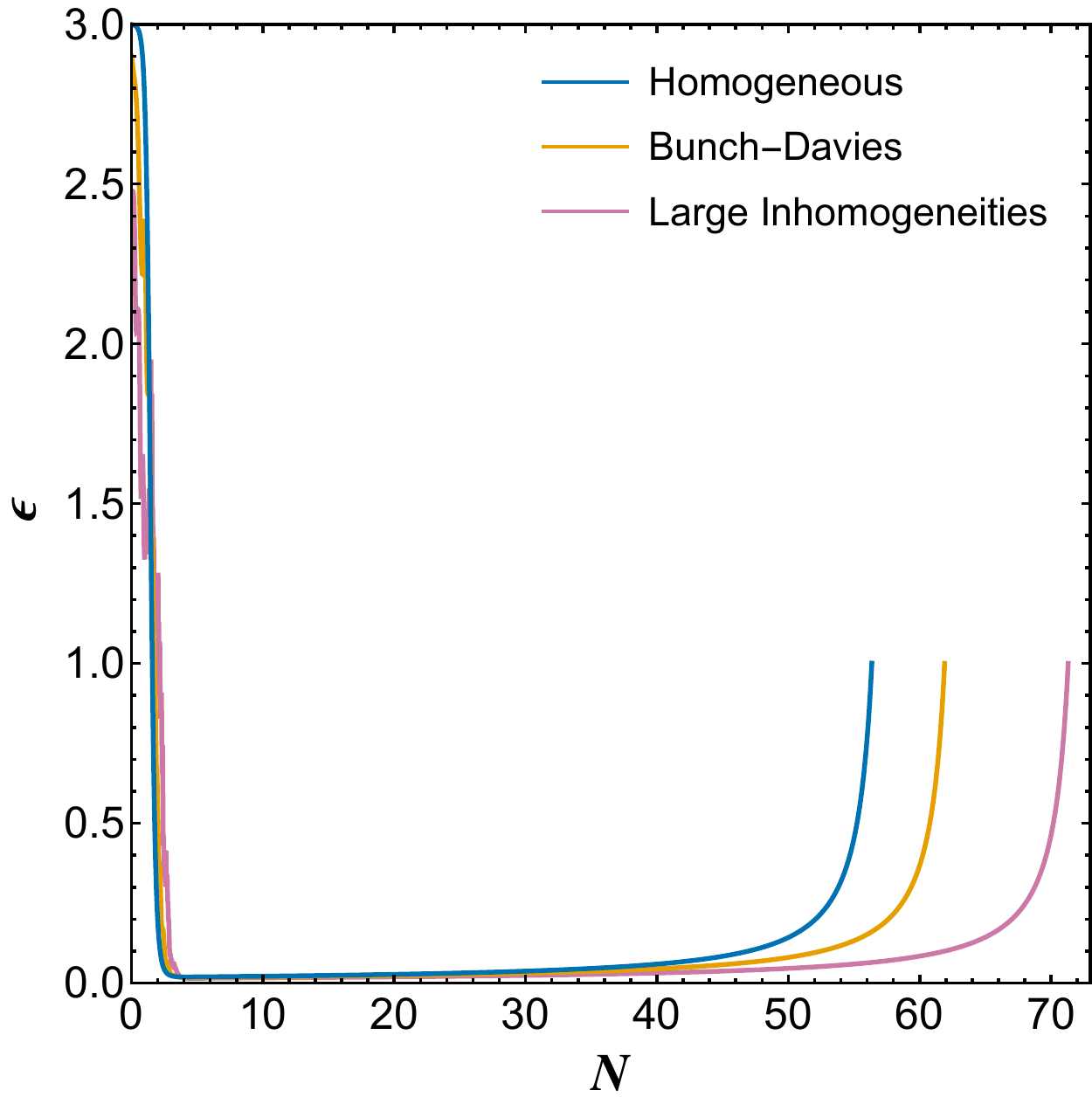}
\caption{\small The evolution of the slow-roll parameter $\epsilon$ versus $N$ with $\lambda = 10^{-10}$ when we neglect fluctuations (blue); when we initialize fluctuations in the Bunch-Davies state (yellow); and for a particular simulation that began with large initial fluctuations, with ${\cal C} = 20$ (pink). }
\label{epsPlotAll}
\end{center}
\end{figure}

Furthermore, as shown in Fig.~\ref{epsPlotAll}, we find that for this set of initial conditions inflation actually persists considerably longer when we include large initial inhomogeneities ($N_{\rm infl} \simeq 69$ efolds of inflation) than when we ignore inhomogeneities ($N_{\rm infl} \simeq 54$ efolds of inflation). (The authors of Ref.~\cite{EastherMultifield} found similar examples in their study of the onset of inflation in multifield models, when fluctuations $\delta \phi (x^\mu)$ and $\delta \psi (x^\mu)$ of the two fields were included.) Remarkably, significant initial inhomogeneities \emph{extended} the duration of inflation in this case, enabling this set of initial conditions $\left( \varphi_{0}, \dot{\varphi}_{0} \right)$ to yield sufficient inflation, with $N_{\rm infl} > 65$.

\subsection{Trajectories in Phase Space}
\label{sec:Trajectories}
We can understand the nontrivial effects of large initial inhomogeneities and their nonlinear backreaction on the evolution of $\varphi (t)$ and $H (t)$ by examining the evolution of the system through the phase space $(\varphi (t),\dot{\varphi} (t))$. As discussed in Refs.~\cite{RemmenCarroll2013,Chowdhury:2019otk}, for single-field models and vanishing spatial curvature $K$, the variables $\varphi (t)$ and $\dot{\varphi} (t)$ define an effective phase space for the evolution of the spatially homogeneous system.  Obviously $(\varphi (t), \dot{\varphi} (t))$ no longer serves as a proper phase space for the full dynamical system when we incorporate the coupled degrees of freedom $\delta \hat{\phi} (x^\mu)$ and $\hat{\Psi} (x^\mu)$, but studying the behavior of the system within $(\varphi (t), \dot{\varphi} (t))$ facilitates comparison with the case in which we neglect fluctuations.

\begin{figure}[ht]
\begin{center}
\includegraphics[width=3in]{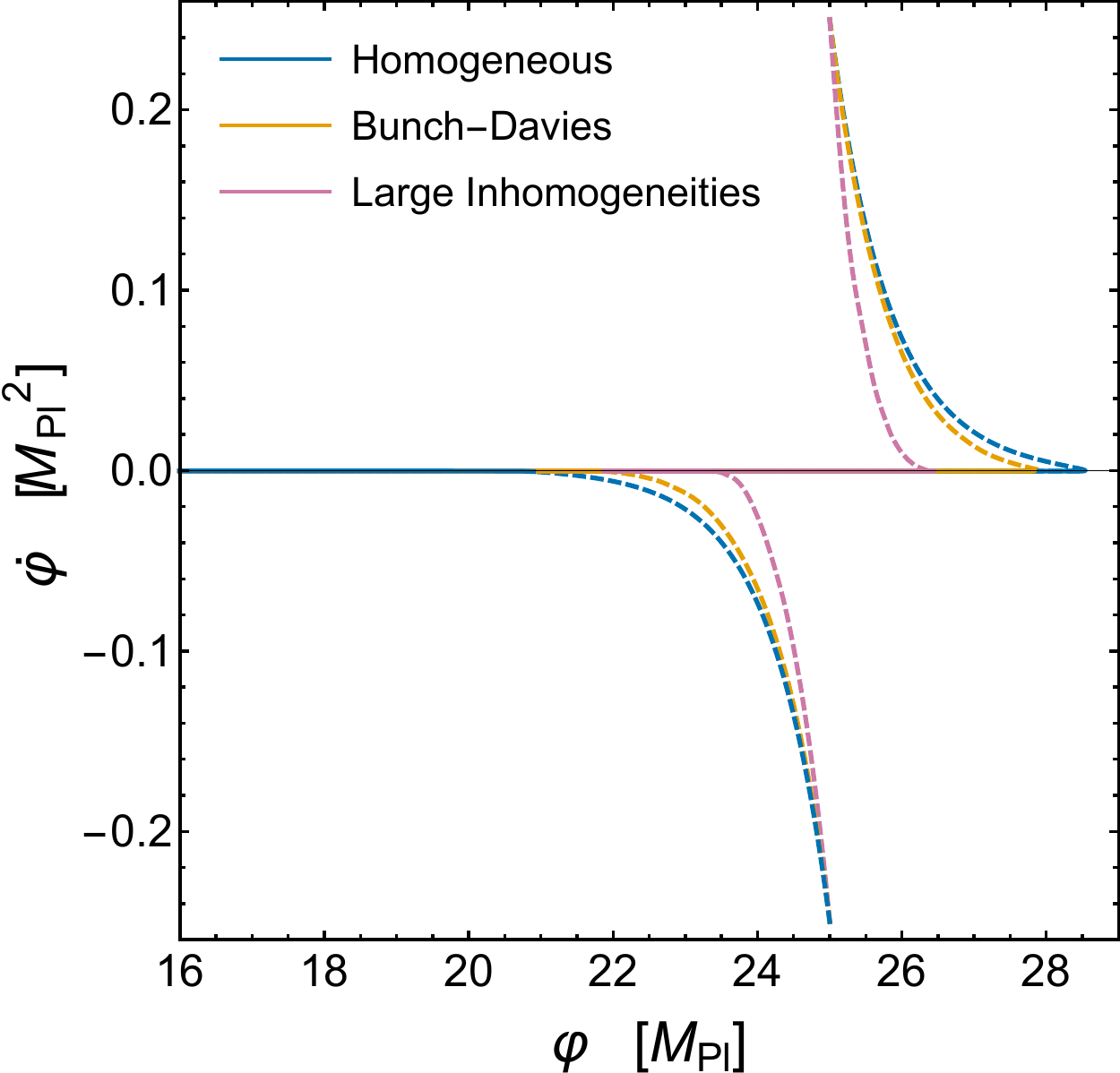}
\caption{ \small The evolution of the system $(\varphi ,\dot{\varphi})$ when we fix $\varphi_0 = 25\, M_{\rm pl}$ and select $\dot{\varphi}_0 = \pm 0.25 \, M_{\rm pl}^2$, with $\lambda = 10^{-10}$. Dashed lines indicate evolution of the system for $\epsilon > 0.1$. The line along $\dot{\varphi} = 0$ corresponds to the slow-roll inflationary attractor.  }
\label{phasePlotAll}
\end{center}
\end{figure}

In Fig.~\ref{phasePlotAll}, we plot the evolution of the system $(\varphi (t), \dot{\varphi} (t))$ when we fix $\varphi_0 = 25 \, M_{\rm pl}$ and select $\dot{\varphi}_0 = \pm 0.25 \, M_{\rm pl}^2$. Dashed lines show the evolution of the system for $\epsilon > 0.1$, before the system reaches the slow-roll inflationary attractor, and solid lines show the evolution once the system has entered slow-roll. We consider cases in which we neglect fluctuations (blue); in which we initialize the fluctuations in the Bunch-Davies state, ${\cal C}_{\rm BD} = 2$ (yellow); and in which we initialize the system with large fluctuations, ${\cal C} = 20$ (pink). As Fig.~\ref{phasePlotAll} makes clear, the value of the field when the system reaches the slow-roll attractor, $\varphi (t_{\rm sr})$, depends on the magnitude of the initial inhomogeneities. In particular, the field $\varphi$ traverses a {\it shorter} distance before arriving at the slow-roll attractor when we incorporate fluctuations, compared to when we neglect fluctuations: either less far ``up the hill" toward higher values of the potential for an initial field velocity $\dot{\varphi}_0 > 0$, or less far ``down the hill" for $\dot{\varphi}_0 < 0$. This effect becomes more pronounced as the size of initial inhomogeneities increases.

We can make sense of this result analytically, using the scaling relations for $H ( t )$ identified in the previous subsection. For these initial conditions and the coupling $\lambda = 10^{-10}$, the system begins with the kinetic energy in $\varphi$ greatly exceeding the potential energy, and hence $\bar{\rho} (t) \propto a^{-6} (t)$ before the system enters slow-roll. This is equivalent to 
\beq
\dot{\varphi} (N) \simeq \frac{ \dot{\varphi_0} }{ a^{3} (N) }= \dot{\varphi_0} \, e^{-3N}
\label{dotvarphiest}
\eeq
at early times. In that limit, Eq.~(\ref{eomvarphi}) reduces to $\ddot{\varphi} + 3 H \dot{\varphi} \simeq 0$, which we may integrate as
\beq
\varphi (t) \simeq \varphi_0 + \dot{\varphi}_0 \int_{a_0}^{a (t) } \frac{ da}{a^4 H} .
\label{eomvarphikinetic}
\eeq
When we neglect fluctuations, $H (t) \propto a^{-3} (t)$ at early times, and we find
\beq
\varphi_h (N) \simeq \varphi_0 + \frac{ \dot{\varphi}_0}{H_0} N ,
\label{varphihN}
\eeq
where the subscript ``$h$" indicates evolution of the homogeneous system. On the other hand, when we include large initial fluctuations with $\delta \rho_{(2)} (t_0) \simeq \bar{\rho} (t_0)$, then $H (t) \propto a^{-2} (t)$ at early times, which yields
\beq
\varphi_q (N) \simeq \varphi_0 + \frac{ \dot{\varphi}_0 }{H_0} \left( 1 - e^{-N} \right) ,
\label{varphiqN}
\eeq
where the subscript ``$q$" indicates the evolution of $\varphi (t)$ when we incorporate effects from the coupled quantum fluctuations. Clearly the field $\varphi$ will traverse a greater distance during early times when the fluctuations are neglected, as in Eq.~(\ref{varphihN}), than when their effects are included, as in Eq.~(\ref{varphiqN}). The solutions for $\dot{\varphi} (N)$ in Eq.~(\ref{dotvarphiest}) and for $\varphi (N)$ in Eqs.~(\ref{varphihN}) or (\ref{varphiqN}) closely match the trajectories shown in Fig.~\ref{phasePlotAll} for the relevant cases, even though the curves in Fig.~\ref{phasePlotAll} come from our full numerical simulations. 

We may use Eqs.~(\ref{varphihN})-(\ref{varphiqN}) to estimate the values $\varphi_h (N_{\rm sr})$ and $\varphi_q (N_{\rm sr})$ at the time $N_{\rm sr}$ when the system reaches the slow-roll attractor. We estimate $N_{\rm sr}$ by setting $\dot{\varphi}^2 (N_{\rm sr} ) / 2 = V (\varphi (N_{\rm sr} ) )$. For $\lambda = 10^{-10}$ and $(\varphi_0, \dot{\varphi}_0) = (25\, M_{\rm pl}, - 0.25 \, M_{\rm pl}^2)$, we have $\bar{H}_0 = [\dot{\varphi}_0^2 / (6 M_{\rm pl}^2 ) ]^{1/2}$ for the homogeneous case and $H_0 = \sqrt{2} \, \bar{H}_0$ for the case with $\delta \rho_{(2)} (t_0) \simeq \bar{\rho} (t_0)$. These yield $\varphi_h (N_{\rm sr}) = 21.4 \, M_{\rm pl}$ and $\varphi_q (N_{\rm sr}) = 23.7 \, M_{\rm pl}$, again closely matching the numerical results shown in Fig.~\ref{phasePlotAll}. 

After the system reaches the slow-roll attractor, $\bar{\rho} (t) \sim {\rm constant}$ while $\delta \rho_{(2)} (t)$ continues to redshift like radiation until most of the modes have crossed outside the Hubble radius, so the dynamics become dominated by $\bar{\rho} \gg \delta \rho_{(2)}$. In that regime, we may use the usual slow-roll approximation to estimate the duration of inflation,
\beq
\begin{split}
N_{\rm infl} &\simeq - \frac{1}{ M_{\rm pl}^2} \int_{\varphi_{\rm sr} }^{\varphi_{\rm end} } d \varphi \left( \frac{ V (\varphi ) }{ V^{(1)} (\varphi) } \right) \\
&= \frac{1}{ 8 M_{\rm pl}^2} \left( \varphi_{\rm sr}^2 - \varphi_{\rm end}^2 \right) ,
\end{split}
\label{Ninflsr}
\eeq
where $\varphi_{\rm sr} = \varphi (N_{\rm sr})$, and $\varphi_{\rm end} = \varphi (N_{\rm end})$ is determined by the condition $\epsilon (N_{\rm end}) = 1$. Again using the usual slow-roll estimate for late times, $\epsilon \simeq (M_{\rm pl}^2 / 2) ( V^{(1)} (\varphi) / V (\varphi) )^2$, we find $\varphi_{\rm end} = \sqrt{8}\, M_{\rm pl}$. Given our estimates of $\varphi_q (N_{\rm sr})$ and $\varphi_h (N_{\rm sr})$, we then find $N_{\rm infl} = 69.2$ efolds of inflation when we incorporate large initial quantum fluctuations, and $N_{\rm infl} = 56.2$ efolds when we neglect fluctuations --- a close match to the behavior shown for the full numerical results in Fig.~\ref{epsPlotAll}.

We can thus understand the most significant effect of the coupled fluctuations on the evolution of the system. Large fluctuations raise the initial value of the Hubble parameter compared to the case with no fluctuations, $H_0 > \bar{H}_0$, thereby increasing the initial Hubble drag on the field $\varphi (t)$. Even more significant, backreaction from the fluctuations changes the scaling of $H(t)$ with $a(t)$ at early times, slowing the rate at which $H(t)$ falls, which further increases the effect of Hubble drag on the evolution of $\varphi (t)$. The backreaction dampens $\varphi$'s motion as the system evolves toward the slow-roll inflationary attractor, such that $\vert \varphi_q (N_{\rm sr}) - \varphi_0 \vert < \vert \varphi_h (N_{\rm sr} ) - \varphi_0 \vert$. Once the system reaches the attractor, the duration of inflation will be governed by the value $\varphi_q (N_{\rm sr})$. For an initial velocity ``up the hill," with $\dot{\varphi}_0 > 0$, $\varphi_q (N_{\rm sr} ) < \varphi_h (N_{\rm sr})$, and the system will spend less time evolving along the inflationary attractor than in the absence of fluctuations. For an initial velocity ``down the hill," with $\dot{\varphi}_0 < 0$, $\varphi_q (N_{\rm sr}) > \varphi_h (N_{\rm sr})$, and the system will spend more time evolving along the inflationary attractor than in the absence of fluctuations.

\subsection{Phase Space of Initial Conditions}

\begin{figure}[t]
\begin{center}
\includegraphics[width=3.41in]{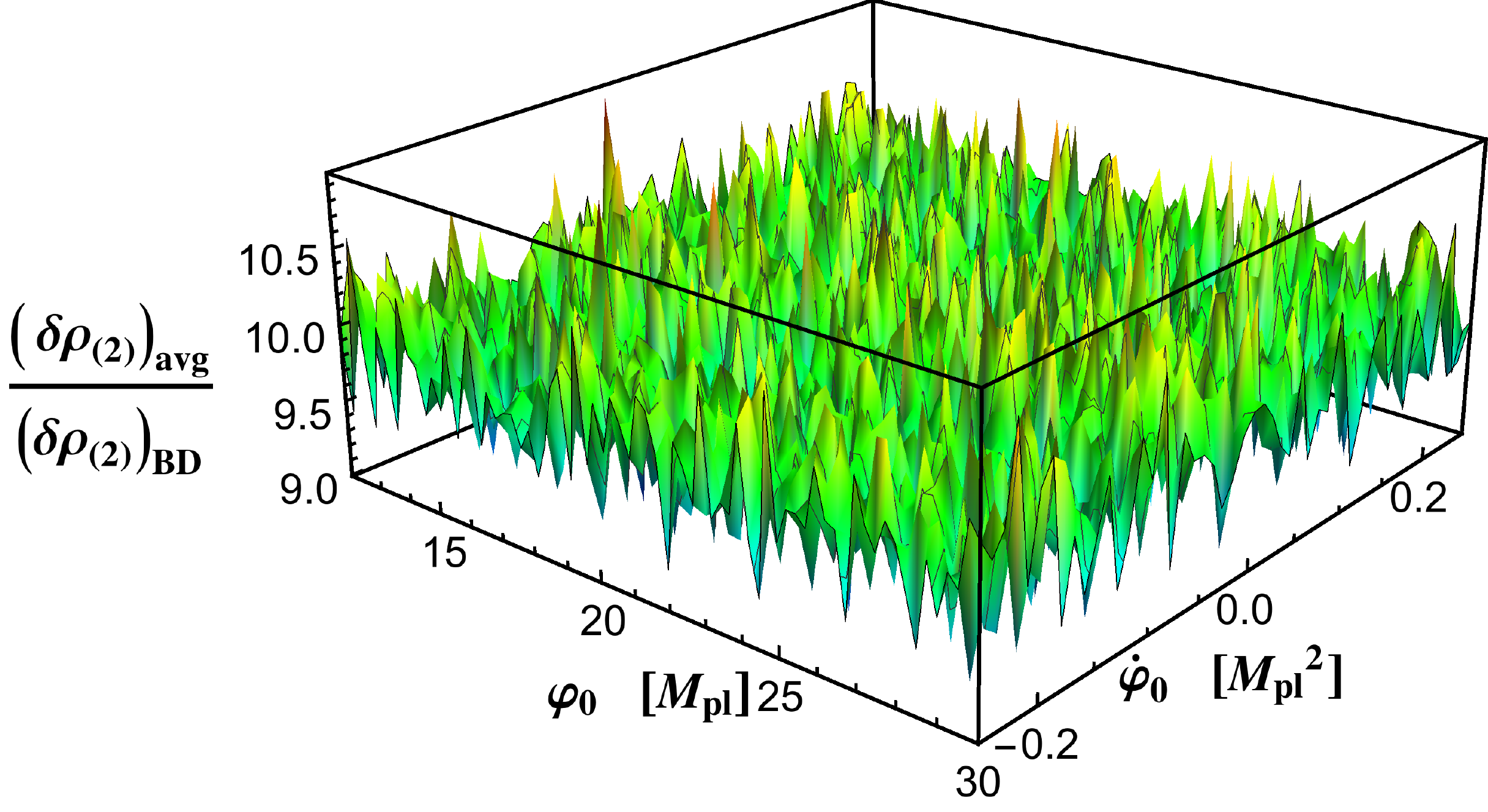}
\caption{ \small The average value of $\delta \rho_{(2)} (t_0)$ at each grid point in $(\varphi_0, \dot{\varphi}_0)$, when quantum fluctuations are initialized with random coefficients drawn from the ranges in Eq.~(\ref{gammadeltaranges}), compared to the value of $\delta \rho_{(2)} (t_0)$ when fluctuations are initialized in the Bunch-Davies state, with $\lambda = 10^{-10}$.}
\label{drho2ratPlot}
\end{center}
\end{figure}

\begin{figure}[t]
\begin{center}
\includegraphics[width=3.41in]{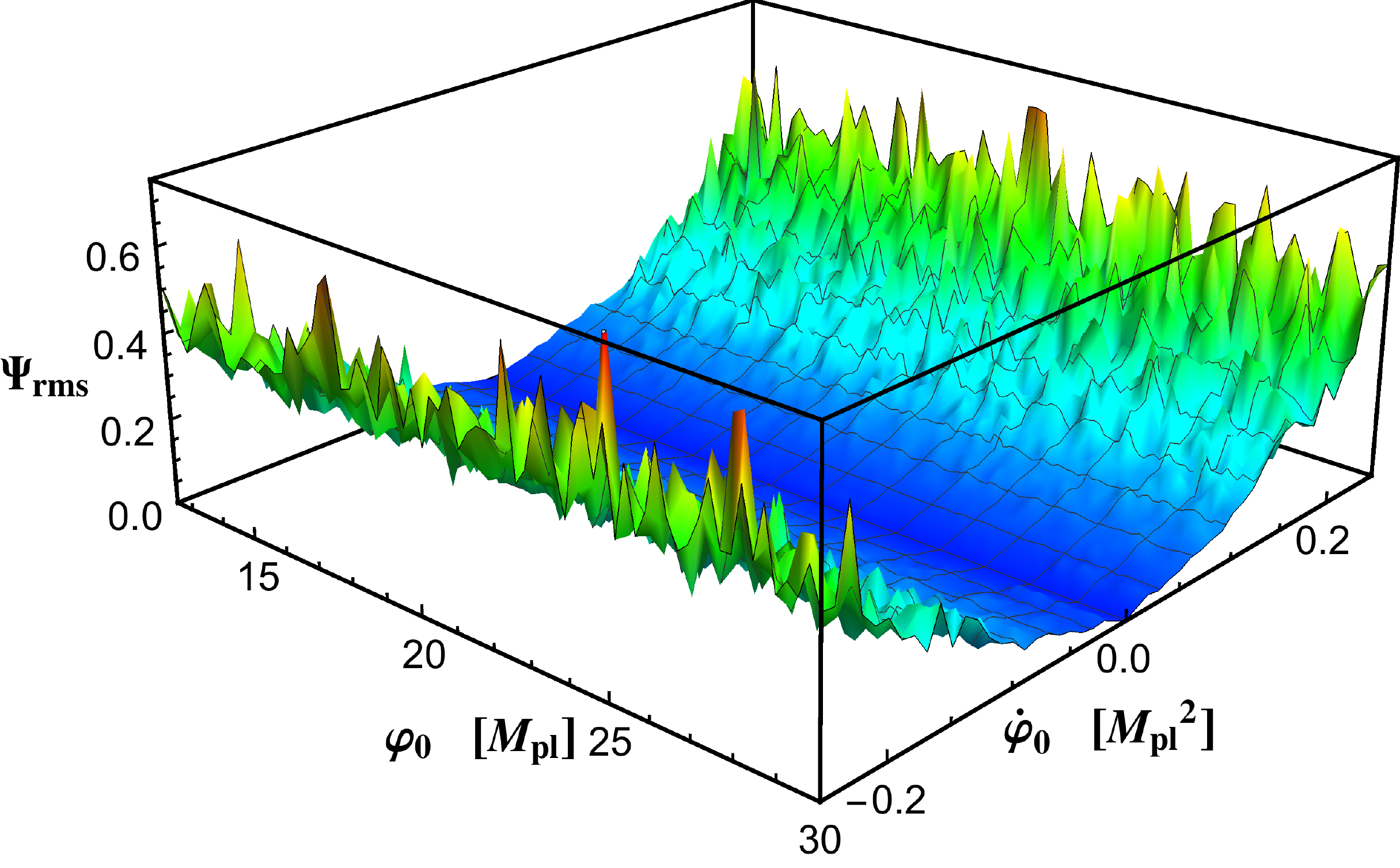}
\caption{\small The average value of $\Psi_{\rm rms} (t_0)$ at each grid point in $(\varphi_0, \dot{\varphi}_0)$ when quantum fluctuations are initialized with random coefficients drawn from the ranges in Eq.~(\ref{gammadeltaranges}), with $\lambda = 10^{-10}$. Note that $\Psi_{\rm rms} (t_0)$ is roughly proportional to $\dot{\varphi}_0^2$.}
\label{PsirmsPlotOn}
\end{center}
\end{figure}

We turn now to discuss the effects of the coupled fluctuations on the evolution of the system across the phase space of initial conditions $(\varphi_0, \dot{\varphi}_0)$ for $\lambda = 10^{-10}$, as we vary $12 \, M_{\rm pl} \leq \varphi_0 \leq 30 \, M_{\rm pl}$ and $- 0.25 \, M_{\rm pl}^2 \leq \dot{\varphi}_0 \leq 0.25 \, M_{\rm pl}^2$. To investigate the phase space behavior for the perturbed initial conditions, we construct averages from the 32 samples evolved at each point in $(\varphi_0, \dot{\varphi}_0)$.
Figure \ref{drho2ratPlot} shows the average value of $\delta \rho_{(2)} (t_0)$ at each grid point compared to $\delta \rho_{(2)} (t_0)$ for Bunch-Davies initial conditions, confirming that for the ranges of coefficients in Eq.~(\ref{gammadeltaranges}), we find initial energy densities about ten times greater than for the Bunch-Davies state. In Fig.~\ref{PsirmsPlotOn}, we plot the average of the initial value $\Psi_{\rm rms} (t_0) \equiv [ \langle \hat{\Psi}^2 (t_0) \rangle ]^{1/2}$ at each grid point, confirming that for large quantum fluctuations, initialized such that ${\cal C} \simeq 20$, the system begins with $\vert \Psi (t_0, {\bf x} ) \vert \lesssim 0.5$.

\begin{figure}
    \centering
    \includegraphics[width=3.41in]{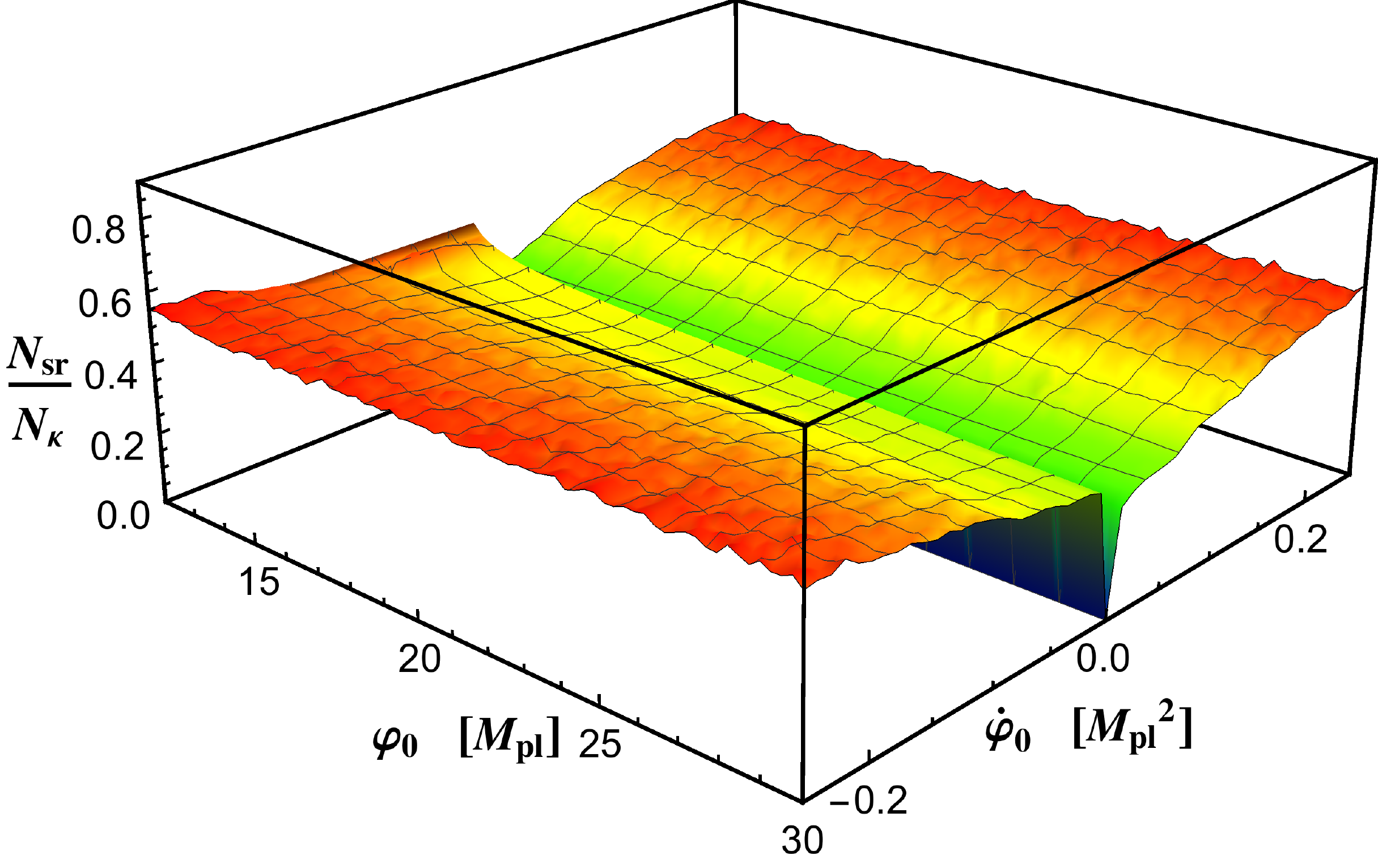}
    \caption{\small Average of the ratio of the time $N_{\rm sr}$ when the system first reaches the slow-roll inflationary attractor with $\epsilon \leq 0.1$, to the time $N_\kappa$ when the mode with comoving wavenumber equal to the UV regulator scale crosses outside the Hubble radius, $\kappa = aH$, for the case of large initial fluctuations (${\cal C} \simeq 20$), with $\lambda = 10^{-10}$. For all simulations across $(\varphi_0, \dot{\varphi}_0)$, we find $N_{sr} / N_\kappa \leq 0.55 \pm 0.06$.}
    \label{trafficlight3d}
\end{figure}

Across $(\varphi_0, \dot{\varphi}_0)$, when the system begins with large initial inhomogeneities, the system reaches the slow-roll inflationary attractor ($\epsilon \leq 0.1$) while significant power remains in fluctuations on sub-Hubble scales. Figure \ref{trafficlight3d} shows the average value of the ratio $N_{\rm sr} / N_{\kappa}$ at each grid point for fluctuations that begin with ${\cal C} \simeq 20$, where $N_{\kappa}$ is the time when the mode with comoving wavenumber equal to the UV regulator scale $\kappa$ crosses outside the Hubble radius, $\kappa = aH$. In all simulations, $N_{\rm sr} / N_\kappa \leq 0.55 \pm 0.06$. The ratio drops to zero at $\dot{\varphi}_0 = 0$, because at those locations in phase space the system begins on the slow-roll attractor, and hence $N_{\rm sr} = 0$.

\begin{figure*}
\begin{minipage}{.32\linewidth}
\centering
\includegraphics[scale=0.31]{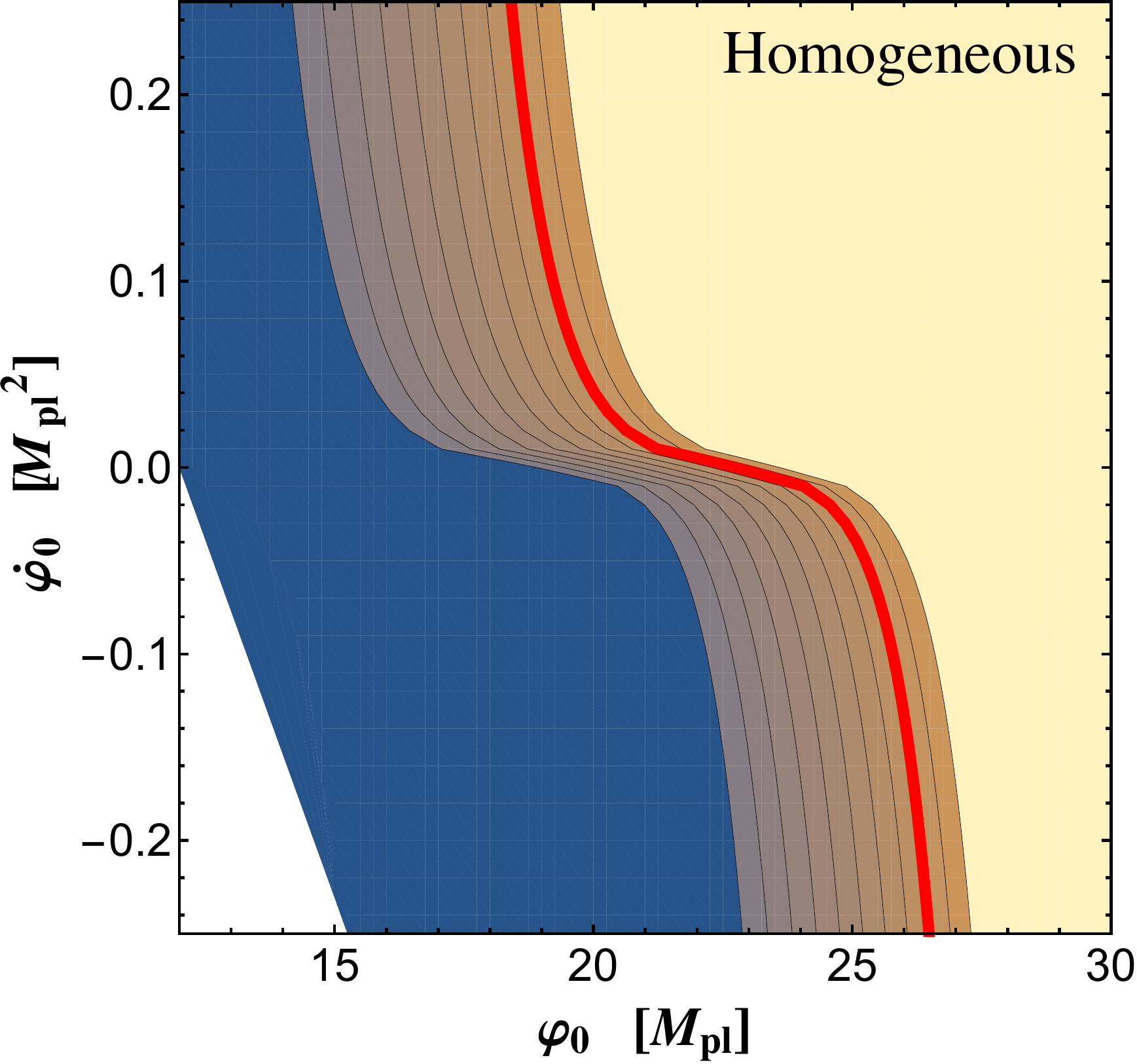}
\end{minipage}
\begin{minipage}{.32\linewidth}
\centering
\includegraphics[scale=0.31]{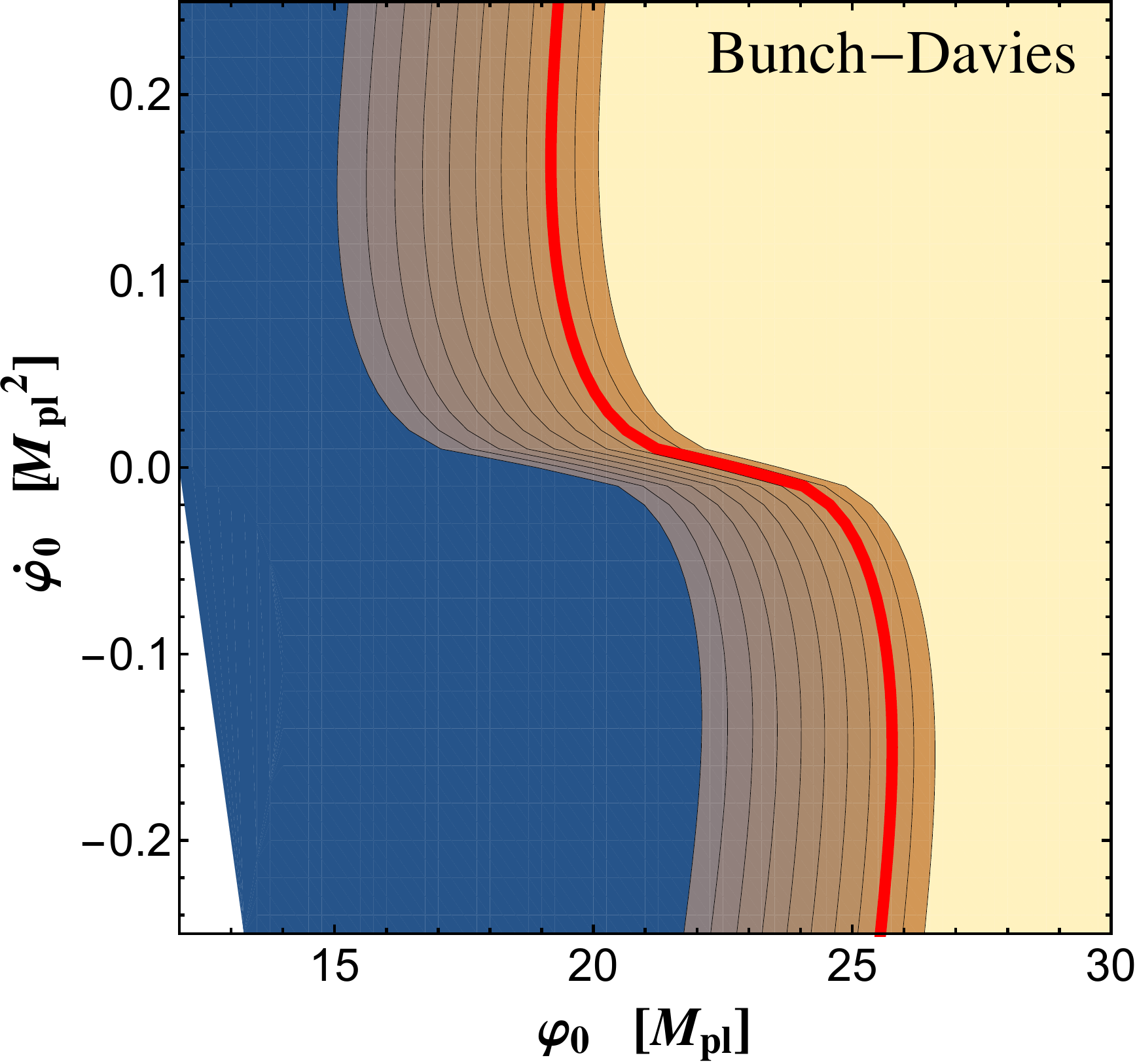}
\end{minipage}
\begin{minipage}{.32\linewidth}
\centering
\includegraphics[scale=0.43]{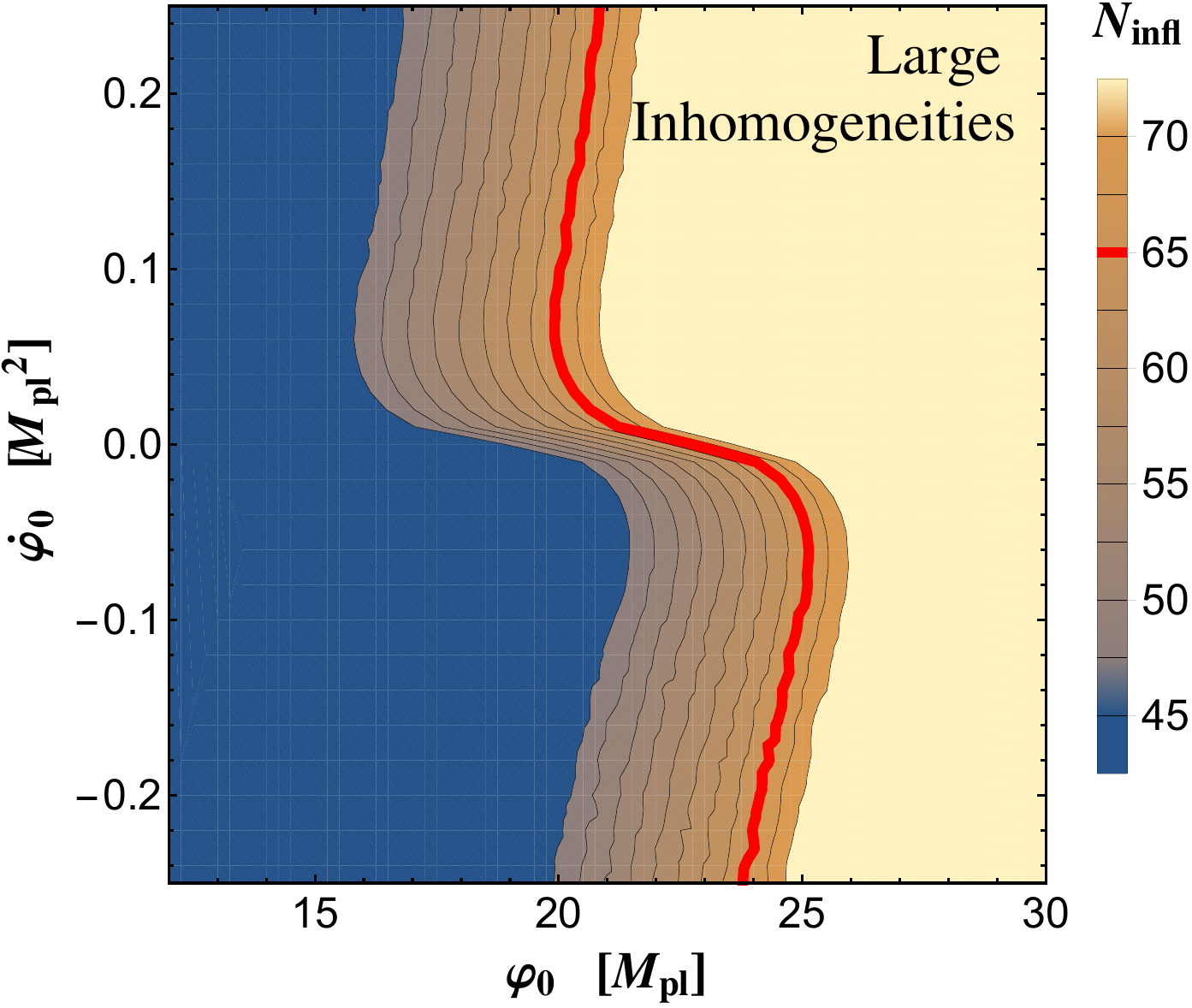}
\end{minipage}
\caption{\small Contours of constant $N_{\rm infl}$ in $(\varphi_0, \dot{\varphi}_0)$ for $\lambda = 10^{-10}$ when the fluctuations are neglected ({\it left}); when the fluctuations are initialized in the Bunch-Davies state, ${\cal C}_{\rm BD} = 2$ ({\it middle}); and when the fluctuations are initialized with random coefficients $\gamma_{n \ell m}$ and $\delta_{n \ell m}$ for each mode drawn from the ranges in Eq.~(\ref{gammadeltaranges}), which yields ${\cal C} \simeq 20$ ({\it right}). For the case of large initial fluctuations ({\it right}), the contours of constant $N_{\rm infl}$ were evaluated by averaging $\numsimflucts$ simulations per grid point. In each plot, regions of dark blue indicate $N_{\rm infl} < 45$ and regions of light yellow indicate $N_{\rm infl}  \geq 70$. In white regions in the lower left, the system never entered slow-roll. The critical lines that yield $N_{\rm infl} = 65$ efolds of inflation are indicated in red. }
\label{Ninflcontours}
\end{figure*}

Results for the duration of inflation $N_{\rm infl}$ across these cases are shown in Fig.~\ref{Ninflcontours}. Consider first the case in which the effects of the coupled fluctuations are neglected. For a given value $\varphi_0$, large initial velocities $\dot{\varphi}_0 > 0$ prolong the duration of inflation compared to the case with $\dot{\varphi}_0 = 0$, whereas large initial velocities $\dot{\varphi}_0 < 0$ decrease the duration of inflation. When initial inhomogeneities are included they backreact on $H(t)$, increasing the effect of Hubble drag on $\varphi (t)$, thereby affecting the shape of the contours of constant $N_{\rm infl}$ within $(\varphi_0, \dot{\varphi}_0)$. In particular, the effects of large $\vert \dot{\varphi}_0 \vert$ are more quickly damped than when the fluctuations are neglected, so that the field $\varphi$ spends less time evolving along the slow-roll inflationary attractor than the corresponding case without fluctuations for $\dot{\varphi}_0 > 0$, and more time along the attractor for $\dot{\varphi}_0 < 0$.

\begin{figure}
\begin{center}
\includegraphics[width=3in]{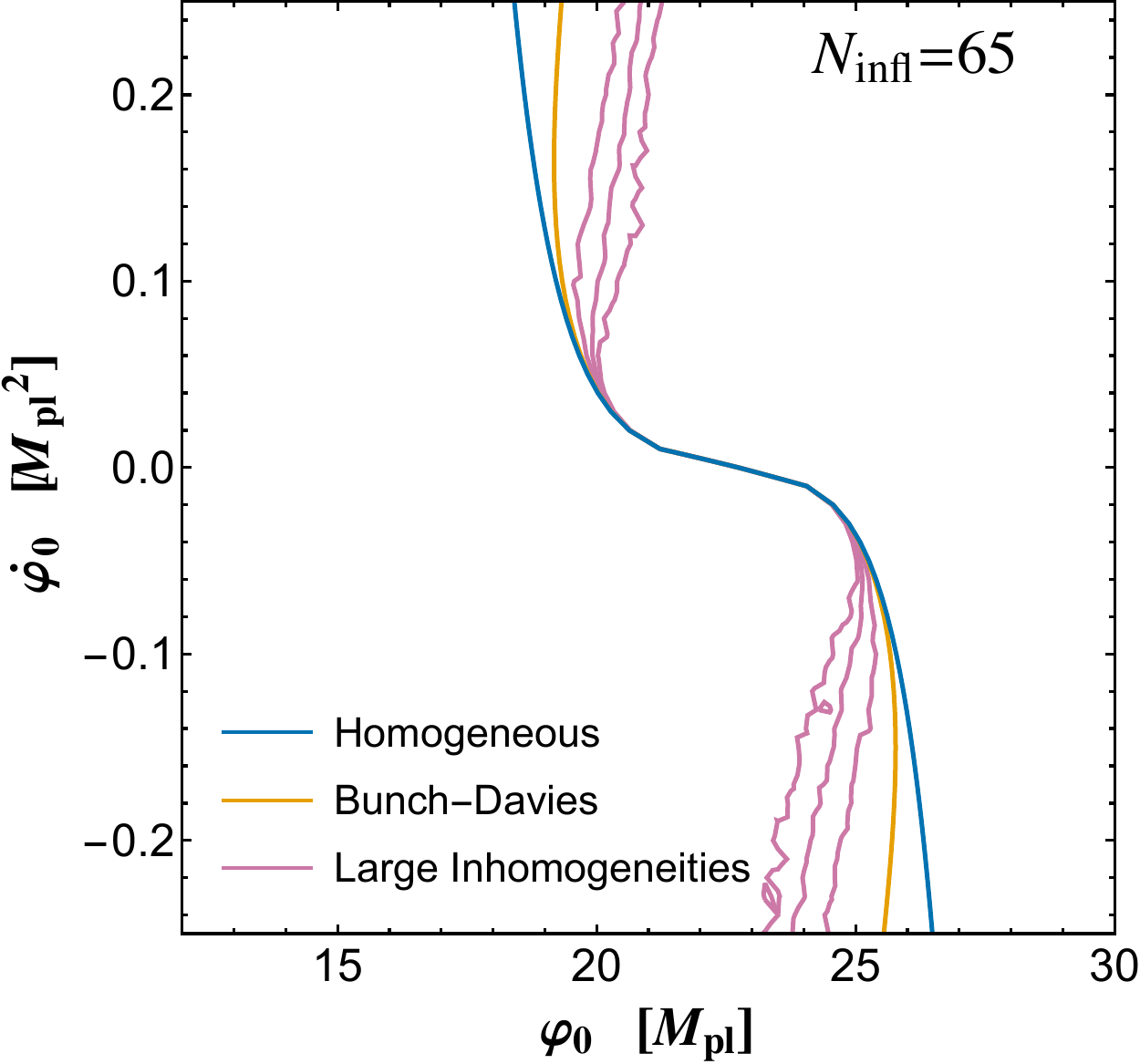}
\caption{\small The critical line in $(\varphi_0, \dot{\varphi}_0)$ that yields sufficient inflation, with $N_{\rm infl} \geq 65$, for $\lambda = 10^{-10}$, for the cases of homogeneous evolution with no fluctuations (blue); fluctuations in the Bunch-Davies initial state (yellow); and large initial fluctuations (pink). (Points to the right of the critical lines achieve sufficient inflation.) For the latter, we show the critical line based on averaging across $\numsimflucts$ simulations per grid point, as well as $\pm 2\sigma$ contours. The jitter in the pink curves arises from stochastic fluctuations, and it is expected that the curves would become smooth as we increase both the resolution of our sampling and the number of simulations at each point in phase space.}
\label{critPlotAll}
\end{center}
\end{figure}

Figure \ref{critPlotAll} shows the critical line in $(\varphi_0, \dot{\varphi}_0)$, to the right of which the system yields $N_{\rm infl} > 65$ efolds of inflation, for the homogeneous system (when we neglect fluctuations), for fluctuations in the Bunch-Davies initial state with ${\cal C}_{\rm BD} = 2$, and for larger initial fluctuations with ${\cal C} \simeq 20$. For the cases with large initial fluctuations, we plot the critical line based on averaging across $\numsimflucts$ simulations per grid point, as well as $\pm 2\sigma$ contours. 

Consistent with the analysis in the previous subsection, the effects of large initial fluctuations are symmetric for $\pm \vert \dot{\varphi}_0 \vert$, and most significant for large $\vert \dot{\varphi}_0 \vert$. As $\vert \dot{\varphi}_0 \vert$ increases, the initial value $\bar{H}_0$ increases; the greater initial energy scale $\bar{H}_0$, in turn, yields more initial energy density in fluctuations, $\delta \rho_{(2)} (t_0)$, which raises $H_0 > \bar{H}_0$ and affects the scaling of $H(t)$ with $a(t)$ at early times. The larger fluctuations also seed larger initial inhomogeneities, $\Psi (t_0, {\bf x})$.

Although the effects of the nonlinear dynamics of the fluctuations are most pronounced at large $\vert \dot{\varphi}_0 \vert$, the volume of the (projected) phase space of initial conditions $(\varphi_0, \dot{\varphi}_0)$ that yields sufficient inflation is {\it conserved}. The same volume of the $(\varphi_0, \dot{\varphi}_0)$ phase space that yields sufficient inflation which is lost in the region with $\dot{\varphi}_0 > 0$, compared to the homogeneous case, is gained in the region with $\dot{\varphi}_0 < 0$. This provides a useful quantitative measure of the robustness of single-field inflation to large initial inhomogeneities.

\subsection{Varying the Coupling Constant}
\label{varycoupling}

\begin{figure}
    \centering
    \includegraphics[width=3.41in]{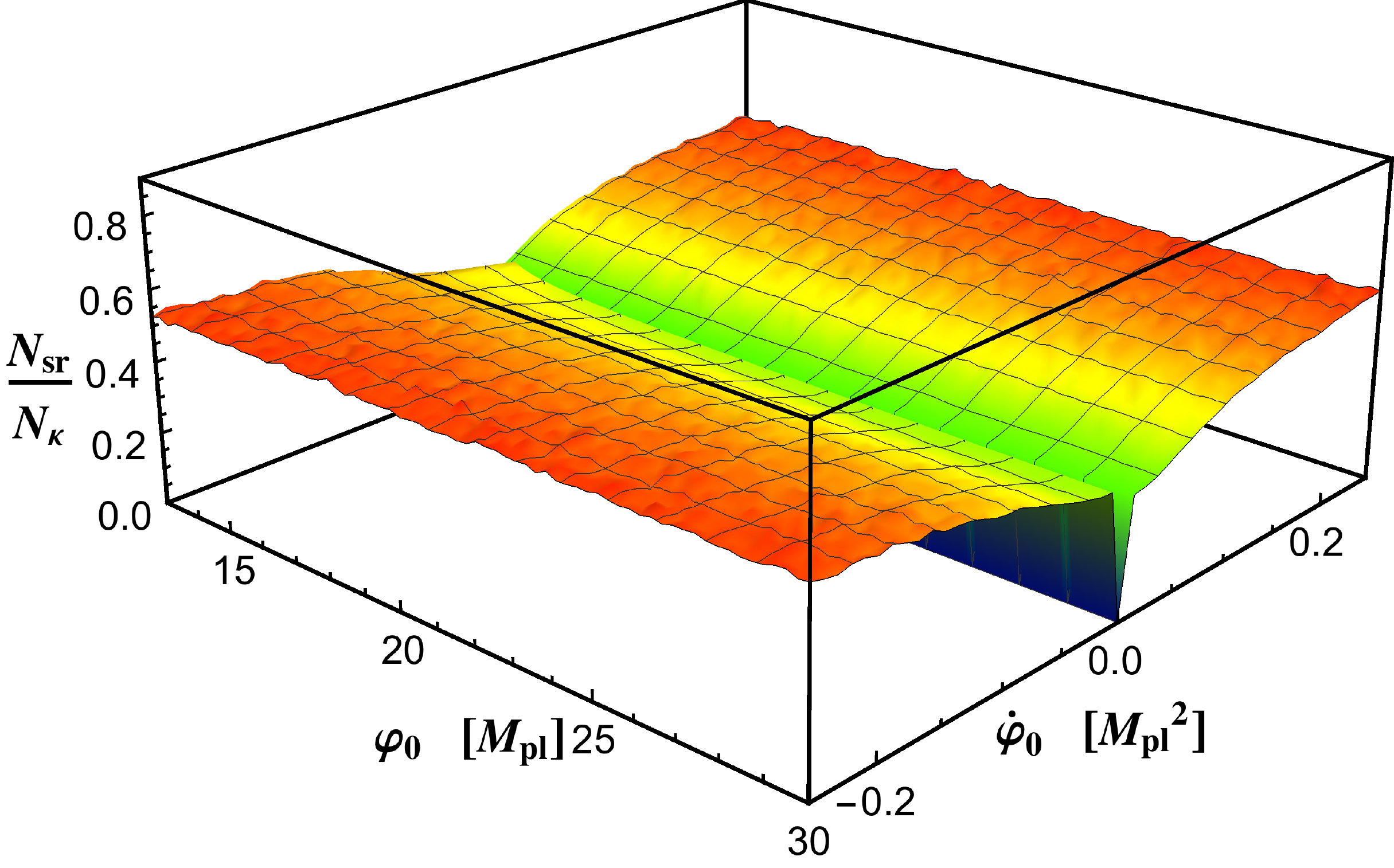}
    \caption{\small Average of the ratio of the time $N_{\rm sr}$ when the system first reaches the slow-roll inflationary attractor to the time $N_\kappa$ when $\kappa = aH$ for the case of large initial fluctuations (${\cal C} \simeq 20)$, with $\lambda = 10^{-12}$. Across all simulations we find $N_{\rm sr} / N_\kappa \leq 0.53 \pm 0.04$.}
    \label{trafficlight3dsmallL}
\end{figure}

The main impact of lowering the coupling constant from $\lambda = 10^{-10}$ to $\lambda = 10^{-12}$ is to increase the ratio $H_0 / H_{\rm infl}$ from ${\cal O} (10^{2})$ to ${\cal O} (10^{3})$. That prolongs the time during which the enhanced Hubble drag from the coupled fluctuations affects the evolution of $\varphi (t)$, compared to the case in which the fluctuations are neglected. We divided the phase space of initial conditions $(\varphi_0, \dot{\varphi}_0)$ into the same grid as for the $\lambda = 10^{-10}$ case, and considered cases in which we neglected fluctuations, began with fluctuations in the Bunch-Davies initial state, and began with larger initial fluctuations, with the random coefficients for each mode $\delta \phi_{n \ell m} (t_0)$ and $\delta \dot{\phi}_{n \ell m} (t_0)$ drawn from the ranges in Eq.~(\ref{gammadeltaranges}). Because the contributions to $\bar{\rho} (t_0)$ and $\delta \rho_{(2)} (t_0)$ that are proportional to $\lambda$ remain subdominant for $\lambda = 10^{-12}$ as for $\lambda = 10^{-10}$, the initial values for $(\delta \rho_{(2)} (t_0) )_{\rm avg} / (\delta\rho_{(2)} (t_0) )_{\rm BD}$ and for $\Psi_{\rm rms} (t_0)$ are little changed from the results shown in Figs.~\ref{drho2ratPlot} and \ref{PsirmsPlotOn}. Likewise, we again find that throughout $(\varphi_0, \dot{\varphi}_0)$, $N_{\rm sr} / N_\kappa \leq 0.53 \pm 0.04$, as shown in Fig.~\ref{trafficlight3dsmallL}.

\begin{figure*}
\begin{minipage}{.32\linewidth}
\centering
\includegraphics[scale=0.31]{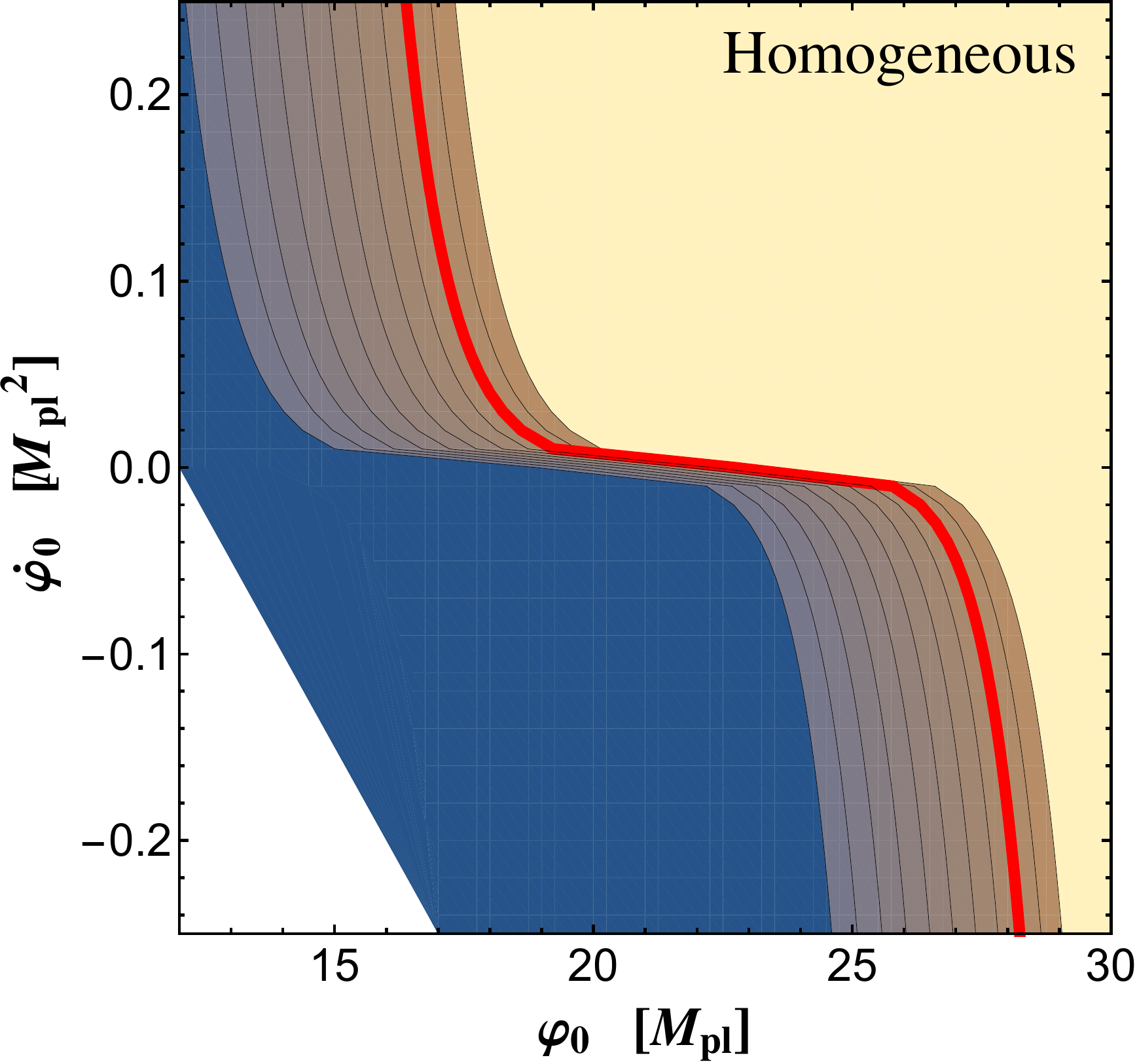}
\end{minipage}
\begin{minipage}{.32\linewidth}
\centering
\includegraphics[scale=0.31]{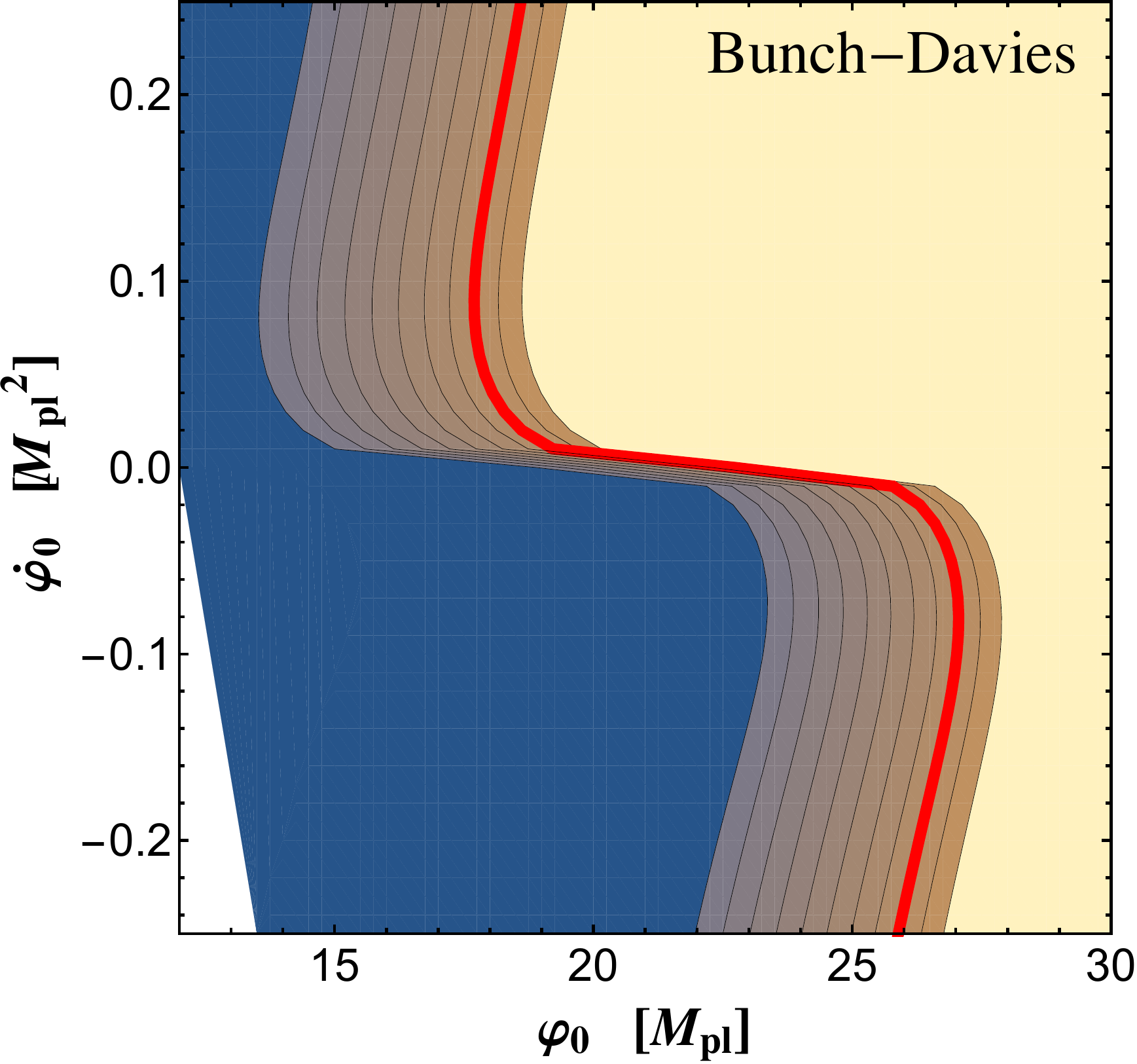}
\end{minipage}
\begin{minipage}{.32\linewidth}
\centering
\includegraphics[scale=0.43]{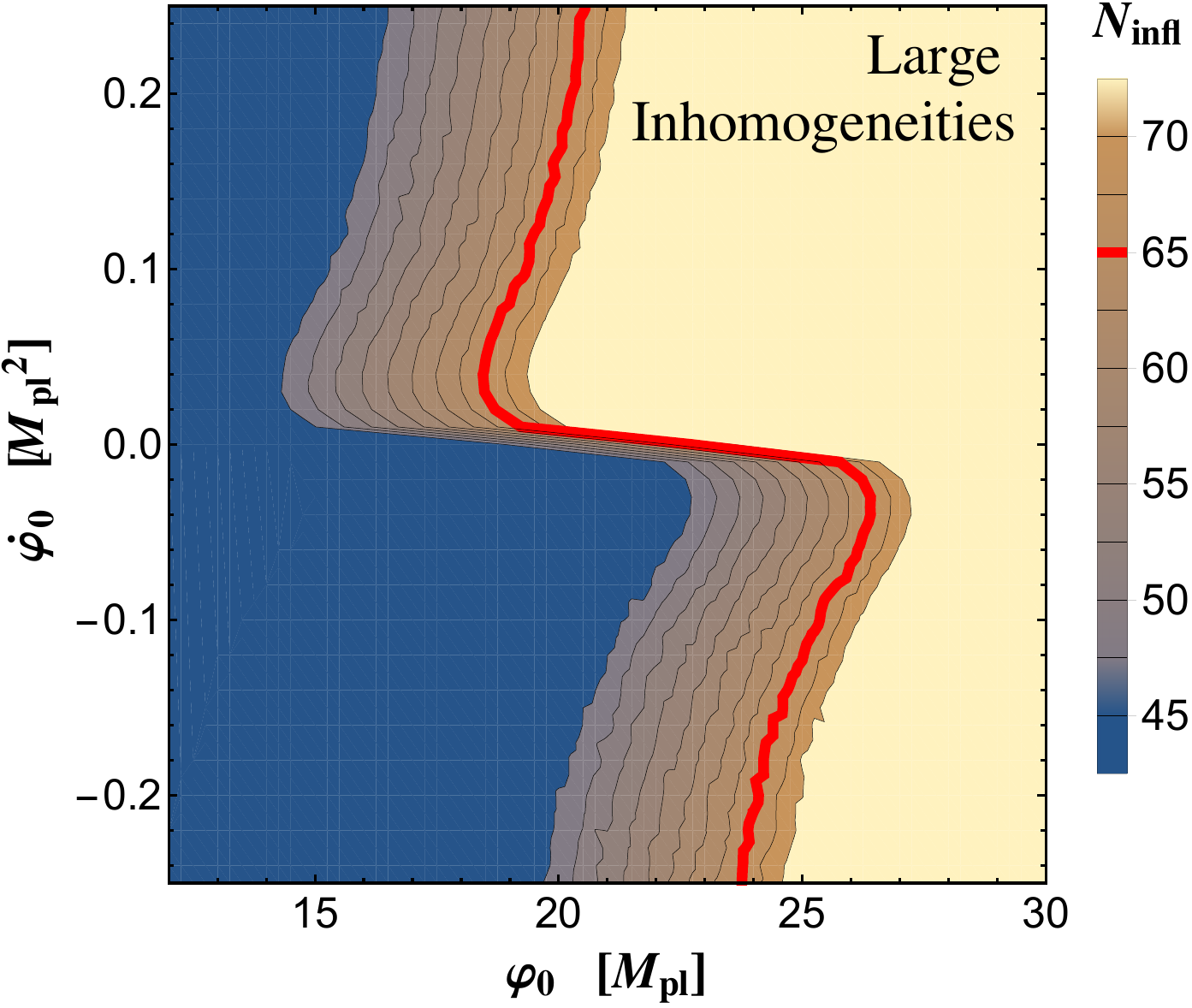}
\end{minipage}
\caption{\small Contours of constant $N_{\rm infl}$ in $(\varphi_0, \dot{\varphi}_0)$ for $\lambda = 10^{-12}$ when the fluctuations are neglected ({\it left}); when the fluctuations are initialized in the Bunch-Davies state with ${\cal C}_{\rm BD} = 2$ ({\it middle}); and when the fluctuations are initialized with random coefficients $\gamma_{n \ell m}$ and $\delta_{n \ell m}$ for each mode drawn from the ranges in Eq.~(\ref{gammadeltaranges}), which yields ${\cal C} \simeq 20$ ({\it right}). For the case of large initial fluctuations ({\it right}), the contours of constant $N_{\rm infl}$ were evaluated by averaging $\numsimflucts$ simulations per grid point. In each plot, regions of dark blue indicate $N_{\rm infl} < 45$ and regions of light yellow indicate $N_{\rm infl}  \geq 70$. In white regions in the lower left, the system never entered slow-roll. The critical lines that yield $N_{\rm infl} = 65$ efolds of inflation are indicated in red. }
\label{Ninflcontoursm12}
\end{figure*}

\begin{figure}
    \centering
    \includegraphics[width=3in]{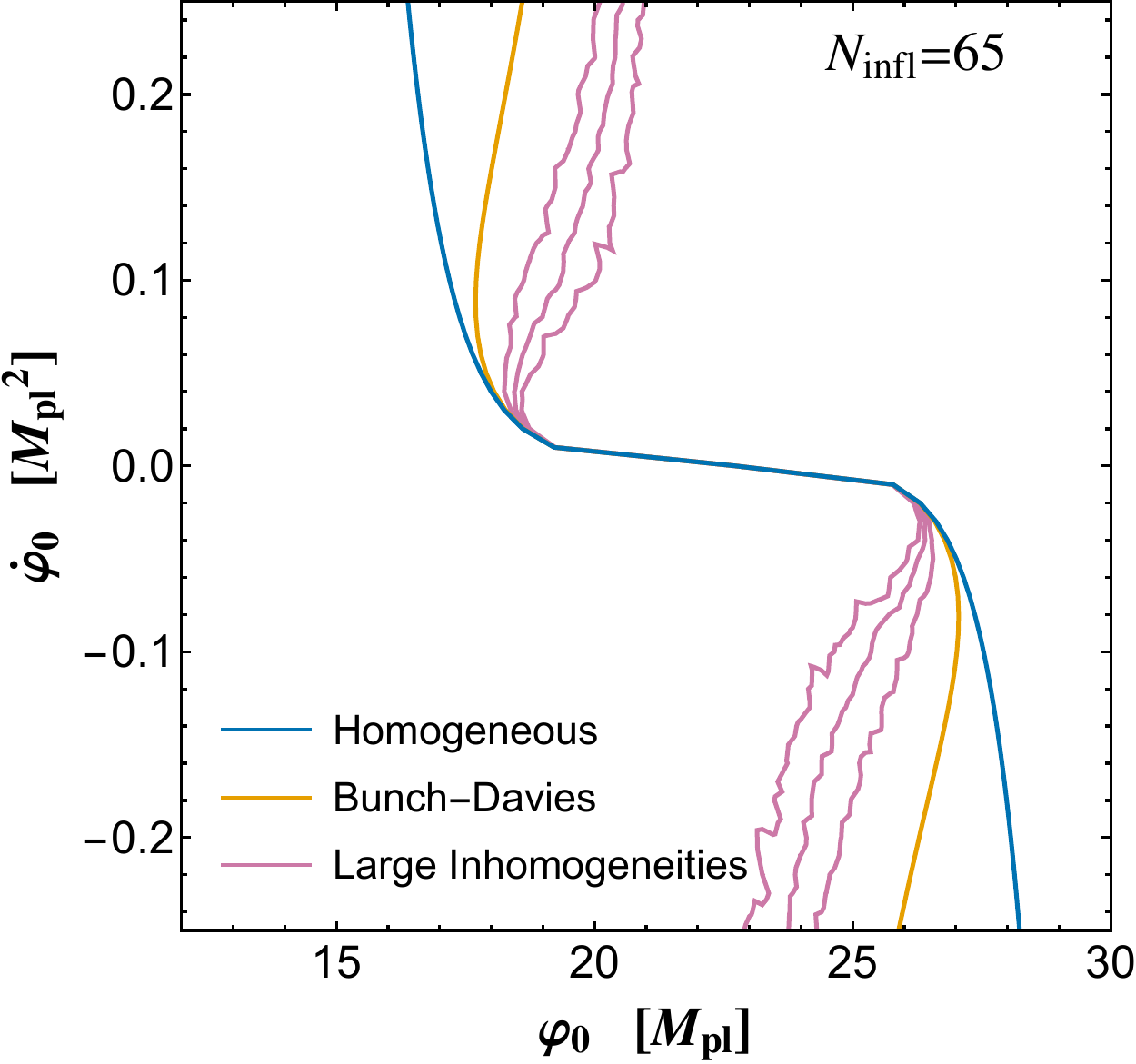}
    \caption{\small The critical line in $(\varphi_0, \dot{\varphi}_0)$ that yields $N_{\rm infl} \geq 65$ efolds of inflation for $\lambda = 10^{-12}$, for the cases of homogeneous evolution (blue); fluctuations in the Bunch-Davies initial state (yellow); and large initial fluctuations (pink). (Points to the right of the critical lines achieve sufficient inflation.) For the latter, we show the critical line based on averaging across $\numsimflucts$ simulations per grid point, as well as $\pm 2\sigma$ contours.}
    \label{CritPlotAllSmallL}
\end{figure}

Figure \ref{Ninflcontoursm12} shows contours of constant $N_{\rm infl}$ in $(\varphi_0, \dot{\varphi}_0)$ with $\lambda = 10^{-12}$ for the three cases of interest: no fluctuations, Bunch-Davies intial state (${\cal C}_{\rm BD} = 2$), and larger initial fluctuations (with ${\cal C}\simeq 20$). In Fig.~\ref{CritPlotAllSmallL} we plot the critical line in $(\varphi_0, \dot{\varphi}_0)$ that yields $N_{\rm infl} = 65$ efolds of inflation for each of the three cases. For the cases with large initial fluctuations, we plot the critical line based on averaging across $\numsimflucts$ simulations per grid point, as well as $\pm 2 \sigma$ contours. The results are comparable to the case with $\lambda = 10^{-10}$. Although the effects of the coupled fluctuations are more dramatic with the smaller coupling --- because the system takes more time to reach the inflationary attractor, and hence the enhanced Hubble drag in the presence of coupled fluctuations acts longer on the evolution of $\varphi (t)$ --- the effects across the projected phase space $(\varphi_0, \dot{\varphi}_0)$ are once again symmetrical for $\pm \vert \dot{\varphi}_0 \vert$, indicating that the total volume of the $(\varphi_0, \dot{\varphi}_0)$ phase space that yields sufficient inflation is conserved.

\section{Conclusions}
\label{sec:Conclusions}

In this paper we have analyzed the onset of inflation for a simple single-field model, $V (\phi) = \lambda \phi^4 / 4$, when the system begins with significant inhomogeneities on length-scales shorter than the initial Hubble radius. We incorporate certain nonlinear interactions among the coupled degrees of freedom by using the nonperturbative Hartree approximation, which resums an infinite set of Feynman diagrams involving the self-interacting quantum fluctuations $\delta \hat{\phi} (x^\mu)$ to construct a dressed propagator. By initializing the system in an excited state, with the energy density of fluctuations approximately ten times greater than in the minimum-energy Bunch-Davies state, our simulations begin with significant spatial inhomogeneities, parameterized by the scalar metric perturbation $\vert \Psi (t_0, {\bf x}) \vert \lesssim 0.5$.

The energy density in fluctuations, $\delta \rho_{(2)} (t)$, backreacts on the evolution of the Hubble parameter $H(t)$, affecting the scaling of $H (t)$ with $a(t)$. This backreaction, in turn, leads to increased Hubble drag on the evolution of the vacuum expectation value of the field, $\varphi (t)$, affecting how quickly $\varphi (t)$ arrives at the slow-roll inflationary attractor, compared to the case in which one neglects fluctuations. 

The impact of inhomogeneities on the evolution of the system is largest for initial conditions that yield the greatest initial value of $H (t_0)$, since the initial energy density in fluctuations scales as $H^4 (t_0)$. Compared to those regions of $(\varphi_0, \dot{\varphi}_0)$ that yield $N_{\rm infl} \geq 65$ efolds of inflation when one neglects inhomogeneities, we find some regions that {\it fail} to yield sufficient inflation when we incorporate inhomogeneities, and an equal volume of regions that {\it succeed} in producing $N_{\rm infl} \geq 65$ but which had failed to do so in the absence of inhomogeneities. (See also Ref.~\cite{EastherMultifield}.) In other words, the total volume of the space $(\varphi_0, \dot{\varphi}_0)$ that yields sufficient inflation is conserved when we incorporate nonlinear backreaction from inhomogeneities, compared to the case in which we neglect inhomogeneities.

Although we have analyzed the system numerically for a specific form of $V (\phi)$, the arguments about the robustness of inflation for such large-field models do not depend on our choice of $V (\phi)$. All that enters into our semi-analytic argument is that the system can begin with large initial quantum fluctuations, such that $\delta \rho_{(2)} (t_0)$ is comparable to (or greater than) the initial energy density associated with the vacuum expectation value, $\bar{\rho} (t_0)$. For weakly coupled models --- as required for large-field inflation, in order to produce a spectrum of primordial density perturbations consistent with observations --- we generically expect that $\delta \rho_{(2)} (t)$ will evolve at early times with an equation of state like that of a gas of (nearly) massless particles, scaling as $\delta \rho_{(2)} (t) \propto a^{-4} (t)$ while most of the power in fluctuations is on sub-Hubble scales. This behavior for the fluctuations contrasts with the scaling of $\bar{\rho} (t)$ at early times, when the energy density associated with $\varphi (t)$ is dominated by kinetic energy, such that $\bar{\rho} (t) \propto a^{-6} (t)$. The backreaction of $\delta \rho_{(2)} (t)$ on $H(t)$ ensures that $\varphi (t)$ will traverse less distance en route to the slow-roll inflationary attractor than in the absence of inhomogeneities, thereby accounting for the differences we observe in the duration of inflation, $N_{\rm infl}$.

In our numerical analysis we initialize fluctuations at $t_0$ across the range of wavenumbers from $k_{\rm min} / a(t_0) \lesssim 2 H_0 / 3$ up to $k_{\rm max} / a(t_0) = 30 \, k_{\rm min} /  a (t_0) \sim M_{\rm pl}$. Including any modes with $k > k_{\rm max}$ in the spectrum would lead to trans-Planckian ambiguities. One could nonetheless imagine initializing additional modes at later times --- to simulate the notion that modes which had begun with $k/ a(t_0) \gg M_{\rm pl}$ at $t_0$ later redshifted to $k / a (t) \leq M_{\rm pl}$ --- but we do not expect such additional, trans-Planckian modes to change the qualitative behavior of the system. In our current framework, the system consistently arrives at the slow-roll inflationary attractor while most of the initial power in sub-Hubble fluctuations remains within the Hubble radius. The energy density associated with any modes that might be initialized at later times $t > t_0$ would be less than $\delta \rho_{(2)} (t_0)$, since $H (t) < H (t_0)$ and $\delta \rho_{(2)} (t) \propto H^4 (t)$. Moreover, after a few efolds of slow-roll inflation, we expect that any newly initialized fluctuations should begin in the Bunch-Davies state, rather than in the more-energetic initial states that we consider here.

For next steps, we aim to generalize our formalism to include the evolution of systems with nonzero spatial curvature $K$, to consider small-field as well as large-field models, and to extend the formalism to multifield models (akin to Ref.~\cite{EastherMultifield}, but incorporating the coupled metric perturbations). Across each of these cases, we believe the approach we have developed here can complement the computationally intensive numerical-relativity approaches of Refs.~\cite{East:2015ggf,Clough:2016ymm,Clough:2017efm}.

\appendix

\section{Discrete Spectrum for Mode Functions}
\label{sec:AppendixADiscrete}

As described in Section \ref{sec:Parameters}, we expand quantized field fluctuations $\delta \hat{ \phi} \left( x^{\mu} \right)$ and metric perturbations $\hat{\Psi} \left( x^{\mu} \right)$ in eigenfunctions $Z_{n \ell m} \left( \mathbf{x} \right)$ of the comoving spatial Laplacian. Within a comoving spatial volume of finite size, Eq.~(\ref{nablaZ}) then takes the form
\begin{equation}
\mathbf{\nabla}^{2} Z_{n \ell m} (r, \theta, \phi) = - k_{n \ell}^{2} \, Z_{n \ell m} (r, \theta, \phi),
\label{nablaZfinite}
\end{equation}
with positive integer $n \geq 1$. We restrict attention to a finite sphere of comoving radius $R$ within a spatially flat FLRW background spacetime ($K = 0$). We select Dirichlet boundary conditions $Z_{n \ell m} (R, \theta, \phi) = 0$, which causes $\mathbf{\nabla}^{2}$ to have a negative definite spectrum as desired, and for the resulting Sturm-Liouville system to yield a complete, orthonormal basis:
\begin{equation}
\int_{r < R} d^{3} \mathbf{x} \sqrt{ h ( \mathbf{x} ) } \, Z_{n \ell m} ( \mathbf{x} ) \, Z^{*}_{n^{\prime} \ell^{\prime} m^{\prime}} ( \mathbf{x} ) = \delta_{n n^{\prime}} \delta_{ \ell \ell^{\prime}} \delta_{m m^{\prime}} .
\label{Zorthofinite}
\end{equation}
In spherical polar coordinates, Eq.~(\ref{nablaZfinite}) becomes
\beq 
\frac{1}{r^2} \frac{\partial}{\partial r} \left( r^2 \frac{ \partial Z_{n \ell m} }{ \partial r} \right) - \frac{1}{ r^2} {\bf L}^2 Z_{n \ell m} + k_{n \ell}^2 Z_{n \ell m} = 0 ,
\label{nablaZfinite2}
\eeq
where 
\beq 
{\bf L}^2 = - \frac{1}{ \sin \theta} \frac{ \partial}{\partial \theta} \left( \sin \theta \frac{ \partial }{\partial \theta} \right) - \frac{1}{ \sin^2 \theta} \frac{ \partial^2}{\partial \phi^2} .
\label{Ldef}
\eeq 
The eigenfunctions of the operator ${\bf L}^2$ are the familiar spherical harmonics $Y_{ \ell m} (\theta, \phi)$, which satisfy ${\bf L}^2 Y_{\ell m} = \ell (\ell + 1) \, Y_{\ell m}$. Solutions to Eq.~(\ref{nablaZfinite2}) may then be written in the form
\beq 
Z_{n \ell m} (r, \theta, \phi) = N_{n \ell m} \, j_{\ell} (k_{n \ell} r) \, Y_{\ell m} (\theta, \phi) ,
\label{Zfinite1}
\eeq 
where $N_{n \ell m}$ is a normalization constant, $j_\ell (x)$ is a spherical Bessel function of order $\ell$, and the boundary conditions require
\beq 
j_\ell (k_{n \ell} R ) = 0 .
\label{jellboundary}
\eeq
The requirement of Eq.~(\ref{jellboundary}) yields a discrete spectrum of allowable wavenumbers,
\beq 
k_{n \ell} (R) = \frac{ x_{n\ell} }{R} 
\label{knl2}
\eeq
where, as noted below Eq.~(\ref{knl}), $x_{n \ell}$ is the $n$th zero of the Bessel function $j_\ell (x)$ for $n \geq 1$. Making use of the orthonormality properties of the $Y_{\ell m} (\theta, \phi)$, Eq.~(\ref{Zorthofinite}) then becomes
\beq 
N_{n \ell m} \, N_{n' \ell m} \int_0^R dr \, r^2 \, j_{\ell} (k_{n \ell} r) \, j_\ell (k_{n' \ell} r) = \delta_{n n'}.
\label{Zorthofinite2}
\eeq
Upon using Eqs.~(11.49) and (11.50) of Ref.~\cite{Arfken}, we find
\beq 
N_{n \ell m} = \frac{ \sqrt{2}}{R^{3/2} } \frac{1}{ \vert j_{\ell + 1} (k_{n \ell} R) \vert} ,
\label{Nnlm}
\eeq 
with (as usual) $j_{\ell} (k_{n \ell} R) = 0$. Note that the normalization $N_{n \ell m}$ is independent of $m$. The basis functions $Z_{n \ell m} (r, \theta, \phi)$ provide spectral convergence to any spatial configuration as we take $n_{\rm max} \to \infty$.

Given the spatial eigenfunctions $Z_{n \ell m} ({\bf x})$ and the properties of the operators $\hat{a}_{n \ell m}$, $\hat{a}^\dagger_{n \ell m}$, we may evaluate various two-point functions. For example, we have
\beq 
\langle 0 \vert \delta \hat{\phi} (t, {\bf x} ) \, \delta \hat{\phi} (t, {\bf y} ) \vert 0 \rangle = \sum_{n \ell m} \vert \delta \phi_{n \ell m} (t) \vert^2 Z_{n \ell m} ({\bf x}) \, Z^*_{n \ell m} ({\bf y}) .
\label{twopoint1}
\eeq 
The term $\langle (\delta \hat{\phi} )^2 \rangle$, which appears throughout the equations of motion in the Hartree approximation, is Eq.~(\ref{twopoint1}) in the limit ${\bf x} \rightarrow {\bf y}$. As our slicing of spacetime has been chosen such that spatial slices are homogeneous and isotropic (this holds even inside the finite sphere due to completeness), we may evaluate Eq.~(\ref{twopoint1}) at ${\bf x} = {\bf y} = {\bf 0}$. From
\beq 
j_{\ell} (z) \rightarrow \frac{ z^\ell}{1 \cdot 3 \cdot 5 \cdot ... \cdot (2 \ell + 1) } \quad {\rm for} \> z \rightarrow 0
\label{jellzsmall}
\eeq
we note that only the $\ell = 0$ mode remains nonzero at the origin, with amplitude $j_0 (0) = 1$. We further note that $Y_{00} (\theta, \phi) = 1/\sqrt{ 4 \pi}$ for $\ell = m = 0$, and hence we find
\beq 
\langle (\delta \hat{\phi} )^2 \rangle = \frac{ \pi}{2R^3} \sum_n n^2 \vert \delta \phi_{n 00} (t) \vert^2 ,
\label{twopoint2}
\eeq
as in Eq.~(\ref{twopointR}), prior to applying the UV regularization. The same steps yield $\langle (\delta \dot{\phi} )^2 \rangle$ as in Eq.~(\ref{twopointR}).

The remaining two-point function of interest is the contribution to $\delta \rho_{(2)}$ from the spatial gradients. We find
\beq 
\begin{split}
    \langle (\nabla \delta \hat{\phi} )^2 \rangle &= \sum_{n \ell m} \vert \delta \phi_{n \ell m} (t) \vert^2 \\
    &\quad \times \bigg\{ h^{rr} \partial_r Z_{n \ell m} ({\bf 0})  \, \partial_r Z^*_{n \ell m} ({\bf 0}) \\
    &\quad\quad\quad  + h^{\theta \theta} \partial_\theta Z_{n \ell m} ({\bf 0}) \, \partial_\theta Z^*_{n \ell m} ({\bf 0}) \\
    &\quad\quad \quad + h^{\phi \phi} \partial_\phi Z_{n \ell m} ({\bf 0}) \, \partial_\phi Z^*_{n \ell m} ({\bf 0} ) \bigg\} .
\end{split}
\label{twopointspatial1}
\eeq
Using Eq.~(\ref{jellzsmall}), we see that only the term with $\ell = 1$ will contribute to the first term in brackets within Eq.~(\ref{twopointspatial1}). From the properties of $Y_{1m} (0, \phi)$ and $N_{n \ell m}$ in Eq.~(\ref{Nnlm}), we then find
\beq 
h^{rr} \partial_r Z_{n \ell m} ({\bf 0}) \, \partial_r Z^*_{n \ell m} ({\bf 0}) = \frac{1}{ 6 \pi R^3} \frac{ k_{n1}^2}{ \vert j_2 (k_{n 1 } R ) \vert^2 } \delta_{\ell, 1} \, \delta_{m,0} .
\label{twopointspatial2}
\eeq
For the last two terms in brackets in Eq.~(\ref{twopointspatial1}), we require that the terms be well-behaved in the vicinity of the origin. Given $h^{\theta \theta} = 1/r^2$ and the properties of the $Y_{\ell m} (\theta, \phi)$, the only contribution to the second term in brackets that will remain regular (and nonzero) near ${\bf x} \rightarrow {\bf 0}$ comes from $\ell = 1$, and we find
\beq 
\begin{split}
\sum_{n \ell m} h^{\theta \theta} \, & \partial_\theta Z_{n \ell m} ({\bf 0} ) \, \partial_\theta Z^*_{n \ell m} ({\bf 0}) \\
&= \sum_n \frac{1}{ 12 \pi R^3} \frac{ k_{n1}^2 }{ \vert j_2 (k_{n 1} R) \vert^2 } \delta_{\ell, 1} \left[ \delta_{m, 1} + \delta_{m, -1} \right] .
\end{split}
\label{twopointspatial3}
\eeq
Proceeding similarly, we find
\beq
\begin{split}
\sum_{n \ell m} h^{\phi \phi} \, &\partial_\phi Z_{n \ell m} ({\bf 0}) \, \partial_\phi Z^*_{n \ell m} ({\bf 0} ) \\
&= \sum_n \frac{1}{ 12 \pi R^3} \frac{ k_{n1}^2 }{ \vert j_2 (k_{n1} R ) \vert^2 } \delta_{\ell, 1} \left[ \delta_{m, 1} + \delta_{m, -1} \right] .
\end{split}
\label{twopointspatial4}
\eeq 
Combining Eqs.~(\ref{twopointspatial1}) - (\ref{twopointspatial4}), we find
\beq
\langle (\nabla \delta \hat{\phi} )^2 \rangle = \frac{1}{ 6 \pi R^3} \sum_n \sum_{m = -1, 0, 1} \frac{ k_{n1}^2} {\vert j_2 (k_{n1} R ) \vert^2 } \vert \delta \phi_{n 1 m} (t) \vert^2 ,
\label{twopointspatial5}
\eeq
as in Eq.~(\ref{twopointR}).

\section{Initial Conditions for the Field Fluctuations}
\label{sec:initialconditions}

To establish the initial conditions for the mode functions $\delta \phi_{n\ell m} (t_0)$ and $\delta \dot{\phi}_{n \ell m} (t_0)$, we first consider the case of $R \rightarrow \infty$, for which the spectrum of allowable wavenumbers is continuous, with $0 \leq k < \infty$. As noted in Section \ref{sec:Parameters}, within the Hartree approximation higher-order interaction terms among the fluctuations $\delta \hat{\phi}$ are replaced by an effective mass, so we may proceed by following most of the steps for quantizing a free, massive scalar field in FLRW spacetime.

The equal-time commutation relation for a free scalar field stipulates
\beq
[ \delta \hat{\phi} (t, {\bf x} ) , \delta \hat{\Pi} (t, {\bf y} ) ] = i \delta^{(3)} ({\bf x} - {\bf y} ) ,
\label{ETCR}
\eeq
where 
\beq
\delta \hat{\Pi} \equiv \frac{ \partial {\cal L} }{ \partial \, \delta \dot{\hat{\phi}} } = a^3 (t) \, \delta \dot{\hat{\phi}}
\label{Pi}
\eeq
is the momentum canonically conjugate to $\delta \hat{\phi}$. Upon expanding $\delta \hat{\Pi} (x^\mu)$ in a series of complex mode functions $\delta \Pi_{k \ell m} (t)$ and creation and annihilation operators akin to Eq.~(\ref{deltaphiexpand}), and using the commutation relations for $\hat{a}_{k \ell m}$ and $\hat{a}^\dagger_{k \ell m}$ in Eq.~(\ref{acommute}), Eq.~(\ref{ETCR}) imposes a constraint on the mode functions:
\beq
\delta \phi_{k \ell m} \, \delta \Pi^*_{k \ell m} - \delta \phi_{k \ell m}^* \, \delta \Pi_{k \ell m} = i .
\label{WronskianPi}
\eeq
We rescale the modes $\delta \phi_{k \ell m} (t) = v_{k \ell m} (t) / a(t)$ and introduce conformal time, $d \tau \equiv dt / a(t)$, so that 
\beq
\delta \Pi_{k \ell m} (\tau) = a (\tau) (v_{k \ell m}' - {\cal H} v_{k \ell m}),
\label{deltaPimode}
\eeq
where primes denote $d / d\tau$ and ${\cal H} \equiv a' / a$. Then Eq.~(\ref{WronskianPi}) becomes
\beq
v_{k \ell m} \, v_{k \ell m}^{* \prime} - v_{k \ell m}^* \, v_{k \ell m}' = i ,
\label{Wronskianv}
\eeq
a Wronskian condition that we will use when setting initial conditions at $\tau_0$.

In terms of $v_{k \ell m} (\tau)$, Eq.~(\ref{eomdeltaphi}) takes the form
\beq
v''_{k \ell m} + \omega_k^2 (\tau) \, v_{k \ell m} = S_{k\ell m} (\tau) ,
\label{veom}
\eeq
where the source term $S_{k \ell m}$ depends on $\Psi_{k \ell m}$ and $\Psi_{k \ell m}'$, and the frequency is given by
\beq
\omega_k^2 (\tau) = k^2 + a^2(\tau) \, m_{\rm eff}^2 (\tau) - \frac{a'' (\tau) }{a (\tau)},
\label{omegadef}
\eeq
with the effective mass $m_{\rm eff}^2$ given in Eq.~(\ref{meff}). The effective mass is suppressed by a small coupling constant $\lambda$ and hence we expect it to remain subdominant around $\tau_0$, while, on dimensional grounds, we expect $a'' / a \sim {\cal O} (H^2)$ around $\tau_0$. When we numerically evolve the modes within a region of finite comoving radius $R$ in our simulation, we introduce an infrared cut-off given by $k_{\rm min} = \pi / R$, with $R \sim 1/[ a (\tau_0) H (\tau_0)]$. Hence even for the longest-wavelength modes in our simulation, we expect $k^2 \gtrsim \{a^2 (\tau_0) \, m_{\rm eff}^2 (\tau_0) , a'' (\tau_0) / a (\tau_0) \}$. For setting initial conditions, we therefore approximate $\omega_k (\tau_0) \sim k$.

We do not assume that the system has attained its minimum energy state at the initial time $\tau_0$, and hence we consider initial conditions for the modes that could depart from the usual Bunch-Davies vacuum state. (Note that as we are describing a weakly interacting system, the Bunch-Davies state that we construct is only an approximation of the lowest-energy state, which we do not attempt to find.) We parameterize the initial conditions for the modes as
\beq
\begin{split}
v_{k \ell m} (\tau_0) &= \frac{1}{\sqrt{2k}} \left( \alpha_{k \ell m} + i \beta_{k \ell m} \right) , \\
v_{k\ell m}' (\tau_0) &= - i \sqrt{\frac{k}{2}} \left(\gamma_{k \ell m} + i \delta_{k \ell m} \right) ,
\end{split}
\label{valpha}
\eeq
where $\alpha_{k \ell m}$, $\beta_{k \ell m}$, $\gamma_{k \ell m}$, and $\delta_{k \ell m}$ are each real-valued, dimensionless constants. (In the limit $\omega_k \rightarrow k$, the Bunch-Davies state corresponds to $\alpha_{k \ell m} = \gamma_{k \ell m} = 1$, $\beta_{k \ell m} = \delta_{k \ell m} = 0$.) The Wronskian condition of Eq.~(\ref{Wronskianv}) then becomes
\beq
\alpha_{k \ell m} \gamma_{k \ell m} + \beta_{k \ell m} \delta_{k \ell m} = 1 .
\label{Wronskianalpha}
\eeq
Without loss of generality, we may set $\beta_{k \ell m} = 0$ for all modes; then Eq.~(\ref{Wronskianalpha}) fixes $\alpha_{k \ell m} = 1/ \gamma_{k \ell m}$. Setting $a (\tau_0) = 1$, Eqs.~(\ref{deltaPimode}) and (\ref{valpha}) yield
\beq
\begin{split}
\delta \phi_{k \ell m} (t_0) &= \frac{ \alpha_{k \ell m} }{\sqrt{2 k} } , \\
\delta \dot{\phi}_{k \ell m} (t_0) &= \sqrt{ \frac{k}{2} } \left( - i \gamma_{k \ell m} + \delta_{k \ell m} - \frac{ \alpha_{k \ell m} \bar{H}_0}{k} \right) .
\end{split}
\label{deltaphiICs}
\eeq
We use $\bar{H}_0$ (rather than $H_0$) in Eq.~(\ref{deltaphiICs}) because, as a practical matter, we must first select a point in $(\varphi_0, \dot{\varphi}_0)$ and then initialize the fluctuations. The parameter $\bar{H}_0$, defined in Eq.~(\ref{barH}), is determined by the selection $(\varphi_0, \dot{\varphi}_0)$; only after the fluctuations are initialized can we evaluate $\delta\rho_{(2)} (t_0)$ and thereby include their contribution to $H_0$. (For the regimes of interest, $H_0 / \bar{H}_0 \lesssim 1.4$, so the difference is not significant.) When we evolve the system within a finite sphere of comoving radius $R$, the continuous variable $k$ in Eqs.~(\ref{Wronskianalpha}) and (\ref{deltaphiICs}) is replaced by the discrete spectrum of wavenumbers $k_{n \ell}$ with $n \geq 1$, which yields Eq.~(\ref{deltaphiICn}).

Given these initial conditions for the modes, we may estimate quantities of interest such as $\langle (\delta \hat{\phi} (t_0) )^2 \rangle$ and $\delta \rho_{(2)} (t_0)$. We derive our expressions in the continuum limit using the UV regulator function $F_{n \ell} (\kappa, R)$ of Eq.~(\ref{Fregulator}). For the initial value of the two-point function, we begin with the expansion for $\delta \hat{\phi} (x^\mu)$ in Eq.~(\ref{deltaphiexpand}), use Eq.~(\ref{acommute}), make use of the properties of $Z_{k \ell m} ({\bf x})$ at ${\bf x} \rightarrow {\bf 0}$, and substitute $\delta \phi_{k \ell m} (t_0)$ from Eq.~(\ref{deltaphiICs}) to write
\beq 
\begin{split}
\langle (\delta \hat{\phi} & (t_0) )^2  \rangle \\
&= \int dk \sum_{\ell, m} \vert \delta \phi_{k \ell m } (t_0) \vert^2 \vert Z_{k \ell m} ({\bf 0} ) \vert^2 \, e^{ - k^2 / (2 \kappa^2) } \\
&=\int_0^\infty \frac{ dk}{2 \pi^2} k^2 \left( \frac{ \alpha_{k 00}^2}{ 2 k} \right) e^{ -k^2 / (2 \kappa^2) } \\
&= \left( \alpha^2 \right)_{\rm avg} \frac{ \kappa^2}{ 4 \pi^2 } . 
\end{split}
\label{twopointalpha1}
\eeq
The third line follows because the random coefficients $\alpha_{k \ell m}$ vary independently of $k$, so we may take the term $\alpha_{k 00}^2$ outside the integral and replace it with an average value; moreover, since we draw the variables $\gamma_{k \ell m} = 1 / \alpha_{k \ell m}$ from the same distribution for all $(k \ell m)$, the same average value $(\alpha^2)_{\rm avg}$ holds for any $k$, $\ell$, and $m$. Upon using $\kappa = b \bar{H}_0$, we arrive at Eq.~(\ref{twopointreg2}).

We proceed similarly to estimate $\delta\rho_{(2)} (t_0)$. Using Eqs.~(\ref{deltaphiexpand}), (\ref{acommute}), (\ref{rhodeltarho}), (\ref{Fregulator}), and (\ref{deltaphiICs}), we find
\beq 
\begin{split}
    \delta &\rho_{(2)} (t_0) \\
    &= \frac{1}{2} \int dk \sum_{\ell, m} \vert \delta \dot{\phi}_{k \ell m} (t_0) \vert^2 \vert Z_{k \ell m} ({\bf 0}) \vert^2 \, e^{ - k^2  /(2 \kappa^2 ) } \\
    &\>\> + \frac{1}{ 2 a^2 (t_0) } \int dk \sum_{\ell, m} \vert \delta \phi_{k \ell m} (t_0) \vert^2 \vert \nabla Z_{k \ell m} ({\bf 0} ) \vert^2 \, e^{ - k^2 / (2 \kappa^2 ) } \\
    &= \frac{1}{2} \int dk \frac{ k^2}{2 \pi^2} \frac{ k}{2} \bigg\{ \gamma_{k00}^2 + \delta_{k00}^2 \\
    &\quad\quad\quad\quad\quad\quad\quad\>  - \frac{2 \alpha_{k00} \delta_{k00} \bar{H}_0 }{k} + \frac{ \alpha_{k00}^2 \bar{H}_0^2}{k^2 } \bigg\} \, e^{-k^2 / (2 \kappa^2)} \\
    &\quad + \frac{1}{ 2 a^2 (t_0) } \int dk \frac{ k^2}{6 \pi^2} \sum_{m = -1, 0, 1} k^2 \left( \frac{ \alpha_{k1m}^2 }{2k}\right)  e^{-k^2 / (2 \kappa^2)} \\
    &= \frac{1}{8 \pi^2} \int dk  \bigg\{ \left( \alpha^2 + \gamma^2 + \delta^2 \right)_{\rm avg}  k^3 \\
    &\quad\quad\quad\quad\quad - 2 \left( \alpha \delta \right)_{\rm avg} \bar{H}_0 k^2 + \left( \alpha^2 \right)_{\rm avg} \bar{H}_0^2 k \bigg\} e^{-k^2 / (2 \kappa)^2} \\
    &= \frac{ \kappa^4}{4 \pi^2} \bigg\{ \left( \alpha^2 + \gamma^2 + \delta^2 \right)_{\rm avg} \\
    &\quad\quad\quad\quad - \left( \alpha \delta \right)_{\rm avg} \sqrt{\frac{\pi}{2}} \left( \frac{ \bar{H}_0}{\kappa} \right) + \frac{ \left( \alpha^2 \right)_{\rm avg} }{2} \left( \frac{ \bar{H}_0 }{\kappa} \right)^2 \bigg\} ,
\end{split}
\label{deltarho2continuum}
\eeq 
where we have used $a (t_0) = 1$ and assumed that $m_{\rm eff}^2 (t_0) \ll H_0^2$ in the regimes of interest, so that the contributions to $\delta \rho_{(2)} (t_0)$ are dominated by the kinetic energy and spatial-gradient energy of the fluctuations. Given $\kappa = b \bar{H}_0 > \bar{H}_0$, we find Eq.~(\ref{deltarhoICn}).

The magnitudes of $\langle (\delta \hat{\phi} (t_0) )^2 \rangle$ and $\delta \rho_{(2)} (t_0)$ depend on the random variables $\alpha_{n \ell m}$, $\gamma_{n \ell m}$, and $\delta_{n \ell m}$ for each mode. We draw $\gamma_{n \ell m}$ and $\delta_{n \ell m}$ from flat distributions within specific ranges; once $\gamma_{n \ell m}$ is selected, $\alpha_{n \ell m}$ is fixed by the quantization condition to be $\alpha_{n \ell m} = 1 / \gamma_{n \ell m}$. If we draw $\gamma_{n \ell m}$ from the range $\{ A, B \}$ and $\delta_{n \ell m}$ from the range $\{ -C, C \}$, then we expect
\beq 
\begin{split} 
\left( \gamma^2 \right)_{\rm avg} &= \frac{1}{ (B - A) } \int_A^B dx \, x^2 = \frac{1}{3} \left( A^2 + AB + B^2 \right) , \\
\left( \alpha^2 \right)_{\rm avg} &= \frac{1}{(B - A) } \int_A^B dx \frac{ 1}{x^2} = \frac{1}{AB} , \\
\left( \delta^2 \right)_{\rm avg} &= \frac{1}{ 2C} \int_{-C}^C dx \, x^2 = \frac{ C^3}{3} .
\end{split}
\label{alphaaverage}
\eeq
In order to study large initial fluctuations, with ${\cal C} \equiv \left( \alpha^2 + \gamma^2 + \delta^2 \right)_{\rm avg} \simeq 20$, we therefore use the ranges in Eq.~(\ref{gammadeltaranges}).

\section{Initializing the Metric Perturbations}
\label{sec:MetricInitial}

In this appendix, we discuss subtleties in the initialization of $\Psi_{n \ell m} (t_0)$, and the techniques we use to avoid these issues numerically.

Given an instantiation of $\delta \phi_{n \ell m} (t_0)$ and $\delta \dot{\phi}_{n \ell m} (t_0)$, the metric perturbations $\Psi_{n\ell m} (t_0)$ are initialized using Eq.~(\ref{constraintPsi}). For our numerical simulations, we evolve the system within a sphere of comoving radius $R$, within which Eq.~(\ref{constraintPsi}) takes the form
\beq
\begin{split}
&\left[ \dot{H} + \frac{2}{3 M_{\rm pl}^2 a^2} \langle ( \nabla \delta \hat{\phi} )^2  \rangle + \frac{1}{ a^2} \left( k_{n \ell}^2 - 3K \right) \right] \Psi_{n \ell m} \\
&\quad\quad = \frac{1}{ 2M_{\rm pl}^2} \left[ \ddot{\varphi} \, \delta \phi_{n \ell m} - \dot{\varphi} \, \delta \dot{\phi}_{n \ell m} \right] .
\end{split}
\eeq
As $\dot{H}$ tends to be large and negative at early times, it is common for there to exist a value for $k_{n \ell}$ at which the coefficient of $\Psi_{n \ell m}$ vanishes at $t_0$: we dub the $k_{n \ell}$ value at which this occurs the ``$\Psi$ pole.'' For $k_{n \ell}$ near the pole, $\Psi_{n \ell m} (t_0)$ would be initialized with an artificially large initial amplitude, which can cause a single mode to dominate the nonlinear contributions.

The pole exists as an artifact of the longitudinal gauge, and arises because the four-dimensional phase space ($\delta \phi_{n \ell m}$, $\delta \dot{\phi}_{n \ell m}$, $\Psi_{n \ell m}$ and $\dot{\Psi}_{n \ell m}$) has two constraints, Eqs.~(\ref{dotPsi}) and (\ref{constraintPsi}). When $k_{n \ell}$ is on the $\Psi$ pole, the constraint surface in phase space forces $\delta \phi_{n \ell m}$ and $\delta \dot{\phi}_{n \ell m}$ to be strictly proportional to each other. In this situation, we do not lose a degree of freedom (the constraint surface remains two-dimensional); it is simply that $\delta \phi_{n \ell m}$ and $\delta \dot{\phi}_{n \ell m}$ cease to be appropriate coordinates on the constraint surface.

We emphasize that this does not arise due to our use of the Hartree approximation, but rather through imposing canonical commutation relations on $\delta \phi_{n \ell m}$ and $\delta \dot{\phi}_{n \ell m}$ at $t_0$ under the assumption that they are free fields. Within linear perturbation theory,
this issue is typically overcome by using Mukhanov-Sasaki variables, $Q_{n \ell m} \equiv \delta \phi_{n \ell m} + (\dot{\varphi} / H) \Psi_{n \ell m}$, which are good coordinates on the phase space. However, the $Q_{n \ell m}$ are no longer gauge-invariant beyond linear order in perturbations. If one were to apply the (nonlinear) Hartree corrections to the $Q_{n \ell m}$, the resulting expressions that went beyond ${\cal O} (\Psi )$ would not bear any clear relationship to an expansion of the Einstein tensor $G_{\mu\nu}$ beyond ${\cal O} (\Psi)$.


The most correct way to approach this issue would be to employ Dirac's constrained Hamiltonian formalism to the constrained phase space, but such an analysis is beyond the scope of this work. For our present purposes, we adopt methods to numerically alleviate the issue of the $\Psi$ pole.

We can compute the boost given to a particular mode $\Psi_{n \ell m} (t_0)$ as a result of being near the pole as follows. Taking $K = 0$, $a (t_0) = 1$, and identifying the pole location as
\begin{align}
k_0^2 \equiv - \dot{H} - \frac{2}{3 M_{\rm pl}^2 } \langle ( \nabla \delta \hat{\phi} )^2 \rangle,
\end{align}
we have
\beq
\begin{split}
\Psi_{n \ell m} (t_0) &= \frac{\bar{H}_0^2}{k_{n \ell}^2 - k_0^2} \frac{1}{2M_{\rm pl}^2 \bar{H}_0^2} \\
&\quad\quad \times \left[ \ddot{\varphi}_0 \, \delta \phi_{n \ell m} (t_0) - \dot{\varphi}_0 \, \delta \dot{\phi}_{n \ell m} (t_0) \right] ,
\end{split}
\eeq
where factors of $\bar{H}_0$ (the relevant scale) have been inserted to make the coefficient dimensionless. We set a threshold such that any instantiation with a boost of $\bar{H}_0^2/(k_{n \ell}^2 - k_0^2) > 10$ for any $\ell = 0$ or $\ell = 1$ modes was too close to the pole for reliable results. For such instantiations, we randomly changed $R = 1.5 \pi / \bar{H}_0$ up or down by up to 10\% (which changes the $k_{n \ell} (R)$ spectrum of the modes), and re-initialized all variables. The new position of the $k_{n \ell}$ values typically meant that no modes fell too close to the pole; if necessary, we repeated the process. We emphasize that these shifts to avoid the $\Psi$ pole are only necessary for setting initial conditions at $t_0$. Dynamically, we evolve the modes $\Psi_{n \ell m} (t)$ according to Eq.~(\ref{dotPsi}), which is well behaved.

When constructing initial surfaces for $\Psi (t_0, {\bf x})$ (as in Fig.~\ref{PsiPos}), this technique for avoiding the pole proved insufficient, as such surfaces required modes for $\ell = 0, 1, \ldots, \ell_{\rm max}$, where $\ell_{\rm max}$ was typically taken to be around 12. (Every mode $\Psi_{n \ell m} (t)$ evolves according to Eq.~(\ref{dotPsi}), though only modes $\delta \phi_{n \ell m}$ with $\ell = 0, 1$ contribute to the Hartree corrections.) The spectrum of all modes in such cases forms a rather dense forest, so shifting the spectrum (by adjusting $R$) simply shifts the pole from one mode to another. To ameliorate the pole in this situation, we multiplied $\bar{H}_0^2/(k_{n \ell}^2 - k_0^2)$ by a regulating function of $k_{n \ell}^2$ with a double zero at $k_0^2$ that takes a value of unity away from the pole. The transition width was chosen to be sufficiently narrow so that the complete factor never grew sufficiently large for one mode to dominate the plots. This is not a physical regularization, but serves to suppress modes near the pole for the purposes of visualization, and is used only to construct $\Psi (t_0, {\bf x})$.

In order to construct an initial spatial representation of $\Psi(t_0, r, \theta, \phi)$, we begin by sampling the initial conditions for $\delta \phi_{n \ell m} (t_0)$ and $\delta \dot{\phi}_{n \ell m} (t_0)$, and use these to construct $\Psi_{n \ell m} (t_0)$ (with regulator, if necessary). We then turn to the mode expansion for $\hat{\Psi} (x^\mu)$,
\beq
\begin{split}
\hat{\Psi} (x^\mu) &= \sum_{n, \ell, m} \left[ \Psi_{n \ell m}(t) \, \hat{a}_{n\ell m} \, Z_{n\ell m}(r, \theta, \phi) + H.c. \right].
\end{split}
\label{discretepsisum}
\eeq
This is a quantum-mechanical expansion (as evidenced by the $\hat{a}_{n \ell m}$ and $\hat{a}^\dagger_{n \ell m}$ operators). To construct a classical realization, we demote the operators to classical complex random variables $\tilde{a}_{n \ell m}$ and $\tilde{a}^*_{n \ell m}$ with the same statistics as the quantum operators.

For a given mode, the quantum operator is $\hat{\Psi}_{n \ell m} = \Psi_{n \ell m}(t) \, Z_{n \ell m} \, \hat{a}_{n \ell m} + H.c.$, so we have $\langle \hat{\Psi}_{n \ell m} \rangle = 0$ and $\langle \hat{\Psi}^2_{n \ell m} \rangle = |Z_{n \ell m}|^2 |\Psi_{n \ell m} (t)|^2$. (We only need to go to the two-point function as we have Gaussian statistics.) Demanding the same statistics for a classical function $\tilde{\Psi}_{n \ell m} = \Psi_{n \ell m} (t) \, Z_{n \ell m} \, \tilde{a}_{n \ell m} + c.c.$, we require
\begin{align}
\langle \tilde{a}_{n \ell m}^2 \rangle = \langle \tilde{a}_{n \ell m}^{*2} \rangle = 0,
\qquad
\langle \tilde{a}_{n \ell m} \tilde{a}^*_{n \ell m} \rangle = \frac{1}{2}.
\end{align}
This is achieved by constructing $\tilde{a}_{n \ell m} = x + i y$ for each mode, where $x$ and $y$ are independent Gaussian random variables with zero mean and standard deviation $\sigma_x = \sigma_y = 1/2$. Constructing the spatial slice is then straightforwardly accomplished by randomly sampling $\tilde{a}_{n \ell m}$, after which Eq.~\eqref{discretepsisum} may be evaluated. Identical techniques can be used to construct initial surfaces for $\delta \phi (t, {\bf x})$.

\section{Numerical Convergence Tests}
\label{sec:NumConvergence}

In this appendix, we discuss the accuracy of our simulations.

First, consider the accuracy of our numerical integration. The RK45 integrator uses a variable time step; early on in the integration, it takes small steps so as to capture the oscillations of the modes, and after the last mode freezes out, it takes increasingly large steps. We cap the maximum time step to have an estimated $\Delta N$ of 0.1, in terms of efolds $N = \ln a$. We employed relative and absolute tolerances of $10^{-10}$. Investigations showed this to give a good trade-off between accuracy, numerical precision and computational time. Our measurements of the duration of inflation, $N_{\rm infl}$, for a given run are completely dominated by uncertainty as to when to start measuring $N_{\rm infl}$ rather than precision from the numerical evolution.

Second, our simulations cannot simulate the entire continuous universe, but only a discrete portion thereof. Our analysis essentially consists of three tuneable parameters describing this discretization: $\kappa$, $k_{\rm max}$ and $R$. The continuum limit takes each of these parameters to approach infinity, but such lies beyond our computational power. As such, we now discuss the convergence of our simulations as each of these parameters is increased. For each of these convergence tests, we look for convergence in the two-point functions and overall efolds for Bunch-Davies simulations, for which comparisons with analytic results are possible.

The outer boundary $R$ controls both the spectrum of the modes and the weight for each mode's contribution to the two-point functions. As the two-point functions are computed by summing over $k$ modes, taking $R \to \infty$ corresponds to taking the limit of the Riemann sum, yielding the continuous limit. Holding the upper limit $k_{\rm max}$ fixed, we expect spectral convergence with increasing $R$ based on the properties of the basis functions we employ. Unfortunately, the presence of the $\Psi$ pole (see Appendix~\ref{sec:MetricInitial}) means that as $R$ increases, modes get pushed closer to the pole, eventually leading to divergent values for the two-point functions. Although our trick of shifting $R$ to avoid the pole provides reasonable initializations (see Appendix \ref{sec:MetricInitial}), we see consistency in our results as $R$ increases, but not convergence. To properly observe convergence will require resolving the issue of the $\Psi$ pole. As such, we need to find a balance between avoiding the $\Psi$ pole and having sufficient modes both inside and outside the horizon at $t_0$. 

\begin{figure}[tb]
    \centering
    \includegraphics[width=3.41in]{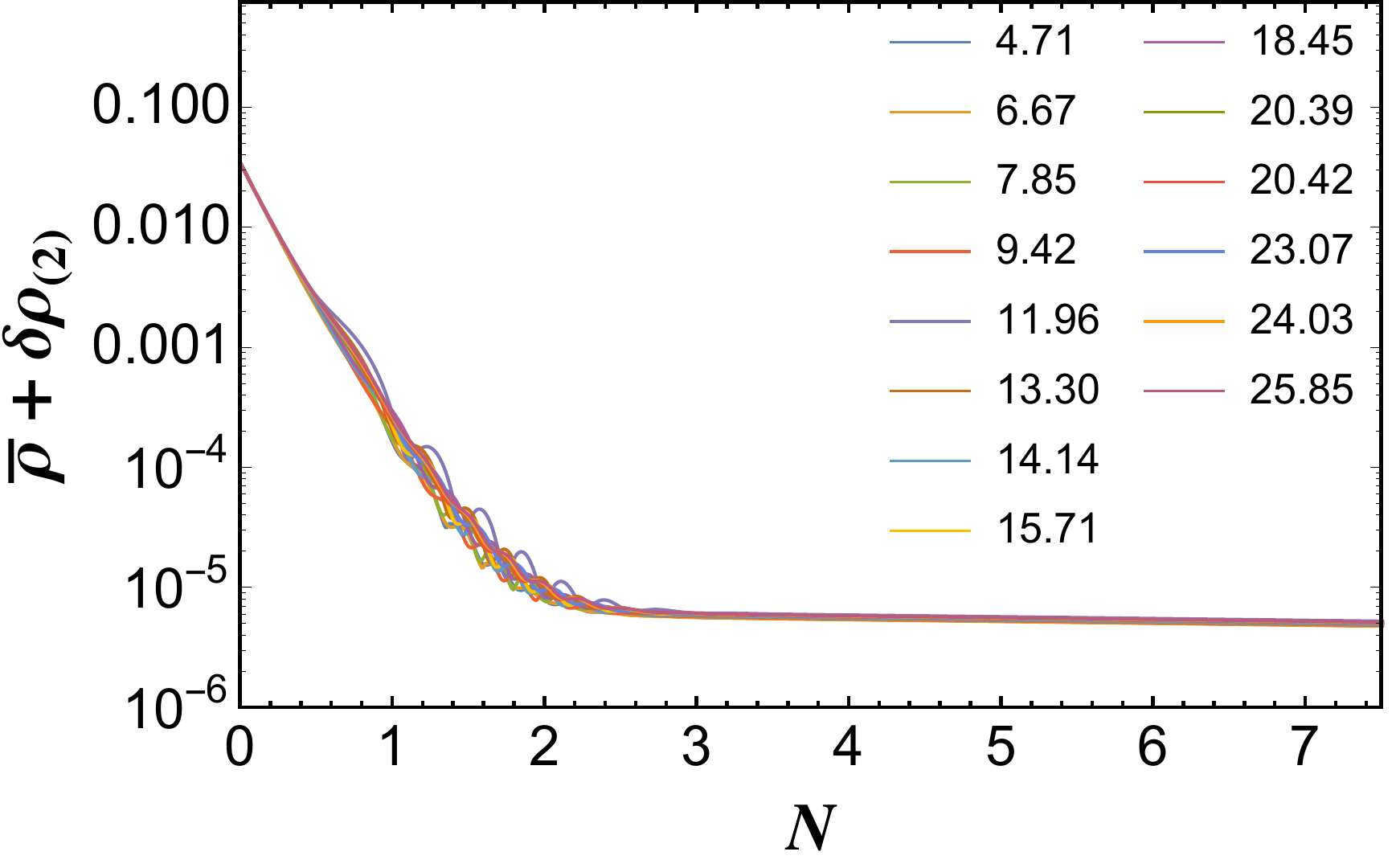}
    \caption{\small The quantity $\rho_{\rm total} = \bar{\rho} + \delta \rho_{(2)}$ versus $N$ as we vary $R = c / \bar{H}_0$. We fix $(\varphi_0, \dot{\varphi}_0) = (25 \, M_{\rm pl}, -0.25 \, M_{\rm pl}^2)$ and initialize the field fluctuations in the Bunch-Davies state. Throughout the simulations described in our paper, we fixed $c = 1.5 \pi \approx 4.71$. We found similar results as we increased $c$. In particular, for these initial conditions we found $N_{\rm total} = 63.01 \pm 0.78$ as we varied $4.71 \leq c \leq 25.85$, indicating minimal dependence of the evolution of the dynamical system on our choice of $R$. }
    \label{fig:Rtest}
\end{figure}

Nonetheless, we can test for the dependence of various numerical quantities on our selection of $R$, keeping $\lambda = 10^{-10}$, $\kappa = 5 \bar{H}_0$, and $k_{\rm max} = 4 \kappa$ fixed. When we initialize the fluctuations in the Bunch-Davies initial state and parameterize $R = c / \bar{H}_0$, we find that quantities such as $\rho_{\rm total} = \bar{\rho} + \delta \rho_{(2)}$ show only modest variation as we vary $c$ between the fiducial value we used throughout our simulations, $c = 1.5 \pi = 4.71$, up through $c = 25.85$, as shown in Fig.~\ref{fig:Rtest}. For initial conditions $(\varphi_0, \dot{\varphi}_0 ) = (25 \, M_{\rm pl}, -0.25 \, M_{\rm pl}^2)$, we found $N_{\rm total} = 63.01 \pm 0.78$ as we varied $4.71 \leq c \leq 25.85$, indicating minimal dependence of the evolution of the dynamical system on our choice of $R$.

The regulator $\kappa = b \bar{H}_0$ is used to constrain the contribution to the two-point functions from small wavelengths. From theoretical analysis, we expect the two-point functions to diverge with increasing $b$, as shown in Section \ref{sec:Parameters}. The expected divergence is observed numerically. In order to demonstrate results that are immune to changing $\kappa$ requires the implementation of a renormalization scheme, which is beyond the scope of this paper.

Finally, the wavenumber cutoff $k_{\rm max}$ represents the shortest length-scale that we allow to contribute to the two-point functions. So long as $k_{\rm max}$ is chosen to be sufficiently larger than $\kappa$, we observe the expected exponential convergence with increasing $k_{\rm max}$. We chose $k_{\rm max} = 4 \kappa$, which corresponds to suppressing the initial contributions of the shortest wavelength modes to the various two-point functions by a factor of $\sim 3 \times 10^{-4}$.

\section*{Acknowledgements}

We are grateful to Katy Clough, John T. Giblin, Jr., Alan H. Guth, Tracy Slatyer, Vincent Vennin, and Matias Zaldarriaga for helpful discussions. Portions of this work were conducted in MIT's Center for Theoretical Physics and supported in part by the U.S. Department of Energy under Contract No.~DE-SC0012567. In addition, PF was partially supported by an NSF Graduate Research Fellowship, and KH by MIT's Undergraduate Research Opportunities Program (UROP).


%

\end{document}